\documentclass{gGAF2e}
\usepackage{color}
\usepackage{nicefrac}
\usepackage{xspace}
\def\ga{\mathrel{\mathchoice {\vcenter{\offinterlineskip\halign{\hfil
$\displaystyle##$\hfil\cr>\cr\sim\cr}}}
{\vcenter{\offinterlineskip\halign{\hfil$\textstyle##$\hfil\cr
>\cr\sim\cr}}}
{\vcenter{\offinterlineskip\halign{\hfil$\scriptstyle##$\hfil\cr
>\cr\sim\cr}}}
{\vcenter{\offinterlineskip\halign{\hfil$\scriptscriptstyle##$\hfil\cr
>\cr\sim\cr}}}}}

\def\la{\mathrel{\mathchoice {\vcenter{\offinterlineskip\halign{\hfil
$\displaystyle##$\hfil\cr<\cr\sim\cr}}}
{\vcenter{\offinterlineskip\halign{\hfil$\textstyle##$\hfil\cr  
<\cr\sim\cr}}}
{\vcenter{\offinterlineskip\halign{\hfil$\scriptstyle##$\hfil\cr
<\cr\sim\cr}}}
{\vcenter{\offinterlineskip\halign{\hfil$\scriptscriptstyle##$\hfil\cr
<\cr\sim\cr}}}}}

\newcommand{\nat}{Nature}  
  
\newcommand{\jcp}{J. Comp. Phys.}  
  
\newcommand{\pre}{Phys. Rev. E}

\newcommand{\apjs}{Astrophys. J. Supp.}
\renewcommand{\vec}[1]{\mbox{\boldmath$#1$}}

\def\beq{\begin{equation}}
\def\eeq{\end{equation}}

\def\berom{\begin{enumerate}[{\rm(i)}]}
\def\eerom{\end{enumerate}}
\def\bRem{\begin{Rem} \setcounter{itemrem}{0}}
         \newcounter{itemrem}
         
\def\eRem{\end{Rem}}

\catcode`\@=11
\@addtoreset{equation}{section}

\newcommand{\SCAL}{{\cdot}}

\newcommand{\DIV}{\nabla\!{\cdot}}   
\newcommand{\ROT}{\nabla{\times}}  
\newcommand{\GRAD}{\nabla}           
\newcommand{\LAP}{{\Delta}}          

\def\Hsp1d{{{   \bf H}^{s+1}   (\Omega)}}

\newcommand{\ie}{i.e.,\@\xspace}

\def\lb{[\![}
\def\rb{]\!]}
\def\jump#1{\lb{#1}\rb}

\begin{document}
\doi{10.1080/0309192YYxxxxxxxx}
\issn{1029-0419}
\issnp{0309-1929}
\jvol{00} \jnum{00} \jyear{2008} \jmonth{January}

\title{Electromagnetic induction in non-uniform domains.}
\author{A. Giesecke$^{\rm a}$\thanks{$^\ast$Corresponding
    author. Email: a.giesecke@fzd.de \vspace{6pt}} , C. Nore$^{\rm
    b,c}$, F. Luddens$^{\rm b,e}$, F. Stefani$^{\rm a}$, G. Gerbeth$^{\rm a}$,
  J. L\'eorat$^{\rm d}$, J.-L. Guermond$^{\rm b,e}$
  \\
  \vspace{6pt} $^{\rm a}${\em{Forschungszentrum Dresden-Rossendorf,
      Dresden, Germany}}
  \\
  \vspace{6pt} $^{\rm b}${\em{Laboratoire d'Informatique pour la
      M\'ecanique et les Sciences de l'Ing\'enieur, CNRS, BP 133,
      91403 Orsay cedex, France}}
  \\
  \vspace{6pt} $^{\rm c}${\em{Universit\'e Paris Sud 11, 91405 Orsay
      cedex, France et Institut Universitaire de France}}
  \\
  \vspace{6pt} $^{\rm d}${\em{Luth, Observatoire de Paris-Meudon,
      place Janssen, 92195-Meudon, France}}
  \\
  \vspace{6pt} $^{\rm e}${\em{Department of Mathematics, Texas A\&M
      University 3368 TAMU, College Station, TX 77843, USA}} }

\maketitle
\begin{abstract}
Kinematic simulations of the induction equation are carried out for different
setups suitable for the von-K\'arm\'an-Sodium (VKS) dynamo experiment.
Material properties of the flow driving impellers are considered by means of
high conducting and high permeability disks that are present in a
cylindrical volume filled with a conducting fluid.
Two entirely different numerical codes are mutually validated by showing
quantitative agreement on Ohmic decay and kinematic dynamo problems using
various configurations and physical parameters.
Field geometry and growth rates are strongly modified by the material
properties of the disks even if the high permeability/high
conductivity material is localized within a quite thin region.
In contrast the influence of external boundary conditions remains small.

Utilizing a VKS like mean fluid flow and high permeability disks yields a reduction
of the critical magnetic Reynolds number for the onset of dynamo action of the
simplest non-axisymmetric field mode. However this decrease is not
sufficient to become relevant in the VKS experiment. 
Furthermore, the reduction of ${\rm{Rm}}^{\rm{c}}$ is
essentially influenced by tiny changes in the flow configuration so that the
result is not very robust against small modifications of setup and properties
of turbulence.
\end{abstract}

\begin{keywords}
{Magnetohydrodynamics, Ohmic decay, kinematic Dynamo, Permeability, VKS dynamo}
\end{keywords}

\section{Introduction}
Magnetic fields of galaxies, stars or planets are produced by homogenous
dynamo action in which a conducting fluid flow provides for generation and
maintenance of field energy. 
During the past decade the understanding of the field generation mechanism has  
considerably benefitted from the examination
of dynamo action in the laboratory.
However, realization of dynamo action at
least requires the magnetic Reynolds number ${\rm{Rm}}=UL/\eta$
to exceed a threshold of the order
of ${\rm{Rm}^{\rm{crit}}}\sim 10...100$. 
From the parameter values  of liquid sodium -- the best known liquid conductor --
at standard laboratory conditions (i.e. $T\approx 200^{\circ}$C,
$\eta=1/\mu_0 \sigma \approx 0.1\mbox{m}^2/\mbox{s}$ and $\mathcal{L}\approx 1\mbox{m}$,
where $\mu_0$ is the vacuum permeability and $\sigma$ the electrical conductivity) it
becomes immediately obvious that self excitation of magnetic fields in the
laboratory needs typical velocity magnitudes of $U\sim
10\mbox{m/s}$, which is already quite demanding.
Therefore, the first successful dynamo experiments performed by
\cite{1963Natur.198.1158L, 1968Natur.219..717L} utilized soft-iron
material so that the magnetic diffusivity is reduced and the magnetic Reynolds
number is (at least locally) increased.
Although these experiments cannot be classified as hydromagnetic dynamos (no
fluid flow and therefore no backreaction of the field on a fluid motion) they
allowed the examination of 
distinct dynamical regimes manifested in steady, oscillating or reversing
fields.
It is interesting to note that these results did not initiate further numerical
studies on induction in the presence of soft iron domains.

A possibility to increase the effective magnetic
Reynolds number in fluid flow driven dynamo experiments arises from the addition of
tiny ferrous particles to the fluid medium leading to an uniform  enlargement
of the relative permeability \citep{2002EPJB...25..399F,
  2003PhRvE..67e6309D}.   
To retain reasonable fluid properties the amount of particles added to
the liquid is limited so that an upper bound for the achievable fluid
permeability is given by $\mu_{\rm{r}}\approx 2$. 
The main effect found in the simulations of \cite{2003PhRvE..67e6309D} was a reduced decay of the
initial field but not a smaller threshold (essentially because of
nonmonotonous behavior of the growth rate in dependence of ${\rm{Rm}}$). 

Another type of ferromagnetic influence on dynamo action is
observed in the von-K\'arm\'an-sodium (VKS) dynamo. 
In the VKS experiment a turbulent flow of liquid sodium is driven
by two counterotating impellers located at the opposite end caps of a
cylindrical domain \citep{2007PhRvL..98d4502M}. 
Dynamo action is only obtained when the impellers
are made of soft-iron with $\mu_{\rm{r}} \sim 100$ \citep{2010NJPh...12c3006V}.
Recently it has been shown in \cite{PhysRevLett.104.044503} that these soft-iron impellers essentially
determine the geometry and the growth rates of the magnetic field by locally
enhancing the magnetic Reynolds number and by enforcing internal 
boundary conditions for the magnetic field at the material interfaces in terms
of jump conditions.  
Furthermore, gradients of the material coefficients $\mu_{\rm{r}}$ and $\sigma$
might support dynamo action because corresponding additional terms in the
induction equation couple toroidal and poloidal fields which is essential for
the occurrence of dynamo action. 
An example for this dynamo type has been presented in 
\cite{1992gafd...64..135B} where it was shown that even a straight flow over an (infinite)
conducting plate with sinusoidal variation of the conductivity is able to
produce dynamo action. 
However, the experimental realization of this setup would require either
an unachievable large magnetic Reynolds number or a rather large variation in the conductivity ($\ga $ factor of 100)
whereas the mean value should not be too far away from the conductivity of the
fluid.
In order to obtain semi-homogenous dynamo action it might be more promising to
replace the conductivity variation by a permeability variation because the relative
permeability of soft-iron alloys easily attains values of several thousands.
Although such dynamos are of little astrophysical relevance 
the experiments of Lowes and Wilkinson and in particular the rich dynamical
behavior in the VKS dynamo demonstrate the usefulness
of such models. 

The scope of the present work is to validate the numerical tool necessary to
establish a basic understanding of the 
influence of material properties on the induction process. 
Emphasis is given to the problem of the free decay in cylindrical geometry
where two disks characterized by high conductivity/permeability and
their thickness are inserted in the interior of a cylindrical container filled
with a conducting fluid.
To demonstrate the reliability of our results
we use two different numerical approaches and show that both methods give
results in agreement. 
The study is completed by an application of a mean
flow as it occurs in the VKS experiment in combination with
two high permeability disks. 
%
%
%
%
%
\section{Induction equation in heterogenous domains}
From Faraday's Law in combination with Ohm's Law one immediately retrieves the
induction equation that determines the temporal behavior of the magnetic flux
density $\vec{B}$ (often abbreviated with magnetic field):
\begin{equation}
\frac{\partial\vec{B}}{\partial t}
= \nabla\times(\vec{u}\times\vec{B}
-\frac{1}{\mu_0\sigma}\nabla\times\frac{\vec{B}}{\mu_{\rm{r}}}).
\label{eq::induction}
\end{equation}
In Eq.~(\ref{eq::induction}) $\vec{u}$ denotes the flow velocity, $\sigma$
the electric conductivity, $\mu_0$ the vacuum permeability and $\mu_{\rm{r}}$ the relative
permeability. 
In case of spatially varying distributions of conductivity and permeability
Eq.~(\ref{eq::induction}) can be rewritten:
\begin{eqnarray}
\frac{\partial\vec{B}}{\partial
  t}&=&\nabla\times(\vec{u}\times\vec{B})+\frac{1}{\mu_0\mu_{\rm{r}}\sigma}\Delta\vec{B}
+\frac{1}{\mu_0\mu_{\rm{r}}\sigma}
\nabla\times(\nabla\ln\mu_{\rm{r}}\times\vec{B})
\\&&
-\frac{1}{\mu_0\mu_{\rm{r}}\sigma}(\nabla\ln
\mu_{\rm{r}}+\nabla\ln\sigma)\times (\nabla\ln\mu_{\rm{r}}\times\vec{B})+\frac{1}{\mu_0\mu_{\rm{r}}\sigma} (\nabla\ln
\mu_{\rm{r}}+\nabla\ln\sigma)\times(\nabla\times\vec{B})\nonumber
\label{eq::general_induction}
\end{eqnarray}
The terms on the right-hand-side that involve gradients of $\mu_{\rm{r}}$
and $\sigma$ couple the toroidal and poloidal field
components which is known to be essential for the existence of a dynamo.
The lack of symmetry between the terms containing $\mu_{\rm{r}}$ and $\sigma$
indicates a distinct impact of $\sigma$ and $\mu_{\rm{r}}$ which is also manifested in the
jump conditions for electric field and magnetic field that have to be
fulfilled at material interfaces.
At interfaces between materials 1 and 2 that exhibit a jump in conductivity
$\sigma$ and/or in relative permeability $\mu_{\rm{r}}$ the normal component
of the magnetic flux density is continuous whereas the tangential components
exhibit a jump described by the ratio of the permeabilities. 
In case of conductivity discontinuities, the tangential components of the electric field are continuous
and the normal component of the electric current is continuous. If there is no contribution of the flow,
the continuity of the normal current leads to the discontinuity of the normal electric
field in the ratio of the conductivities.
Mathematically these jump conditions are given by (see
e.g. \citeauthor{1975clel.book.....J}, \citeyear{1975clel.book.....J}): 
\begin{equation}
\begin{array}{ccl}
\vec{n}\cdot(\vec{B}_1-\vec{B}_2)&=&0,\\[0.4cm]
\displaystyle
\vec{n}\times\left(\frac{\vec{B}_1}{\mu_{{\rm{r}},1}}-\frac{\vec{B}_2}{\mu_{\rm{r},2}}\right)&=&0,\\[0.6cm]
\vec{n}\cdot(\vec{j}_1-\vec{j}_2)&=&0,\\[0.5cm]
\vec{n}\times(\vec{E}_1-\vec{E}_2)&=&0,
\label{eq::jumpconditions}
\end{array}
\end{equation}
where $\vec{n}$ denotes the unit vector in the normal direction on the interface
between materials 1 and 2. Although these transmission conditions are standard,
their dynamical consequences in flows at large ${\rm{Rm}}$ are largely unknown.
\section{Numerical schemes}
Two different numerical algorithms and codes are used for the numerical
solution of problems involving the kinematic
induction equation (\ref{eq::induction}). 
A combined finite volume/boundary element method (FV/BEM) is a grid based approach
which provides a flexible scheme that utilizes a local discretization and intrinsically maintains
the solenoidal character of the magnetic field.

The second solution method is based on a
Spectral/Finite Element approximation technique denoted
SFEMaNS for Spectral/Finite Elements for Maxwell and Navier-Stokes equations. 
Taking advantage of the cylindrical
symmetry of the domains, Fourier modes are used in the azimuthal
direction and finite elements are used in the meridional plane. 
For each Fourier mode this leads to independent two-dimensional-problems in
the meridian plane.   
\subsection{Hybrid finite volume/boundary element method}
We start with the induction equation in conservative form:
\begin{equation}
\frac{\partial\vec{B}}{\partial t} + \nabla\times\vec{E}=0 \label{Faraday}
\end{equation}
where the electric field $\vec{E}$ is given by 
\begin{equation}
\vec{E}=-\vec{u}\times\vec{B}+\eta\nabla\times\frac{\vec{B}}{\mu_{\rm{r}}} \label{Ohm}
\end{equation}
and $\eta=1/\mu_0 \sigma$ is the magnetic diffusivity.
For the sake of simplicity we give a short sketch for the treatment of
inhomogeneous conductivity and permeability only in Cartesian coordinates. The
scheme can easily be adapted to different (orthogonal) coordinate systems
(e.g. cylindrical or spherical coordinate system) making use of generalized
coordinates \citep{1992ApJS...80..753S, 1992ApJS...80..791S}. 

\begin{figure}[h!]
\includegraphics[width=10cm,angle=0]{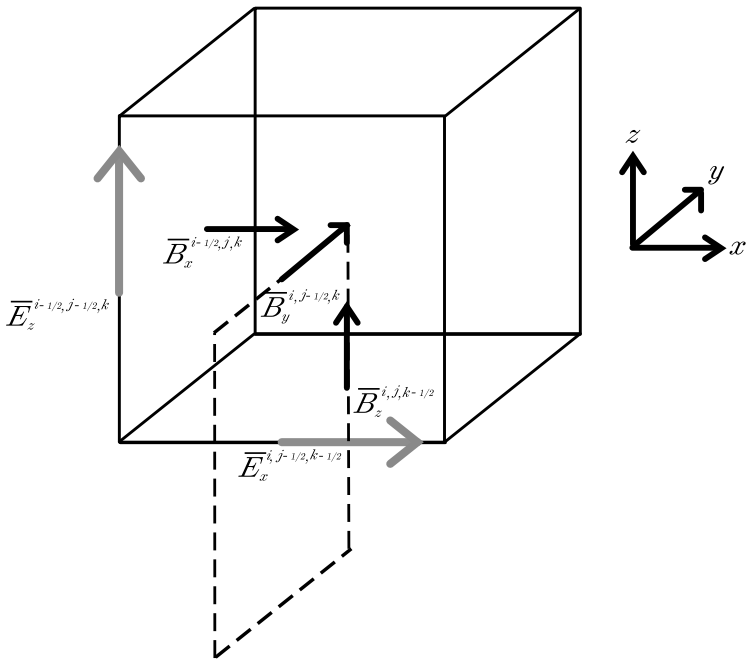}
\caption{Localization of vector quantities on a grid cell $ijk$ with the cell
  center located at $(x_i, y_j, z_k)$. The dotted curve denotes the path along
  which the integration of $\vec{B}$ is executed for the computation of
  ${\overline{E}}_x^{i, j-\frac{1}{2}, k-\frac{1}{2}}$.}\label{fig1}
\end{figure}

In the finite volume scheme the grid representation of the magnetic field is
given by a staggered collocation of the field components that are interpreted
as an approximation of the (cell-)face average:
\begin{equation}
\overline{B}^{i-\frac{1}{2},j,k}_x\approx \frac{1}{\Delta y \Delta
  z}\int\limits_{\Gamma_{yz}}B_x(x_{i-\frac{1}{2}},y,z)dydz 
\end{equation}
where the integration domain $\Gamma$ corresponds to the surface of a single cell-face:
$\Gamma_{yz}=[y_{j-\frac{1}{2}},y_{j+\frac{1}{2}}]\times[z_{k-\frac{1}{2}},z_{k+\frac{1}{2}}]$
(see Fig.~\ref{fig1}). 
A comparable definition is applied for the electric field that is localized 
at the center of a cell edge and which is defined as the line average (see Fig.~\ref{fig1}):
\begin{equation}
\overline{E}_x^{i,j-\frac{1}{2},k-\frac{1}{2}}\approx
\frac{1}{\Delta x}
\int\limits_{x^{i-\frac{1}{2}}}^{x^{i+\frac{1}{2}}}E_x(x,y_{j-\frac{1}{2}},z_{k-\frac{1}{2}})dx
\end{equation}
Similar definitions hold for the components ${\overline{B}}^{i,j-\frac{1}{2},k}_y$
and ${\overline{B}}^{i,j,k-\frac{1}{2}}_z$, respectively for
${\overline{E}}^{i-\frac{1}{2},j,k-\frac{1}{2}}_y$  and ${\overline{E}}^{i-\frac{1}{2},j-\frac{1}{2},k}_z$. 

The finite volume discretization of the induction equation reads 
\begin{eqnarray}
\frac{d}{dt}{\overline{B}}^{i-\frac{1}{2},j,k}_x&=&-\left(\frac{{\overline{E}}^{i-\frac{1}{2},j+\frac{1}{2},k}_z(t)
-\overline{E}_z^{i-\frac{1}{2},j-\frac{1}{2},k}(t)}{\Delta y}-
\frac{{\overline{E}}^{i-\frac{1}{2},j,k+\frac{1}{2}}_y(t)
-\overline{E}_y^{i-\frac{1}{2},j,k-\frac{1}{2}}(t)}{\Delta z}\right)
\end{eqnarray}
and it can easily been
shown that this approach preserves the $\nabla\cdot\vec{B}$ constraint for all
times (to machine accuracy) if the initial field is divergence free.

\subsubsection{Material coefficients}
In the following we only discuss the treatment of the diffusive part of the
electric field, $\vec{E}=+\eta\nabla\times{\vec{B}}/{\displaystyle\mu_r}$
because the advective contributions $(\propto -\vec{u}\times\vec{B})$ do not
involve the material properties and can be treated separately in the framework
of an operator splitting scheme (see
e.g. \citeauthor{2004JCoPh.197..540I} \citeyear{2004JCoPh.197..540I},
\citeauthor{2008giesecke_maghyd} \citeyear{2008giesecke_maghyd}, \citeauthor{1999CoPhC.116...65Z}
\citeyear{1999CoPhC.116...65Z}).
To obtain the computation directive for the electric field the
magnetic field has to be integrated along a (closed
path) around ${\overline{E}}_{x(,y,z)}$ at the edge of a grid cell (see dotted
curve in Fig.~\ref{fig1}). 
\begin{equation}
\overline{E}_x\approx
\frac{1}{\Gamma}\int\limits_{\Gamma_{yz}} {E}_x dA=
\frac{1}{\Gamma}\int\limits_{\Gamma_{yz}}\eta\left(\nabla\times\frac{\vec{B}}{\mu_{\rm{r}}}\right)dA
\approx\frac{\overline{\eta}}{\Delta y\Delta z}\int\limits_{\partial\Gamma_{yz}}\frac{\vec{B}}{\mu_{\rm{r}}}d\vec{l}
\label{eq::fv_ediff}
\end{equation}
where $\Gamma=\Delta y\Delta z$ is the surface surrounded by the path
$\Gamma_{yz}$ and $\overline{\eta}$ is the average diffusivity
($\eta=(\mu_0\sigma)^{-1}$) "seen" by the electric field.
Unlike vectorial quantities the material coefficients are scalar quantities
that are localized in the center of a grid cell. 
The consideration of spatial variations and/or jumps in conductivity respectively permeability is
straightforward if corresponding averaging procedures for $\sigma$ or
$\mu_{\rm{r}}$ are applied \citep{2001siam...22..1943H}. 
For the component ${\overline{E}}_x$ the discretization of
Eq.~(\ref{eq::fv_ediff}) leads to: 
\begin{equation}
\overline{{E}}_x^{i,j-\frac{1}{2},k-\frac{1}{2}}=
\overline{\eta}_{i,j-\frac{1}{2},k-\frac{1}{2}}\left[
\frac{1}{\Delta y}\!\left(
\frac{\overline{{B}}_z^{i,j,k-\frac{1}{2}}}{(\overline{\mu}_{{\rm{r}}})_{i,j,k-\frac{1}{2}}}
-\frac{\overline{{B}}_z^{i,j-1,k-\frac{1}{2}}}{(\overline{\mu}_{{\rm{r}}})_{i,j-1,k-\frac{1}{2}}}
\!\right)
\!-\frac{1}{\Delta z}\!\left(
\!\frac{\overline{{B}}_y^{i,j-\frac{1}{2},k}}{(\overline{\mu}_{{\rm{r}}})_{i,j-\frac{1}{2},k}}
-\frac{\overline{{B}}_y^{i,j-\frac{1}{2},k-1}}{(\overline{\mu}_{{\rm{r}}})_{i,j-\frac{1}{2},k-1}}
\!\right)\right].\label{eq::disc_efield}
\end{equation}
In Eq.~(\ref{eq::disc_efield}), $\overline{\eta}_{i,j-\frac{1}{2},k-\frac{1}{2}}$ represents the
diffusivity that is seen by the
electric field component ${\overline{E}}_x^{i,j-\frac{1}{2},k-\frac{1}{2}}$ at the edge of
the grid cell $(ijk)$ and which is
given by the arithmetic average of the diffusivity of the four adjacent cells:
\begin{equation}
{\overline{\eta}}_{i,j-\frac{1}{2},k-\frac{1}{2}}=\frac{\eta_{i,j,k}+\eta_{i,j-1,k}
+\eta_{i-1,j,k}+\eta_{i-1,j-1,k}}{4}.
\label{eq::eta_average}
\end{equation}
Similarly, ${\overline{\mu_{\rm{r}}}}$ denotes the relative
permeability that is seen by the magnetic field
components ($\overline{B}_y$ and $\overline{B}_z$)  at the interface between
two adjacent grid cells which for example reads for the case given in Eq.~(\ref{eq::disc_efield}):
\begin{eqnarray}
{\mbox{for $\overline{B}_y^{i,j-\frac{1}{2},k}$: }}\quad{(\overline{\mu}}_{\rm{r}})_{i, j-\frac{1}{2},k}&=&
\frac{2(\mu_{\rm{r}})_{i,j,k}(\mu_{\rm{r}})_{i,j-1,k}}{(\mu_{\rm{r}})_{i,j,k}+(\mu_{\rm{r}})_{i,j-1,k}},
\nonumber\\
{\mbox{for $\overline{B}_z^{i,j,k-\frac{1}{2}}$: }}\quad{(\overline{\mu}}_{\rm{r}})_{i, j,k-\frac{1}{2}}&=&
\frac{2(\mu_{\rm{r}})_{i,j,k}(\mu_{\rm{r}})_{i,j,k-1}}{(\mu_{\rm{r}})_{i,j,k}+(\mu_{\rm{r}})_{i,j,k-1}}.
\label{eq::mu_average}
\end{eqnarray}
For the computation of ${\overline{E}}_y^{i-\frac{1}{2},j,k-\frac{1}{2}}$ and
${\overline{E}}_z^{i-\frac{1}{2},j-\frac{1}{2},k}$  Eq.~(\ref{eq::eta_average}) and Eq.~(\ref{eq::mu_average}) have to be
adjusted according to the localization and the involved field components. 
Applying the averaging rules (\ref{eq::eta_average}) and
(\ref{eq::mu_average}) for the computation of the "diffusive" part of the
electric field results in a scheme that intrinsically fulfills the jump
conditions~(\ref{eq::jumpconditions}) at material interfaces. The scheme is
robust and simple to implement, however, the averaging procedure results in a
artificial smoothing of parameter jumps at interfaces and in concave corners additional difficulties might occur caused by ambiguous
expressions for $\mu_{\rm{r}}$. Furthermore in the simple realization as presented above,
the parameter range is restricted. 
For larger jumps of $\mu_{\rm{r}}$ or
$\sigma$ a more careful treatment of the discontinuities at
the material interfaces is necessary which would require a more elaborate
field reconstruction that makes use of slope limiters.
\subsubsection{Boundary conditions}
In numerical simulations of laboratory dynamo action insulating boundary
conditions are often treated adopting simplified 
expressions like vanishing tangential fields (VTF, sometimes also called pseudo
vacuum condition).
This ambiguous denomination may be understood only for convecting flows at
small magnetic Reynolds number where the induced fields remain small compared
to a given field.
In fact, a restriction of the boundary magnetic field to its normal component
resembles an 
artifical but numerically convenient setup where the exterior of the
computational domain is 
characterized by an infinite permeability.
In case of VTF boundary conditions the calculation of field growth rates for typical
dynamo problems are overestimated.
Therefore a more elaborate treatment of the field behavior on the boundary is
recommended which is nontrivial in non-spherical coordinate systems.
Insulating domains are characterized by a vanishing current $\vec{j}\propto
  \nabla\times 
  \vec{B}=0$ so that $\vec{B}$ can be expressed as the gradient of a scalar field
$\varPhi$ (in the case of a simply connected vacuum)
which fulfills the Laplace equation:
\begin{equation}
\vec{B}=-\nabla\varPhi \quad\mbox{  with  }\quad \Delta\varPhi =0, \quad
\varPhi \rightarrow O(r^{-2}) \mbox{ for } r\rightarrow\infty.
\label{eq::laplace}
\end{equation}
Integrating  $\Delta \varPhi=0$ and adoption of Green's 2nd theorem leads to 
\begin{equation}
\varPhi(\vec{r})=2\int\limits_{\Gamma}
  G(\vec{r},\vec{r}')\underbrace{\frac{\partial \varPhi(\vec{r}')}{\partial 
  n}}_{\displaystyle -B^{\rm{n}}(\vec{r}')}-\varPhi(\vec{r}')\frac{\partial
  G(\vec{r},\vec{r}')}{\partial n}d\Gamma(\vec{r}'). \label{eq::bie_phi}
\end{equation}
where $G(\vec{r}, \vec{r}')=-(4\pi\left|\vec{r}-\vec{r}'\right|)^{-1}$ is the Greens function 
(with $\Delta G(\vec{r},\vec{r}')=-\delta(\vec{r}-\vec{r}')$) and
$\nicefrac{\partial}{\partial n}$ represents the derivative in the normal
direction on the surface element  
$d\Gamma$ so that
$\partial_n\varPhi=-B^{\rm{n}}$ yields the normal component of
$\vec{B}$ on $d\Gamma$.
From Eq.~(\ref{eq::bie_phi}) 
the tangential components of the magnetic field on the boundary
$B^{\rm{\tau}}=\vec{e}_{\tau}\cdot\vec{B}=-\vec{e}_{\tau}\cdot\nabla\varPhi(\vec{r})$
are computed by:
\begin{equation}
{B}^{\tau}=2\int\limits_{\Gamma}\vec{e}_{\tau}\cdot\left(\varPhi(\vec{r}')\nabla_{r}\frac{\partial
  G(\vec{r},\vec{r}')}{\partial
  n}+B^{\rm{n}}(\vec{r}')\nabla_{r}G(\vec{r},\vec{r}')\right)d\Gamma(\vec{r}')
\label{eq::bie_b}
\end{equation}
where $\vec{e}_{\tau}$ represents the tangential unit vector on the surface
element $d\Gamma(\vec{r}')$.
After the subdivision of the surface
$\Gamma$ in boundary elements $\Gamma_j$ with $\Gamma=\cup\Gamma_j$
the potential $\varPhi_i=\varPhi(\vec{r_i})$ and 
the tangential field
${B}^{{\tau}}_i={B}^{{\tau}}(\vec{r}_i)=-\vec{e}_{\tau}\cdot(\nabla\varPhi_i)$
in discretized form are given by  
\begin{eqnarray}
\frac{1}{2}\varPhi_i&=&-\sum\limits_j\left({\int\limits_{\Gamma_j}\frac{\partial G}{\partial
    n}(\vec{r}_i,\vec{r}'){\rm{d}}\Gamma_j'}\right)\varPhi_j,
-{\sum\limits_j}\left({\int\limits_{\Gamma_j}G(\vec{r}_i,\vec{r}'){\rm{d}}\Gamma_j'}\right)B^{\rm{n}}_j\nonumber
\\[-0.5cm]
&&\label{eq::discretebem}
\\[-0cm]
B^{{\tau}}_i&=&\sum\limits_j\left({\int\limits_{\Gamma_j}2\vec{e}_{\tau}\cdot\nabla_{\!r}\frac{\partial
  G}{\partial n}(\vec{r}_i,\vec{r}'){\rm{d}}\Gamma_j'}\right)\varPhi_j
+\sum\limits_j{\left({\int\limits_{\Gamma_j}2\vec{e}_{\tau}
\cdot\nabla_{\!r}G(\vec{r}_i,\vec{r}'){\rm{d}}\Gamma_j'}\right)} B^{\rm{n}}_j\nonumber.
\end{eqnarray}
The solution of the system of equations~(\ref{eq::discretebem}) gives a
linear, non local relation for the tangential field components on the
boundary in terms of the normal components and closes the problem of magnetic
induction in finite (connected) domains with insulating boundaries
\citep{2005GApFD..99..481I}. 
A more detailed description of the scheme can be found in \citet{2008giesecke_maghyd}.
\subsection{Spectral/Finite Elements for Maxwell equations}
The conducting part of the computational domain
is denoted $\Omega_c$, the non-conducting part (vacuum) is denoted
$\Omega_v$, and we set $\Omega:=\Omega_c\cup\Omega_v$. We use the
subscript $c$ for the conducting part and $v$ for the vacuum. We
assume that $\Omega_c$ is partitioned into subregions
$\Omega_{c1},\cdots,\Omega_{cN}$, so that the magnetic permeability in
each subregion $\Omega_{ci}$, say $\mu^{ci}$, is smooth.  We denote
$\Sigma_\mu$ the interface between all the conducting subregions.  We
denote $\Sigma$ the interface between $\Omega_c$ and $\Omega_v$.

The electric and magnetic fields in $\Omega_c$ and $\Omega_v$ solve the
following system: 
\begin{align}
\frac{\partial(\mu^c\vec{H}^c)}{\partial t} & =  -\nabla\times\vec{E}^c, &
 \frac{\partial(\mu^v\vec{H}^v)}{\partial t} & =  -\nabla\times\vec{E}^v, 
\label{eq::hvfar}\\
 \DIV \mu^c\vec{H}^c & =  0, &
 \DIV \mu^v\vec{H}^v & =  0, \\
 \vec{E^c} & =-\vec{u}\times \mu^c\vec{H^c} + \frac{1}{\sigma} \ROT \vec{H^c}, &
 \ROT\vec{H}^v & =  0. \label{eq::hvgrad}
\end{align}
and the following transmission conditions hold across  $\Sigma_\mu$ and $\Sigma$:
\begin{align}
\vec{H}^{ci}\times\vec{n}^{ci} + \vec{H}^{cj}\times\vec{n}^{cj} & =  0, &
 \vec{H}^c\times\vec{n}^c + \vec{H}^v\times\vec{n}^v & =  0,
\label{eq::intcond1}\\
\mu^{ci}\vec{H}^{ci}\cdot\vec{n}^c + \mu^{cj}\vec{H}^{cj}\cdot\vec{n}^{cj} & =  0, &
\mu^c\vec{H}^c\cdot\vec{n}^c + \mu^v\vec{H}^v\cdot\vec{n}^v & =  0 \\
\vec{E}^{ci}\times\vec{n}^{ci} + \vec{E}^{cj}\times\vec{n}^{cj} & =  0, & 
\vec{E}^c\times\vec{n}^c + \vec{E}^v\times\vec{n}^v & =  0,  
\label{eq::intcond3}
\end{align}
where $\vec{n}^c$ (resp. $\vec{n}^v$) is the outward unit normal on
$\partial\Omega_c\cap\Sigma$
(resp. $\partial\Omega_v\cap\Sigma$), and $\vec{n}^{ci}$ is the unit outward normal
on $\partial\Omega_{ci}\cap\Sigma_\mu$.

\subsubsection{Weak formulation}
The finite element solution is computed by solving a weak form of the
system \eqref{eq::hvfar}-\eqref{eq::intcond3}.
We proceed as follows in $\Omega_{ci}$.  Multiplying the induction
equation in $\Omega_{ci}$ by a test-function $\vec{b}$, integrating
over $\Omega_{ci}$ and integrating by parts gives
\begin{align}
  0 &= \int_{\Omega_{ci}}
  \frac{\partial(\mu^{ci}\vec{H}^{ci})}{\partial t}\cdot
  \vec{b} + \int_{\Omega_{ci}} \nabla\times\vec{E}^{ci}\cdot \vec{b} \nonumber\\
  & = \int_{\Omega_{ci}}
  \frac{\partial(\mu^{ci}\vec{H}^{ci})}{\partial t}\cdot \vec{b} +
  \int_{\Omega_{ci}} \vec{E}^{ci}\cdot\nabla\times\vec{b} +
  \int_{\partial\Omega_{ci}} (\vec{n}^{ci}\times\vec{E}^{ci})\cdot\vec{b} \nonumber\\
 & =\int_{\Omega_{ci}}
  \frac{\partial(\mu^{ci}\vec{H}^{ci})}{\partial t}\cdot \vec{b} +
  \int_{\Omega_{ci}}
  \left(-\vec{u}\times\mu^{ci}\vec{H}^{ci}+\frac{1}{\sigma}\nabla\times\vec{H}^{ci}\right)\cdot\nabla\times\vec{b}
  + 
  \int_{\partial\Omega_{ci}} \vec{E}^{ci}\cdot(\vec{b}\times \vec{n}^{ci})\label{eq::wfc}
\end{align}

We proceed slightly differently in $\Omega_v$. From \eqref{eq::hvgrad}
we infer that $\vec{H}^v$ is a gradient for a simply connected vacuum, \ie $\vec{H}^v =
\nabla\phi^v$. Thus taking a test-function of the form $\nabla\psi$,
where $\psi$ is a scalar potential defined on $\Omega_v$, multiplying
\eqref{eq::hvfar} by $\nabla\psi$ and integrating over $\Omega_v$, we
obtain
\begin{equation}
  \int_{\Omega_{v}} \frac{\partial(\mu^v\nabla\phi^v)}{\partial t}\cdot \nabla\psi 
  + \int_{\Sigma} \vec{E}^v\cdot\nabla\psi\times\vec{n}^v
  + \int_{\partial\Omega} \vec{E}^v\cdot\nabla\psi\times\vec{n}^v = 0
\label{eq::wfv}
\end{equation}
We henceforth assume that $\vec{a} := \vec{E}_{|\partial\Omega}$ is a
data. Since only the tangential parts of the electric field are
involved in the surface integrals in \eqref{eq::wfc} and
\eqref{eq::wfv}, we can use the jump conditions \eqref{eq::intcond3}
to write
\begin{align*}
  \int_{\Sigma_\mu} \vec{E}^{ci}\cdot\vec{b}\times\vec{n}^{ci}
   =  \int_{\Sigma_\mu} \{\vec{E}^c\}\cdot\vec{b}\times\vec{n}^{ci}, \qquad 
  \int_{\Sigma} \vec{E}^v\cdot\nabla\psi\times\vec{n}^v  =
  \int_{\Sigma} \vec{E}^c\cdot\nabla\psi\times\vec{n}^v,
\end{align*}
where $\{\vec{E}^c\}$ is defined on $\Sigma_\mu$ by $\{\vec{E}^c\} =
\frac 12 \left(\vec{E}^{ci}+\vec{E}^{cj}\right)$.  We now add
\eqref{eq::wfc} (for $i=1, \ldots, N$) and \eqref{eq::wfv} to obtain
\begin{multline*}
 \int_{\Omega_{c}} \frac{\partial(\mu^c\vec{H}^c)}{\partial t}\cdot \vec{b} 
+ \int_{\Omega_{v}} \frac{\partial(\mu^v\nabla\phi^v)}{\partial t}\cdot \nabla\psi 
+ \int_{\cup_{i=1}^N \Omega_{ci}}\left(\frac{1}{\sigma}\nabla\times\vec{H}^{ci}-\vec{u}\times\mu^{ci}\vec{H}^{ci}\right)\cdot\nabla\times\vec{b}  \\
+ \int_{\Sigma_\mu}\{\vec{E}^c\}\cdot \jump{\vec{b}\times\vec{n}} 
+\int_{\Sigma}\vec{E}^c\cdot\left(\vec{b}\times\vec{n}^c+\nabla\psi\times\vec{n}^v\right)  
=-\int_{\partial\Omega}\vec{a}\cdot\nabla\psi\times\vec{n}^v,
\end{multline*}
where we have set $\jump{\vec{b}\times\vec{n}} := \left(\vec{b}_i\times\vec{n}^{ci}+\vec{b}_j\times\vec{n}^{cj}\right)$
with $\vec{b}_i:=\vec{b}|_{\Omega_{ci}}$ and $\vec{b}_j:=\vec{b}|_{\Omega_{cj}}$.
We finally get rid of $\vec{E}^c$ by using Ohm's law in the conductor:
\begin{align}
 \int_{\Omega_{c}} &\frac{\partial(\mu^c\vec{H}^c)}{\partial t}\SCAL
  \vec{b} + \int_{\Omega_{v}}
  \frac{\partial(\mu^v\nabla\phi^v)}{\partial t}\SCAL \nabla\psi +
  \int_{\cup_{i=1}^N
    \Omega_{ci}}\left(\frac{1}{\sigma}\ROT\vec{H}^{ci}-\vec{u}{\times}\mu^{ci}\vec{H}^{ci}\right)
  \SCAL\ROT\vec{b} \label{eq::weakform} \\
  & +\int_{\Sigma_\mu}\{\frac{1}{\sigma}\ROT\vec{H}^c-\vec{u}{\times}\mu^c\vec{H}^c\}\SCAL
  \jump{\vec{b}{\times}\vec{n}}
  +\int_{\Sigma}\left(\frac{1}{\sigma}\ROT\vec{H}^c-\vec{u}{\times}\mu^c\vec{H}^c\right)\SCAL\left(\vec{b}{\times}\vec{n}^c
+\nabla\psi{\times}\vec{n}^v\right)
  =-\int_{\partial\Omega}\vec{a}\SCAL\nabla\psi{\times}\vec{n}^v \nonumber
\end{align}
This formulation is the starting point for the finite element discretization.

\subsubsection{Space discretization}
As already mentioned, SFEMaNS takes advantage of the cylindrical
symmetry.  We denote $\Omega_v^{2d}$ and $\Omega_{ci}^{2d}$ the
meridian sections of $\Omega_v$ and $\Omega_{ci}$, respectively. These
sections are meshed using quadratic triangular meshes (we assume that
$\Omega_v^{2d}$ and the sub-domains $\Omega_{c1}^{2d} \ldots
\Omega_{cN}^{2d}$ have piecewise quadratic boundaries). We denote
$\{\mathcal F_h^v\}_{h>0}$, $\{\mathcal F_h^{c1}\}_{h>0}\ldots
\{\mathcal F_h^{cN}\}_{h>0}$ the corresponding regular families of
non-overlapping quadratic triangular meshes. For every $K$ in the mesh
we denote $T_K : \hat{K} \longrightarrow K$ the quadratic
transformation that maps the reference triangle
$\hat{K}:=\{(\hat{r},\hat{z})\in \mathbb R^2,\ 0 \le \hat{r},\ 0 \le
\hat{z},\ \hat{r} + \hat{z}\le 1\}$ to $K$. Given $\ell_H$ and
$\ell_\phi$ two integers in $\{1,2\}$ with $\ell_\phi\geqslant\ell_H$
we first define the meridian finite elements spaces
\begin{align*}
&
  \vec{X}_h^{H,2d}:=\left\{\vec{b}_h\in \vec{L}^1(\Omega_c)\;/\;
    \vec{b}_h|_{\Omega_{ci}}\in\mathcal C^0(\overline\Omega_{ci}),\;  \forall i=1,\ldots, M,\;  
    \vec{b}_h(T_K)\in \mathbb{P}_{\ell_H}, \ \forall K\in \cup_{i=1}^N \mathcal{F}_h^{ci}\right\},
\\
&  X_h^{\phi,2d}:=\left\{\psi_h\in\mathcal C^0(\overline\Omega_{v})\;/\;\psi_h(T_K)\in\mathbb{P}_{\ell_\phi}, \ \forall K\in \mathcal F_h^{v}\right\},
\end{align*}
where $\mathbb P_k$ denotes the set of (scalar or vector valued) bivariate
polynomials of total degree at most $k$. Then, using the complex notation
$\mathrm{i}^2=-1$, , the magnetic field and the scalar potential are
approximated in the following spaces: 
\begin{align*}
  & \vec{X}_h^H :=\left\{ \vec{b}_h = \sum_{m=-M}^M
    \vec{b}_h^m(r,z)\mathrm{e}^{\mathrm{i}m\theta} ;\; \forall
    m=0,\ldots,M,\; \vec{b}_h^m\in\vec{X}_h^{H,2d} \text{ and }
    \vec{b}_h^m
    = \overline{\vec{b}_h^{-m}} \right\},\\
  & X_h^\phi:= \left\{ \psi_h = \sum_{m=-M}^M
    \psi_h^m(r,z)\mathrm{e}^{\mathrm{i}m\theta} ;\; \forall
    m=0,\ldots,M,\; \psi_h^m\in\vec{X}_h^{H,2d} \text{ and } \psi_h^m
    = \overline{\psi_h^{-m}} \right\},
\end{align*}
where $M+1$ is the maximum number of complex Fourier modes.

\subsubsection{Time discretization}
We approximate the time derivatives using the second-order Backward
Difference Formula (BDF2). The terms that are likely to mix Fourier
modes are made explicit. Let $\Delta t$ be the time step and set
$t^n:=n\Delta t,\; n\geqslant 0$. After proper initialization at $t^0$
and $t^1$, the algorithm proceeds as follows. For $n\geqslant 1$ we set
\[
\vec{H}^* = 2\vec{H}^{c,n} - \vec{H}^{c,n-1},\quad \text{and} \quad
\left\{\begin{aligned}
    D\vec{H}^{c,n+1} & :=  \frac 12 \left(3\vec{H}^{c,n+1}-4\vec{H}^{c,n}+\vec{H}^{c,n-1}\right) \\
    D\phi^{v,n+1} & := \frac 12
    \left(3\phi^{v,n+1}-4\phi^{v,n}+\phi^{v,n-1}\right),
\end{aligned}\right.
\]
and the discrete fields $\vec{H}^{c,n+1}\in\vec{X}_h^H$ and $\phi^{v,n+1}\in X_h^\phi$ are
computed so that the following holds for all $\vec{b}\in\vec{X}_h^H,\; \psi\in X_h^\phi$:
\begin{equation}
  \mathcal{L}\left((\vec{H}^{c,n+1},\phi^{v,n+1}),(\vec{b},\psi)\right) = \mathcal{R}(\vec{b},\psi),
\end{equation}
where the linear for $\mathcal{R}$ is defined by
\begin{multline*}
\mathcal{R}(\vec{b},\psi)
  = -\int_{\partial\Omega}\vec{a}\SCAL\nabla\psi{\times}\vec{n}^v
  + \int_{\Omega_{c}}\vec{u}{\times}\mu^c\vec{H}^*\SCAL\ROT\vec{b} 
  +\int_{\Sigma_\mu}\{\vec{u}{\times}\mu^c\vec{H}^*\}\SCAL\jump{\vec{b}{\times}\vec{n}}
  +\int_{\Sigma}\vec{u}{\times}\mu^c\vec{H}^*\SCAL\left(\vec{b}{\times}\vec{n}^c
    +\nabla\psi{\times}\vec{n}^v\right),
\end{multline*}
the bilinear form $\mathcal{L}$ is defined by
\begin{multline*}
  \mathcal{L}\left((\vec{H}^{c,n+1},\phi^{v,n+1}),(\vec{b},\psi)\right)
  := \int_{\Omega_{c}} \mu^c\frac{D\vec{H}^{c,n+1}}{\Delta t}\SCAL
  \vec{b} + \int_{\Omega_{v}} \mu^v\frac{\nabla
    D\phi^{v,n+1}}{\Delta t}\SCAL \nabla\psi
  + \int_{\Omega_{c}}\frac{1}{\sigma}\ROT\vec{H}^{c,n+1}\SCAL\ROT\vec{b} \\
  + g\left((\vec{H}^{c,n+1},\phi^{v,n+1}),(\vec{b},\psi)\right) +
  \int_{\Sigma_\mu}\{\frac{1}{\sigma}\ROT\vec{H}^{c,n+1}\}\SCAL\jump{\vec{b}{\times}\vec{n}}
  +\int_{\Sigma}\frac{1}{\sigma}\ROT\vec{H}^{c,n+1}\SCAL\left(\vec{b}{\times}\vec{n}^c
    +\nabla\psi{\times}\vec{n}^v\right),
\end{multline*}
and the bilinear form $g$ is defined by
\begin{eqnarray*}
  g((\vec{H}_h,\psi_h),(\vec{b}_h,\psi_h)) 
  &:=& \beta_1 h_F^{-1}\int_{\Sigma_\mu} \left(\vec{H}_{h,1}{\times}\vec{n}_1^c 
    +\vec{H}_{h,2}{\times}\vec{n}_2^c  \right)\SCAL\left(\vec{b}_{h,1}{\times}\vec{n}_1^c 
    +\vec{b}_{h,2}{\times}\vec{n}_2^c\right) \\
  &\;+&\beta_2 h_F^{-1}\int_{\Sigma} \left(\vec{H}_{h}{\times}\vec{n}^c 
    +\nabla\psi_{h}{\times}\vec{n}^v  \right)\SCAL\left(\vec{b}_{h}{\times}\vec{n}^c 
    +\nabla\psi_{h}{\times}\vec{n}^v\right),
\end{eqnarray*}
where $h_F$ denotes the typical size of $\partial K\cup\Sigma_\mu$ or
$\partial K\cup\Sigma$ for all $K$ in the mesh such that $\partial
K\cup\Sigma_\mu$ or $\partial K\cup\Sigma$ is not empty. The constant
coefficients $\beta_1$ and $\beta_2$ are chosen to be of order 1. The purpose
of the bilinear form $g$ is to penalize the tangential jumps
$\jump{\vec{H}^{c,n+1}{\times}\vec{n}}$ and
$\vec{H}^{c,n+1}{\times}\vec{n}^c +\nabla\psi^{v,n+1}{\times}\vec{n}^v$,
so that they converge to zero when the mesh-size goes to zero.

\subsubsection{Addition of a magnetic pressure}
The above time-marching algorithm is convergent on finite time
intervals but may fail to provide a convergent solution in a steady state
regime since errors may accumulate on the divergence of the magnetic
induction. 
We now detail the technique which is employed to control the
divergence of $\vec{B}^c$ on arbitrary time intervals.

To avoid non-convergence properties that could occur in non-smooth
domains and discontinuous material properties, we have designed a non
standard technique inspired from \cite{AB_JLG_FL} to control $\DIV\vec{B}$. 
We replace the induction equation in $\Omega_{ci}$, $i=1,\ldots,N$,
by the following 
\begin{equation}
\frac{\partial(\mu^{ci}\vec{H}^{ci})}{\partial t} = -\nabla\times\vec{E}^{ci}
+ \mu^{ci}\nabla p^{ci},\quad (-\Delta_0)^{\alpha}p^{ci} = \DIV\mu^{ci}\vec{H}^{ci}, \quad p^{ci}|_{\partial\Omega_{ci}} = 0.
\label{eq::inductionpmag} 
\end{equation} 
where $\alpha$ is a real parameter, $\LAP_0$ is the Laplace
operator on $\Omega _{ci}$, and $p^{ci}$ is a new scalar unknown. A
simple calculation shows that $p^{ci}=0$ if the initial magnetic
induction is solenoidal; hence, \eqref{eq::inductionpmag} enforces
$\DIV\mu^{ci}\vec{H}^{ci}=0$.  Taking $\alpha=0$ amounts to penalizing
$\DIV\mu^{ci}\vec{H}^{ci}$ in $\vec{L}^2(\Omega_{ci})$, which turns
out to be non-convergent with Lagrange finite elements when the
boundary of $\Omega_{ci}$ is not smooth, (see \cite{Costabel_1991} for details).
The mathematical analysis shows that the method converges with
Lagrange finite elements when $\alpha\in(\frac12,1)$.  In practice we
take $\alpha=0.7$.

We introduce new finite elements spaces to approximate the new scalar unknown $p^c$
\begin{align*}
  X_h^{p,2d} &:= \left\{ p_h\in L^1(\Omega_c)\;/\; p_h \in
    \mathcal{C}^0(\overline{\Omega_{ci})},\; \forall i=0,\ldots,N,\;
    p_h(T_K)\in\mathbb{P}_{\ell_p},\; \forall K \in \cup_{i=1}^N
    \mathcal{F}_h^{ci},\;
    p_h=0\text{ on }\partial\Omega_{ci}\right\},\\
  X_h^p &:= \left\{ p = \sum_{m=-M}^M p^m(r,z)\text{e}^{\mathrm{i}m\theta}
    \;/\; \forall m=1\ldots,M,\; p^m\in X_h^{p,2d}\text{ and } p^m =
    \overline{p^{-m}} \right\}
\end{align*}
Here $\ell_p$ is an integer in \{1,2\}. The final form of the
algorithm is the following: after proper initialization, we solve for
$\vec{H}^{c,n+1}\in\vec{X}_h^H$, $\phi^{v,n+1}\in X_h^\phi$ and
$p^{n+1}\in X_h^p$ so that the following holds for all
$\vec{b}\in\vec{X}_h^H,\; \psi\in X_h^\phi,\; q\in X_h^p $
\begin{equation}
  \mathcal{L}\left((\vec{H}^{c,n+1},\phi^{v,n+1}),(\vec{b},\psi)\right)
+\mathcal{D}\left((\vec{H}^{c,n+1},p^{c,n+1},\phi^{v,n+1}),(\vec{b},q,\psi)\right) 
+\mathcal{P}(\phi^{v,n+1},\psi)
= \mathcal{R}(\vec{b},\psi)
\end{equation}
with
\begin{equation}
\mathcal{D}\left((\vec{H},p,\phi),(\vec{b},q,\psi)\right) := \sum_{i=1}^N
  \int_{\Omega_{ci}} \left( \mu^c\vec{b}\SCAL \nabla p  - \mu^c\vec{H}\SCAL\GRAD q
  + h^{2\alpha} \DIV\mu^c\vec{H}\;\DIV \mu^c\vec{b}+ 
h^{2(1-\alpha)} \GRAD p\SCAL\GRAD q \right)
\end{equation}
where $h$ denotes the typical size of a mesh element. The term
$h^{2\alpha}\int_{\Omega_c} \DIV\mu^c\vec{H}^{c,n+1}\;
\DIV\mu^c\vec{b}$ is a stabilization quantity which is added in to
have discrete well-posedness of the problem irrespective of the
polynomial degree of the approximation for $p^c$. The additional
stabilizing bilinear form $\mathcal{P}$ is defined by
\[
\mathcal{P}(\phi,\psi) = \int_{\Omega_v} \GRAD\phi\SCAL \GRAD\psi - \int_{\partial\Omega_v} \psi \vec{n}\SCAL \GRAD \psi.
\] 
This bilinear form is meant to help ensure that $\LAP\phi^{v,n+1}=0$ for all times.

\subsubsection{Taking advantage from the cylindrical symmetry for Maxwell and Navier-Stokes equations}
SFEMaNS is a fully nonlinear code integrating coupled Maxwell and Navier-Stokes equations
(\citet{MR2290574,GLLN09}).
As mentioned before, any term that could mix different Fourier modes
has been made explicit. Owing to this property,
there are $M+1$ independent linear systems to solve at each time step ($M+1$ being the
maximum number of complex Fourier modes). This immediately provides a
parallelization strategy. In practice we use one processor per Fourier
mode. The computation of the nonlinear terms in right-hand side is done using a parallel Fast
Fourier Transform.
Note that, in the present paper, we use only the kinematic part of the code with an axisymmetric
steady flow. Typical time step is $\Delta t= 0.01$ and typical mesh size is $h=1/80$
with refined meshes $h=1/400$ on curved $\Sigma_\mu$ interfaces.
\section{Ohmic decay in heterogenous domains}
The inspection of equations (\ref{eq::jumpconditions}) shows that even in the absence of flow,
heterogeneous domains can lead to non trivial Ohmic decay problems.
Therefore the reliability and the application range of both numerical schemes are
first examined by studying pure Ohmic decay problems in absence of flow.
A cylindrical geometry is chosen with radius $R=1.4$ and height $H=2.6$
which is in accordance with
setting of the VKS experiment. 
The cylinder is filled with a
conductor with diffusivity $\eta=(\mu_0\sigma)^{-1}=1$ and relative
permeability $\mu_{\rm{r}}=1$.  
\begin{figure}[h!]
\vspace*{-1.5cm}
\includegraphics[width=8cm]{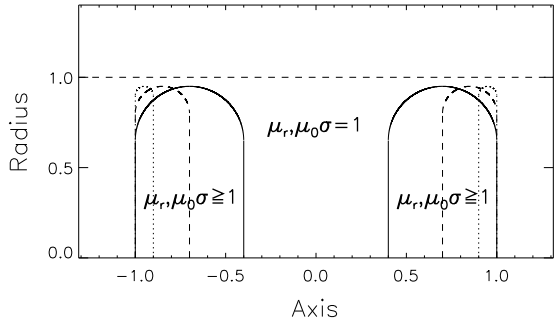}
\caption{Sketch of the set up. Two disks with thickness $d=0.6, 0.3, 01$ (solid,
  dashed, dotted curve) are introduced in
  a cylinder with height $H=2.6$ and radius $R=1.4$.
  In all runs the location of the backside of each
  disk is fixed at $z=\pm 1$. At the outer disk edge a circular
  shape is applied with a curvature radius corresponding to half of the disk thickness. The radial
  extension of the disks is fixed and given by $R_{\rm{disk}}=0.95$. The
  dashed horizontal line denotes the inner boundary that separates the dynamical
  active region from the stagnant outer layer in the runs with ${\rm{Rm}} > 0$
  (see Sec.~\ref{sec::mnd}).} \label{fig::setup_freedecay} 
\end{figure}  
Inside the domain two disks are introduced, characterized by thickness $d$,
conductivity $\sigma$ and permeability
$\mu_{\rm{r}}$ (see Fig.~\ref{fig::setup_freedecay}). 
The thickness $d=0.1$ is
representative of the VKS impellers but the other $d$ have been tested to
study the scaling law with
an effective permeability or an effective conductivity
and also to estimate the impact of the numerical resolutions.

As long as $\mu_{\rm{r}}$ and $\sigma$ are axisymmetric, in a freely decaying
system the axisymmetric mode ($m=0$) can be split into decoupled poloidal
($B_r, B_z$) and toroidal ($B_{\varphi}$) components which decay independently
and exhibit two distinct decay rates. 
The components of azimuthal modes with $m\ge 1$ are coupled and exhibit a single
eigenstate and decay rate. 
In the following we limit our examinations to the decay of the axisymmetric
mode ($m=0$) and the
simplest non-axisymmetric mode, the ($m=1$)-mode ($B\propto \cos\varphi$).
\subsection{External boundary conditions and field pattern}
\begin{figure}[h!]
\hspace*{3cm}$H_r$\hspace*{5cm}$H_{\varphi}$\hspace*{5cm}$H_z$
\\
\begin{minipage}{16cm}
\includegraphics[height=3cm]{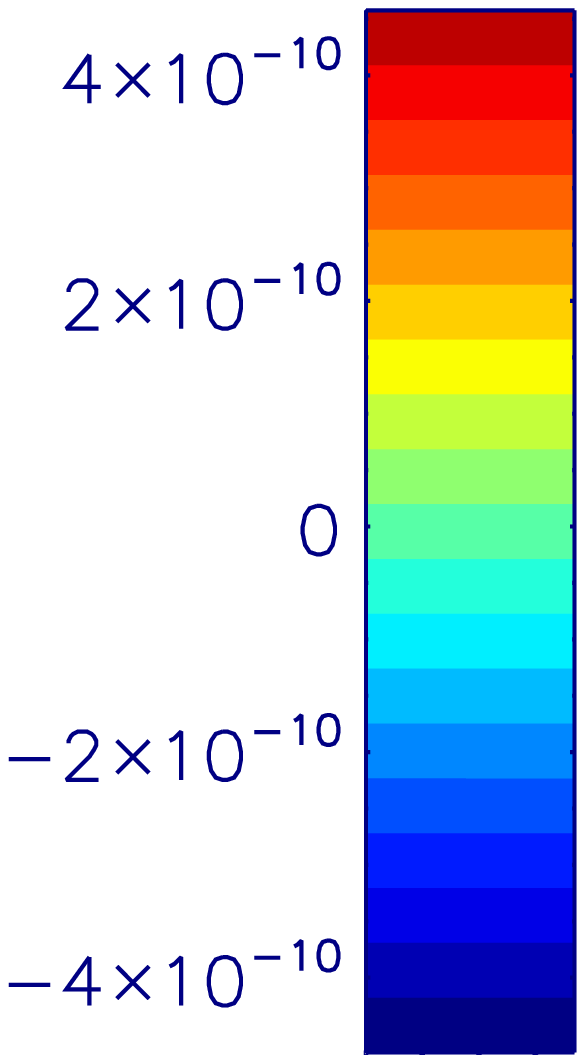}
\hspace*{-3.5cm}
\includegraphics[height=3cm]{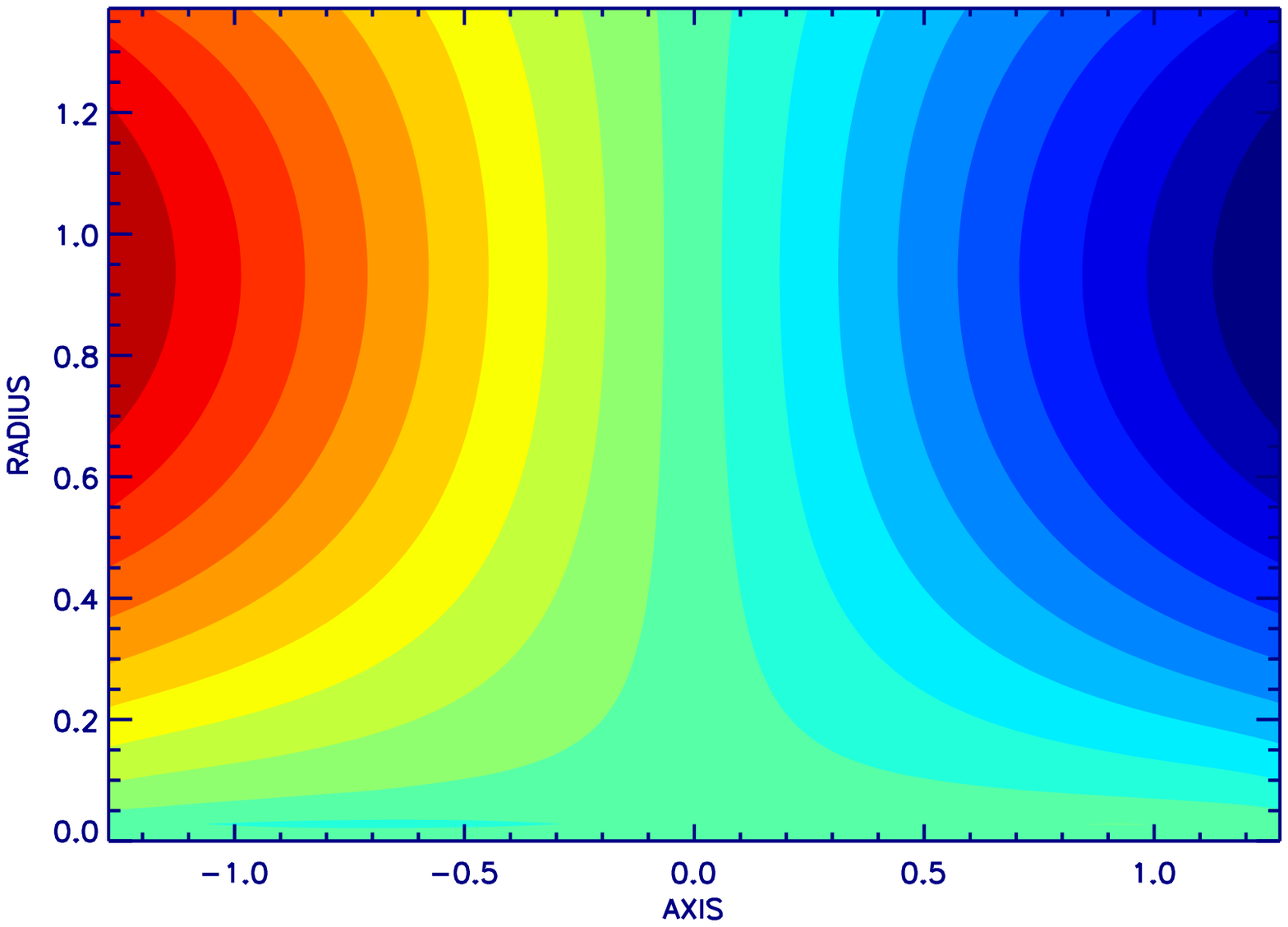}
\hspace*{0.1cm}
\includegraphics[height=3cm]{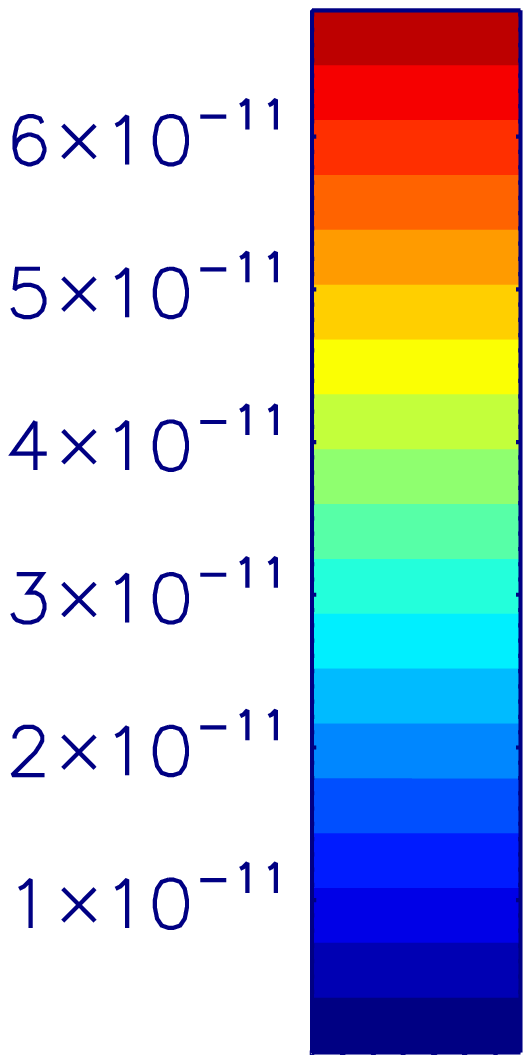}
\hspace*{-3.5cm}
\includegraphics[height=3cm]{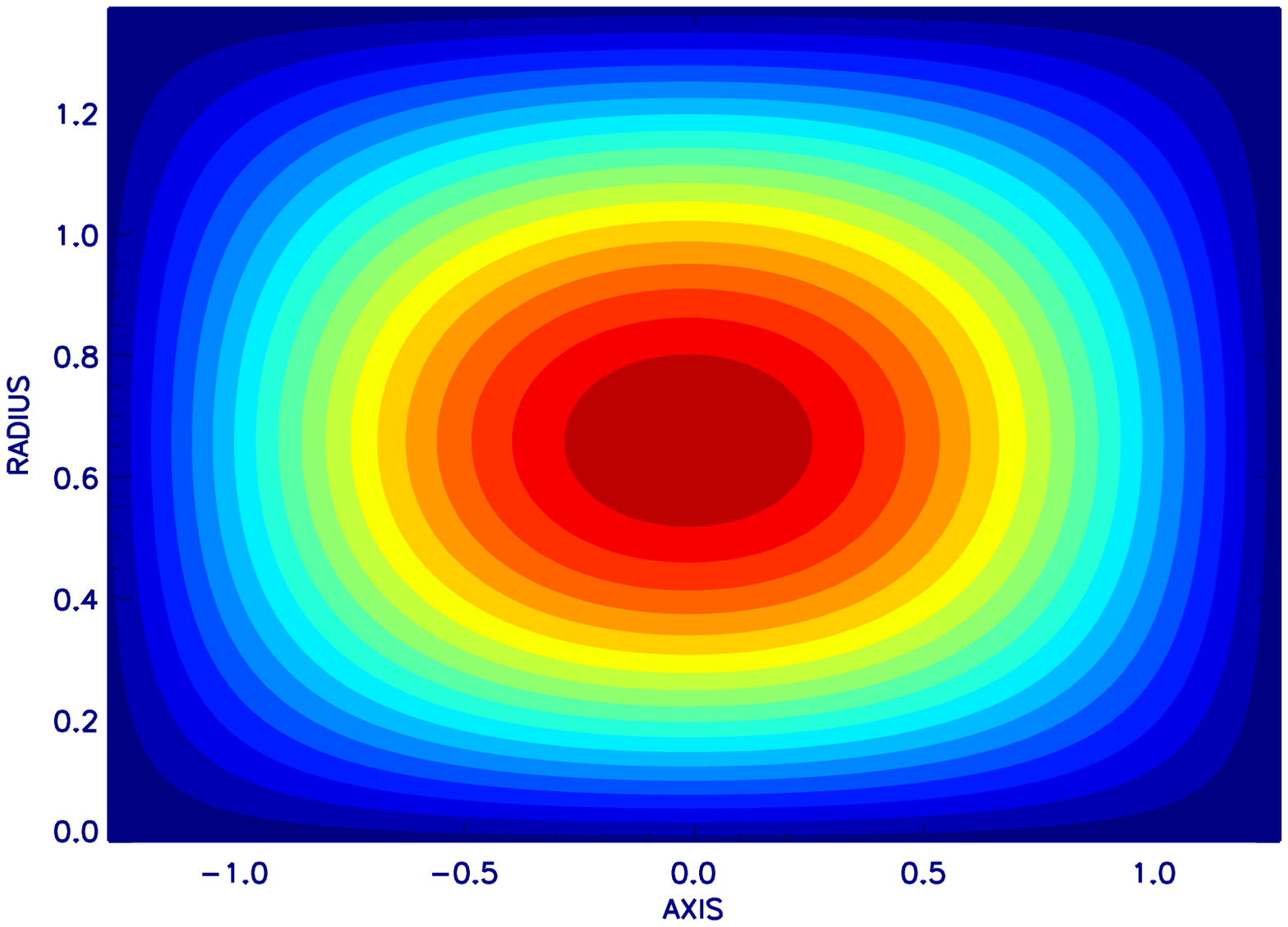}
\hspace*{0.1cm}
\includegraphics[height=3cm]{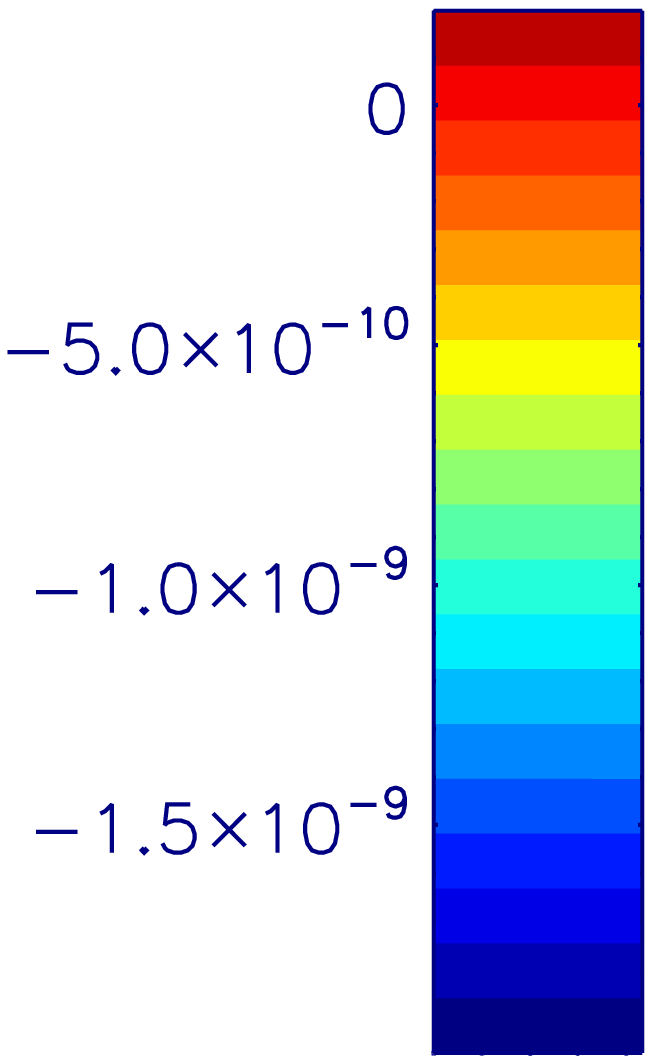}
\hspace*{-3.5cm}
\includegraphics[height=3cm]{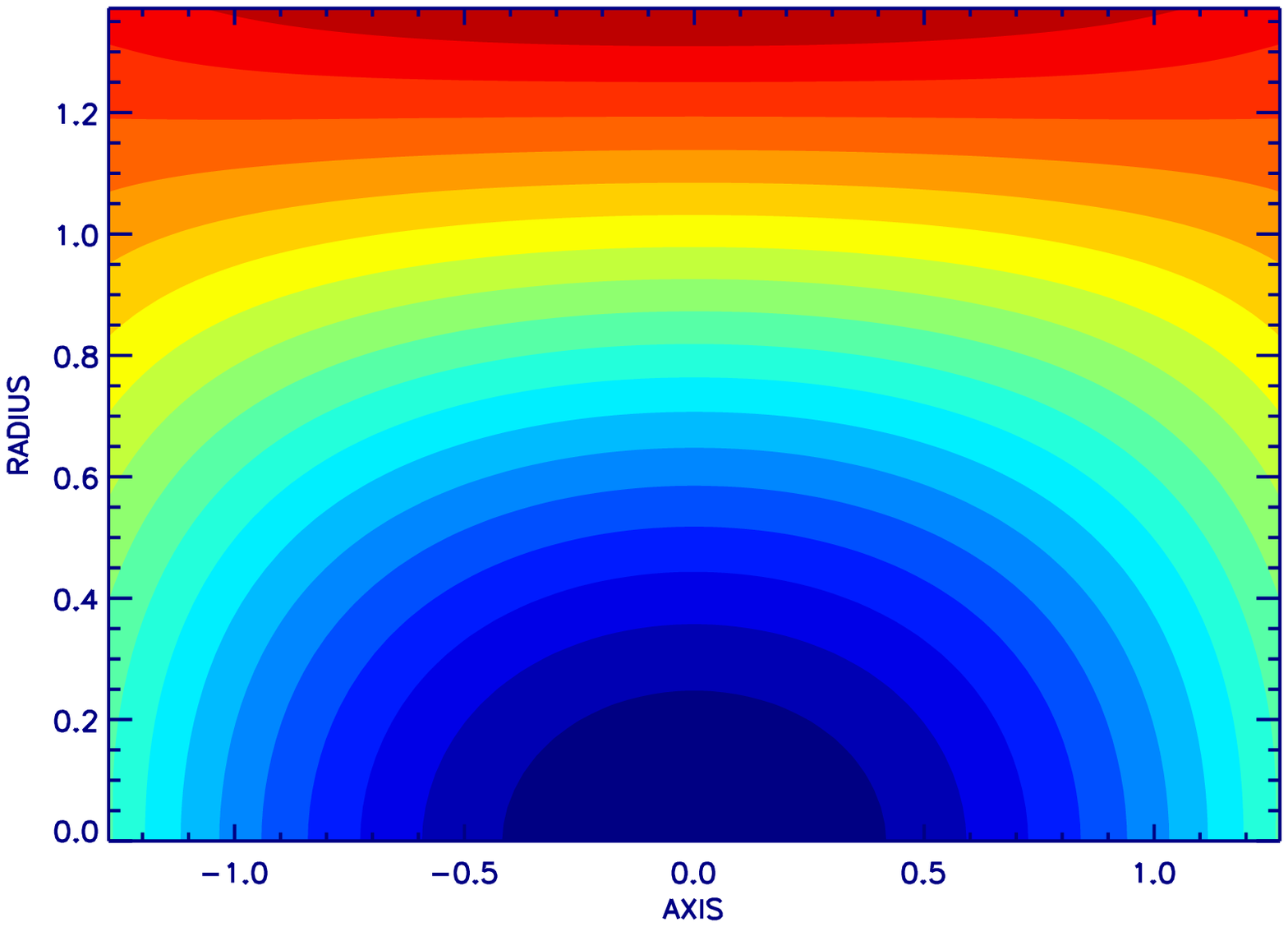}
\end{minipage}
\begin{minipage}{1.5cm}
\vspace*{-1cm}
\begin{eqnarray}
\mu_{\rm{r}}&=&1\nonumber\\
\mu_0\sigma&=&1\nonumber
\end{eqnarray}
\\[-1.4cm]
\begin{center}
Vacuum
\end{center}
\end{minipage}
\\
\begin{minipage}{16cm}
\includegraphics[height=3cm]{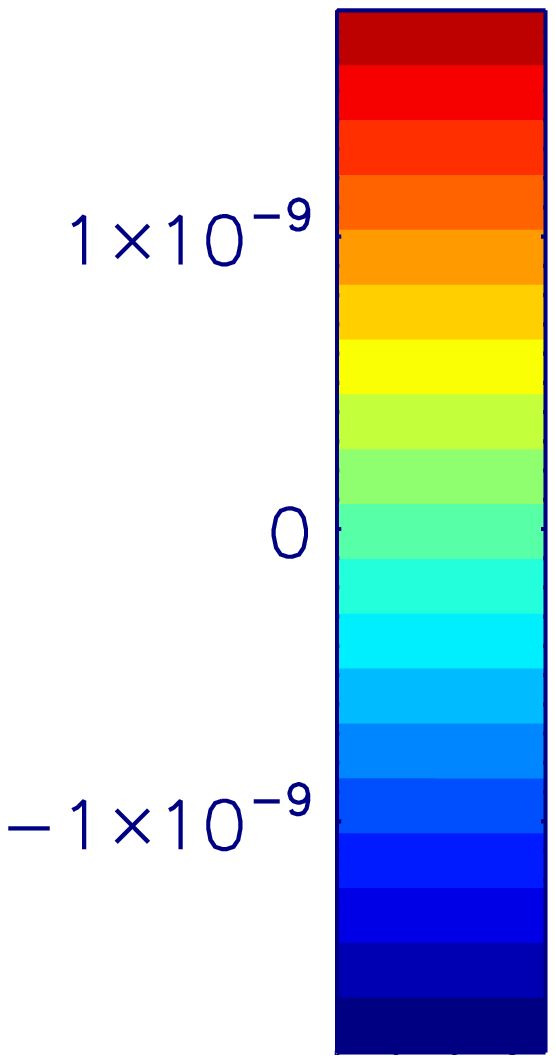}
\hspace*{-3.5cm}
\includegraphics[height=3cm]{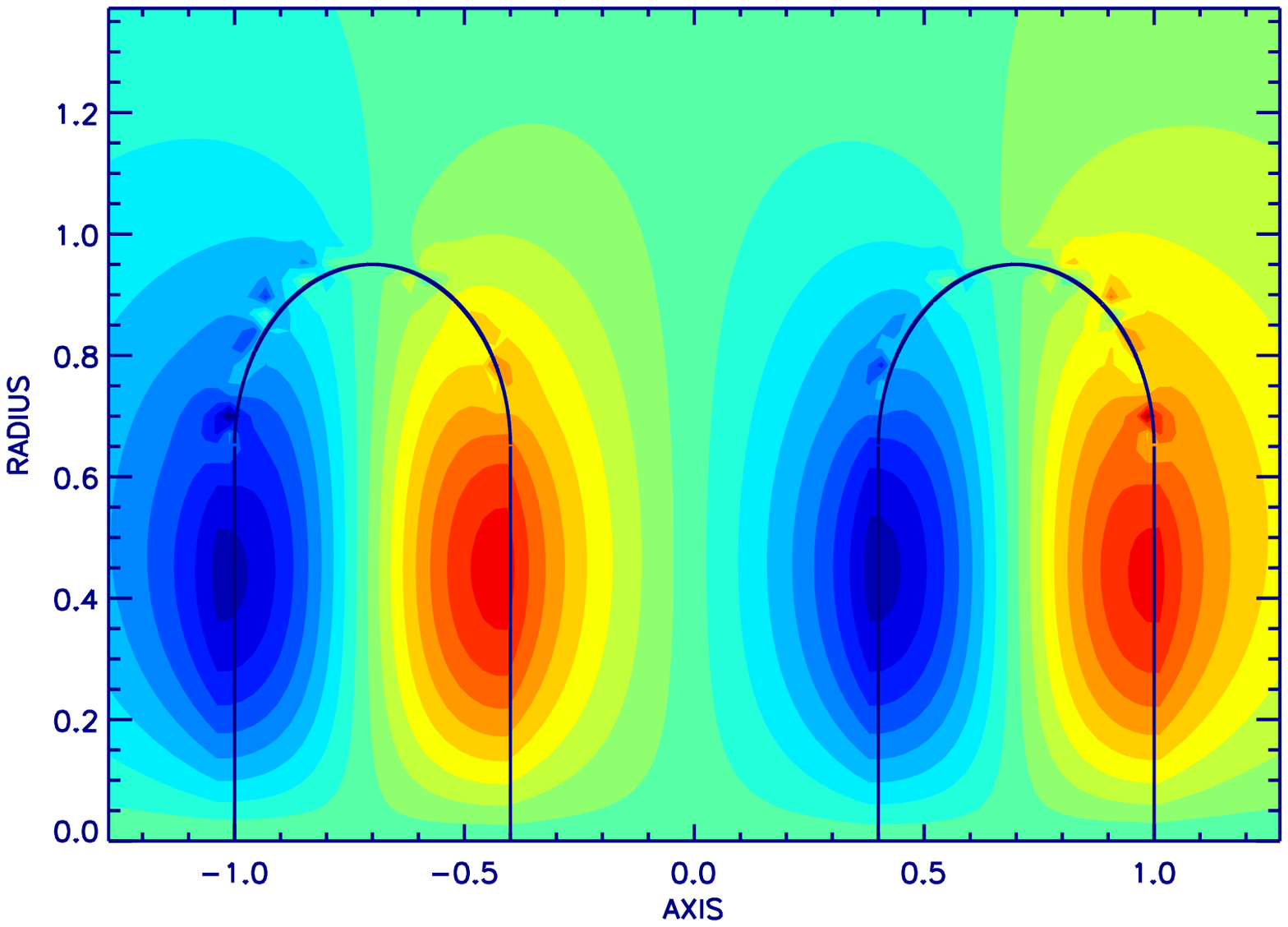}
\hspace*{0.1cm}
\includegraphics[height=3cm]{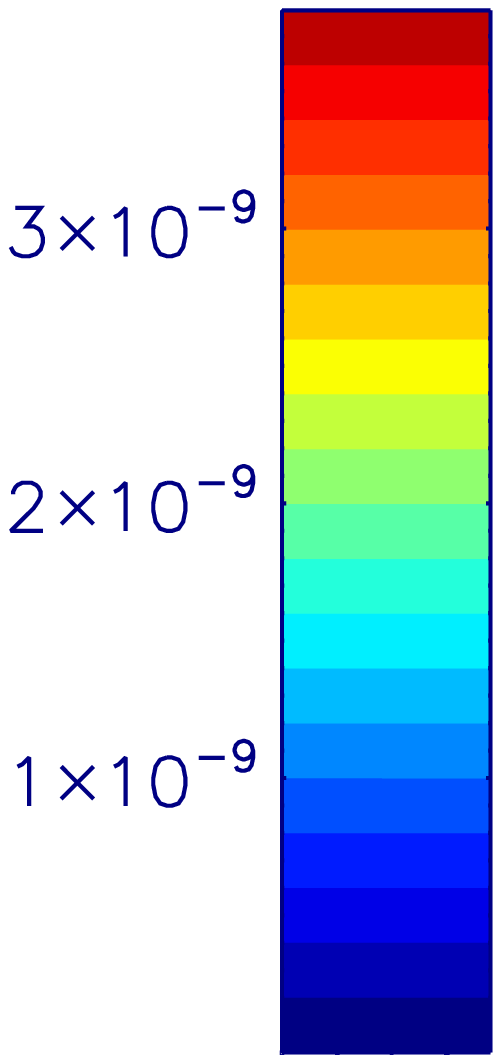}
\hspace*{-3.5cm}
\includegraphics[height=3cm]{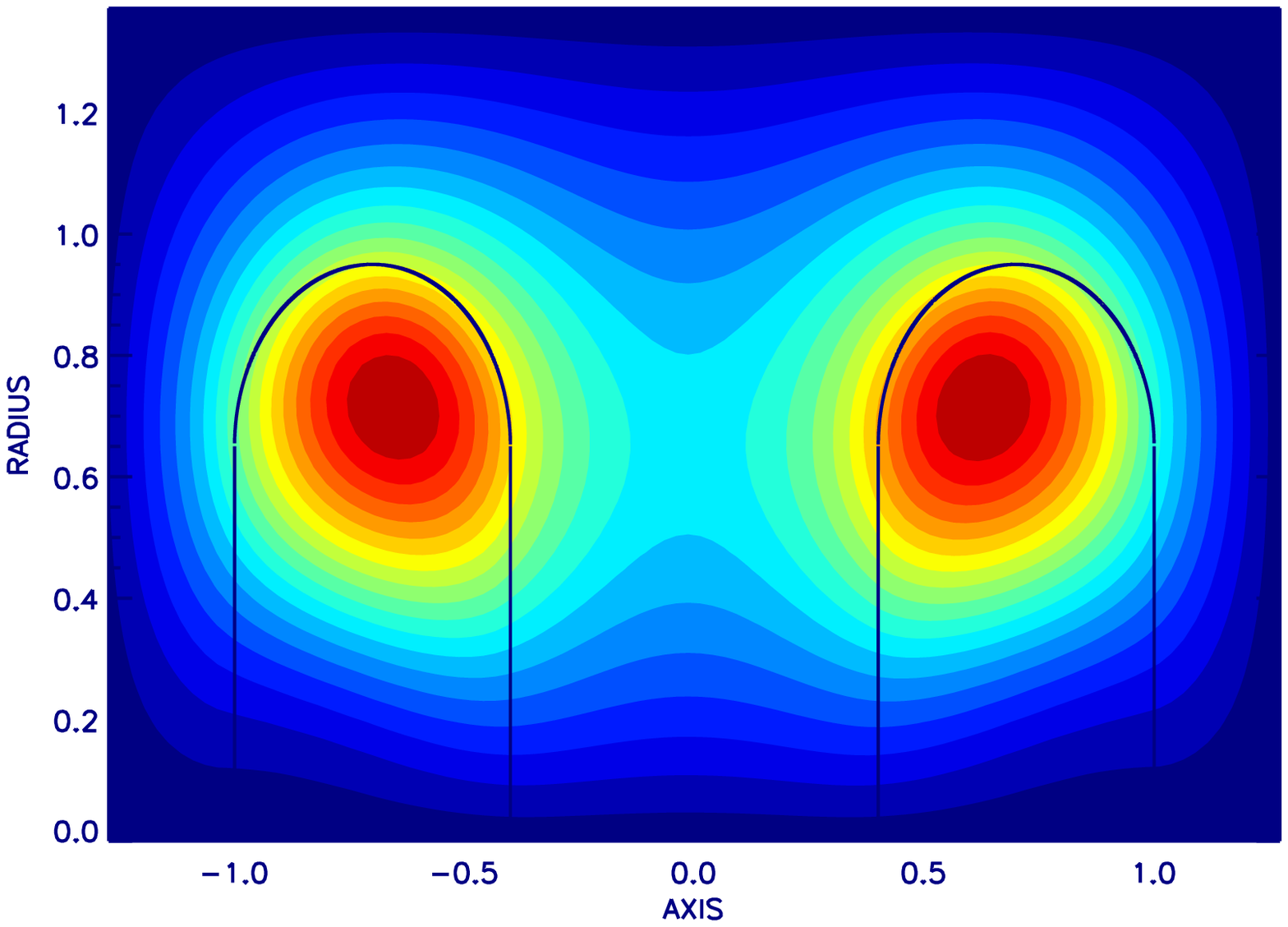}
\hspace*{0.1cm}
\includegraphics[height=3cm]{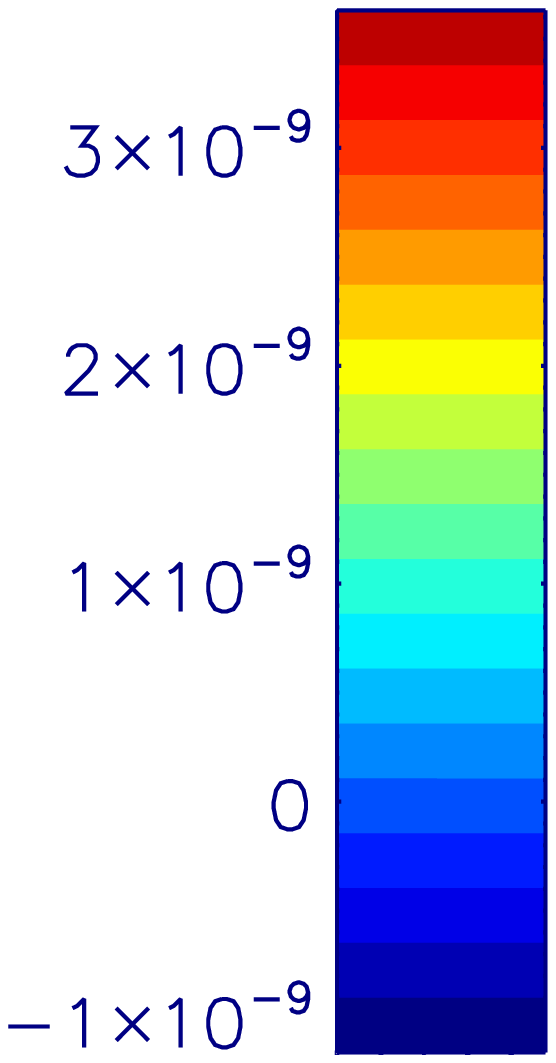}
\hspace*{-3.5cm}
\includegraphics[height=3cm]{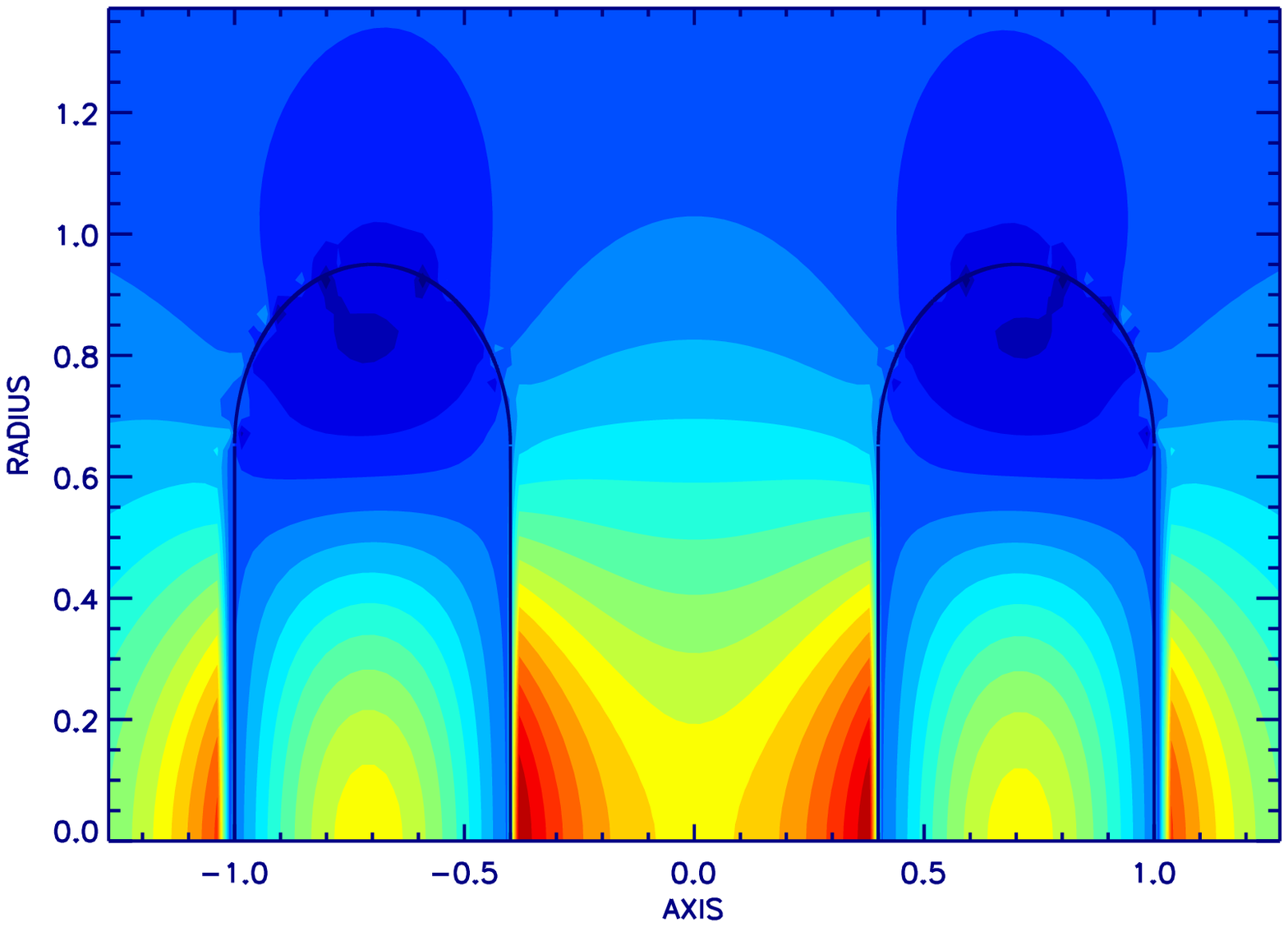}
\end{minipage}
\begin{minipage}{1.5cm}
\vspace*{-1cm}
\begin{eqnarray}
\mu_{\rm{r}}&=&100\nonumber\\
d&=&0.6\nonumber
\end{eqnarray}
\\[-1.4cm]
\begin{center}
Vacuum
\end{center}
\end{minipage}
\\
\begin{minipage}{16cm}
\includegraphics[height=3cm]{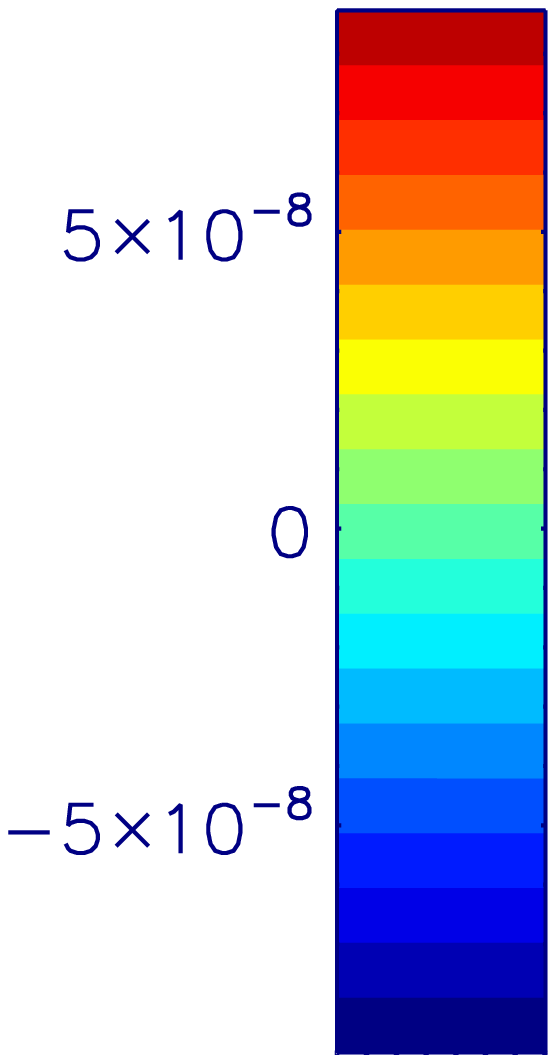}
\hspace*{-3.5cm}
\includegraphics[height=3cm]{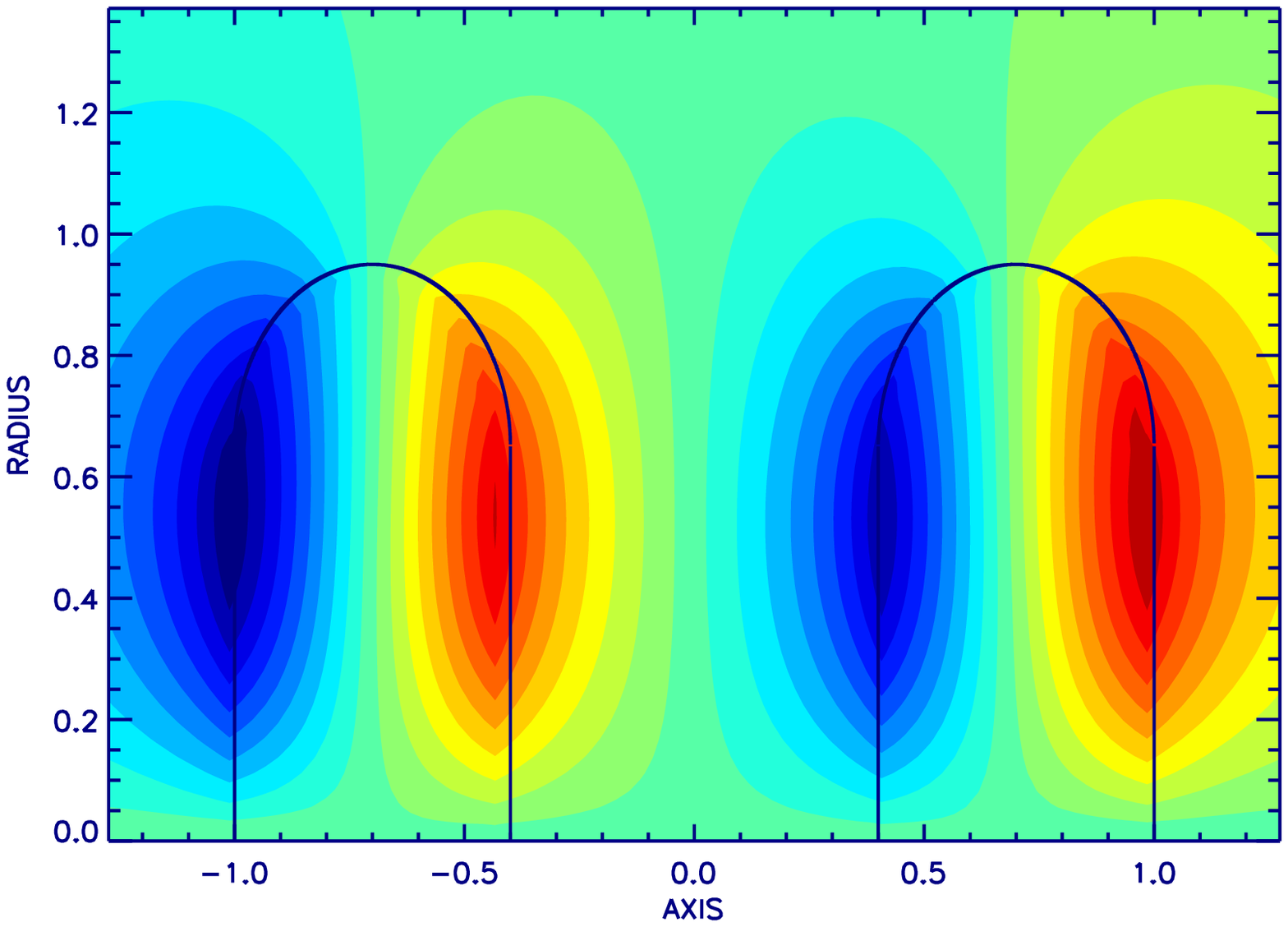}
\hspace*{0.1cm}
\includegraphics[height=3cm]{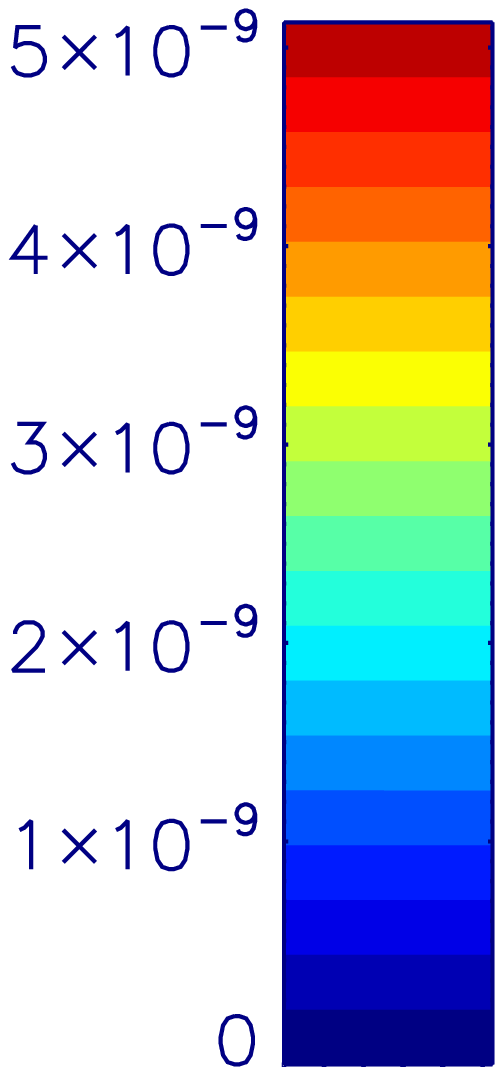}
\hspace*{-3.5cm}
\includegraphics[height=3cm]{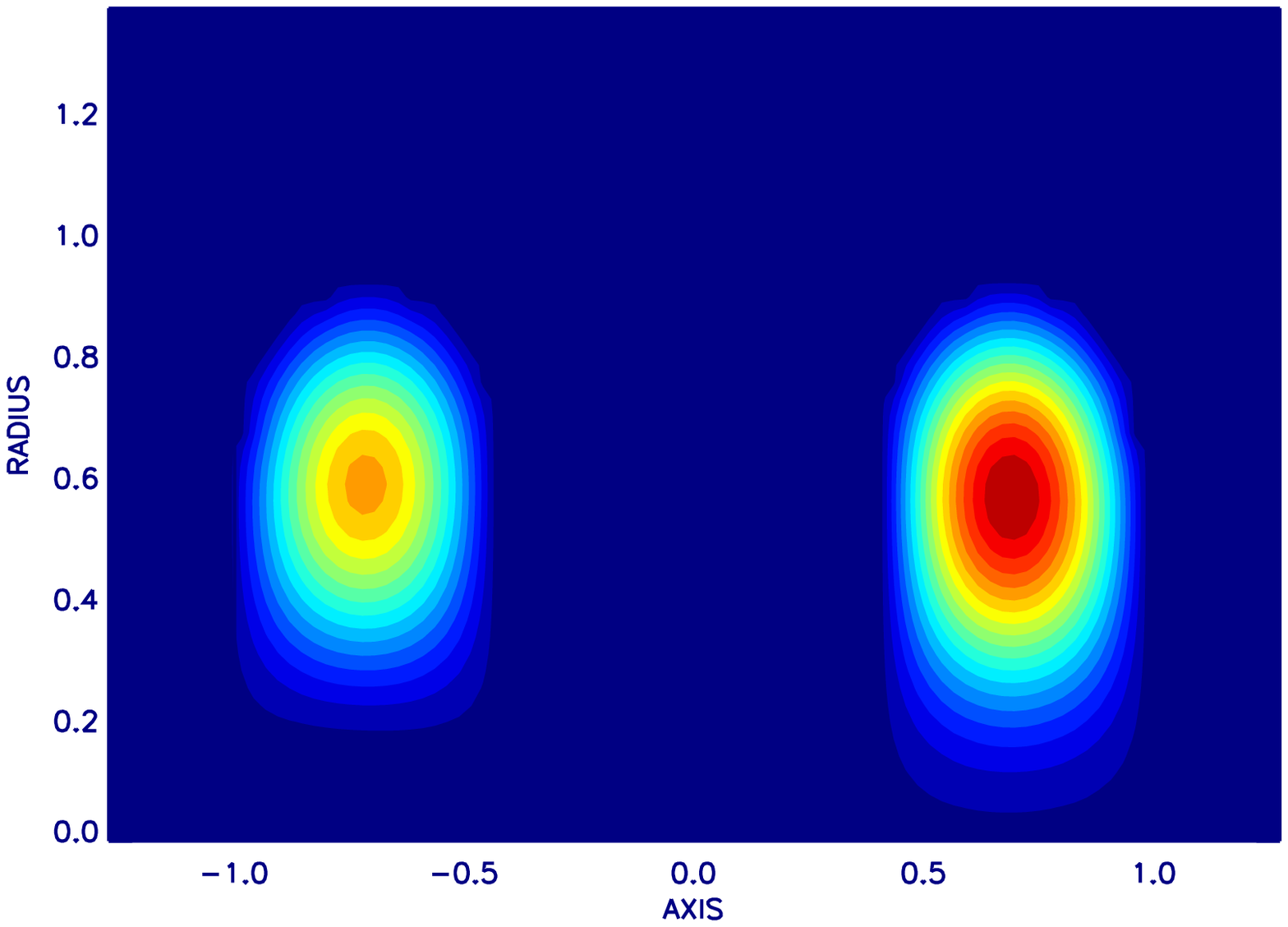}
\hspace*{0.1cm}
\includegraphics[height=3cm]{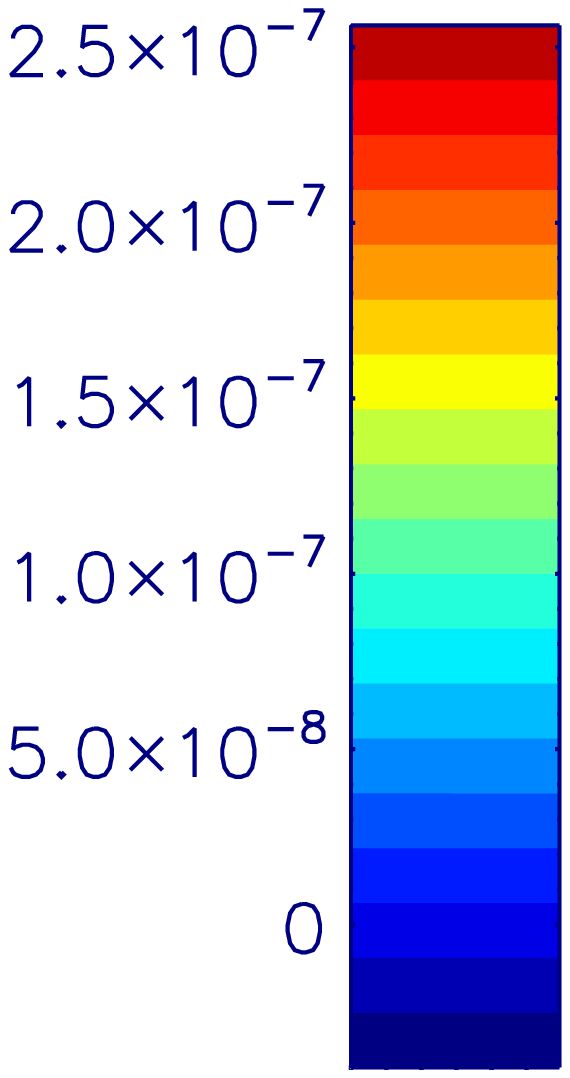}
\hspace*{-3.5cm}
\includegraphics[height=3cm]{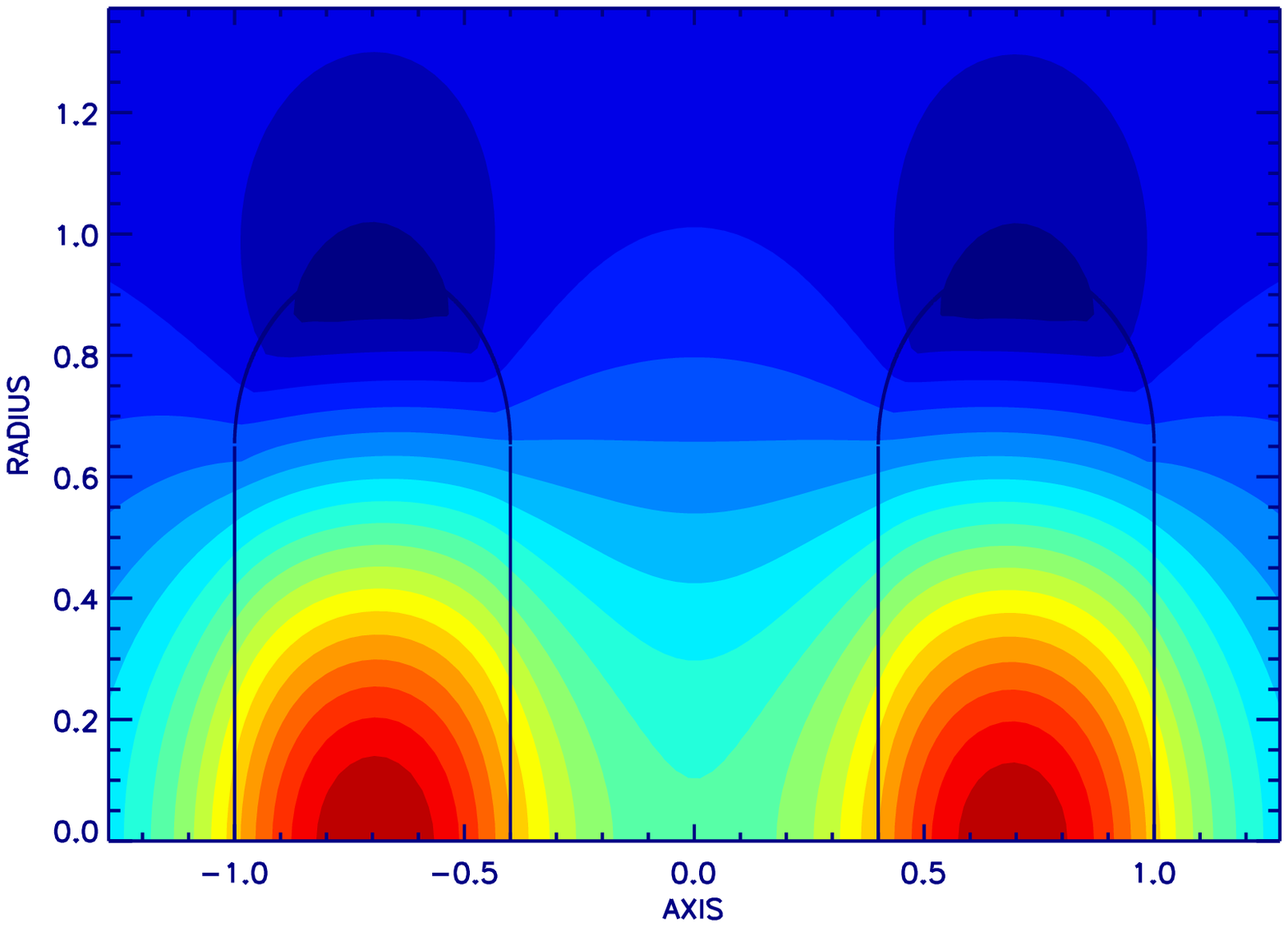}
\end{minipage}
\begin{minipage}{1.5cm}
\vspace*{-1cm}
\begin{eqnarray}
\mu_0\sigma&=&100\nonumber\\
d&=&0.6\nonumber
\end{eqnarray}
\\[-1.4cm]
\begin{center}
Vacuum
\end{center}
\end{minipage}
\\
\begin{minipage}{16cm}
\includegraphics[height=3cm]{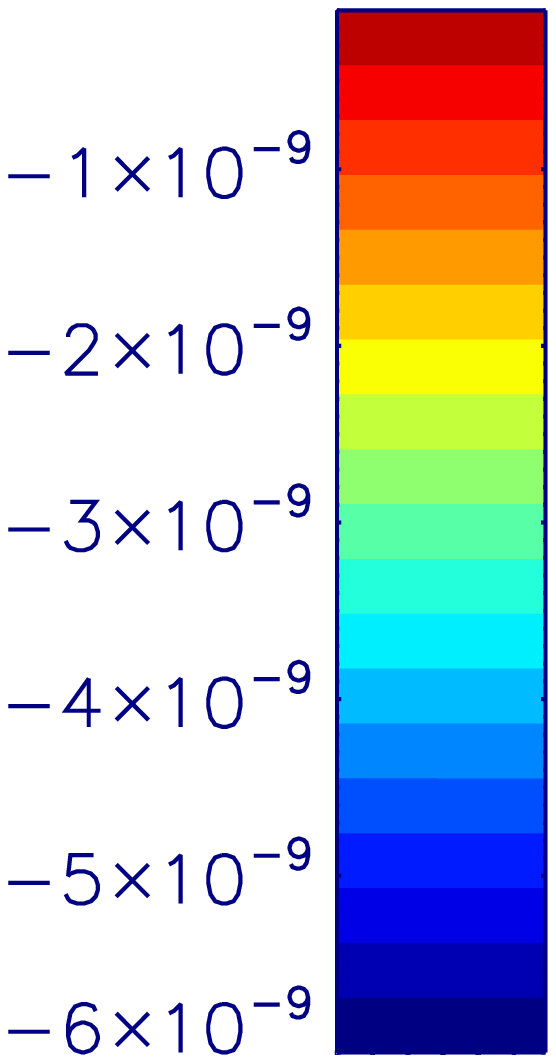}
\hspace*{-3.5cm}
\includegraphics[height=3cm]{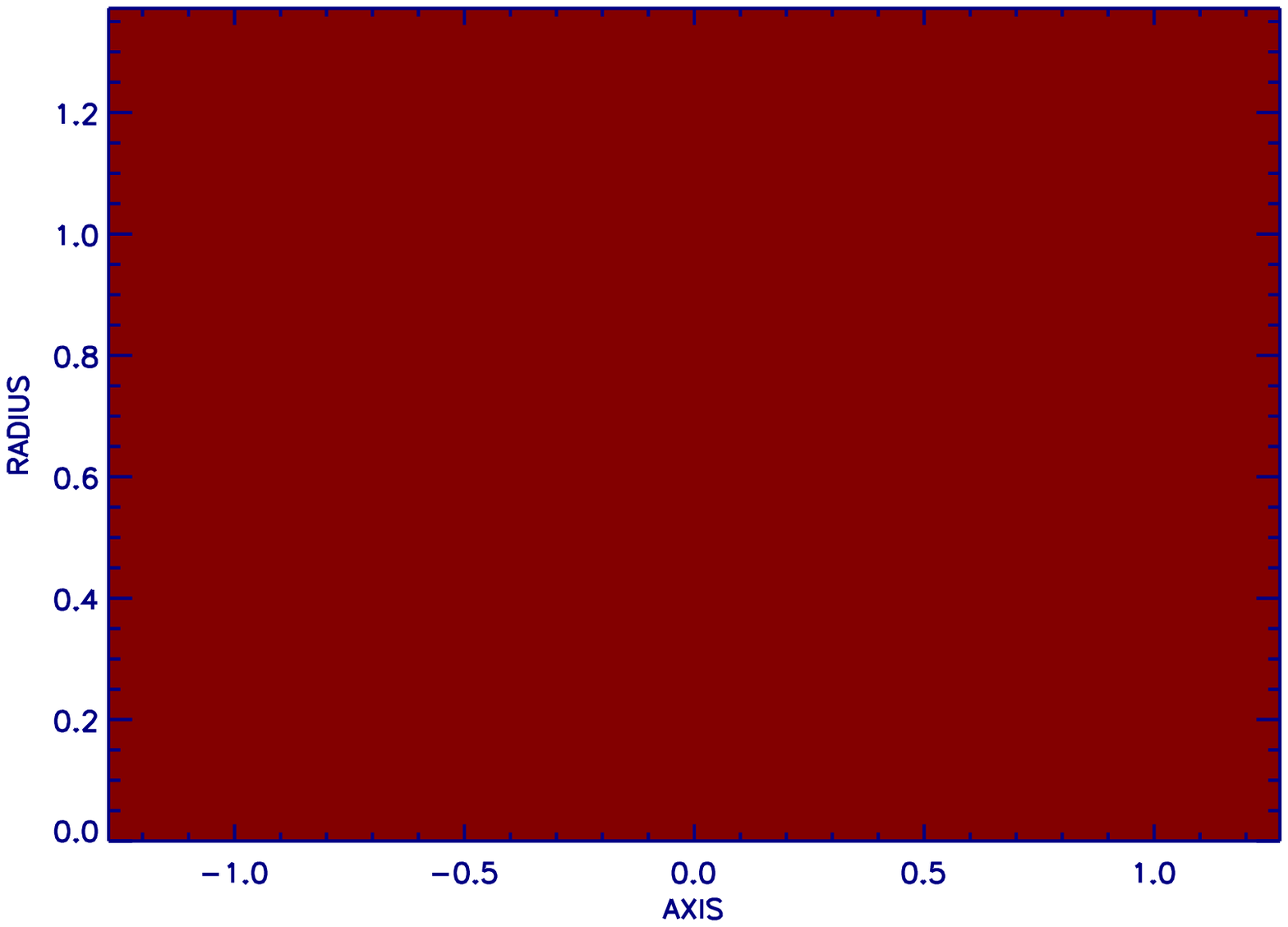}
\hspace*{0.1cm}
\includegraphics[height=3cm]{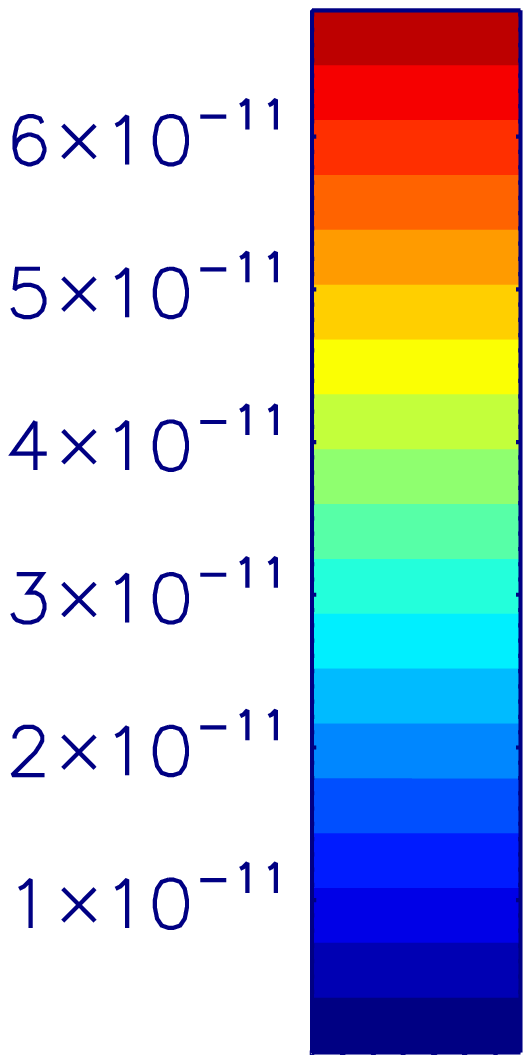}
\hspace*{-3.5cm}
\includegraphics[height=3cm]{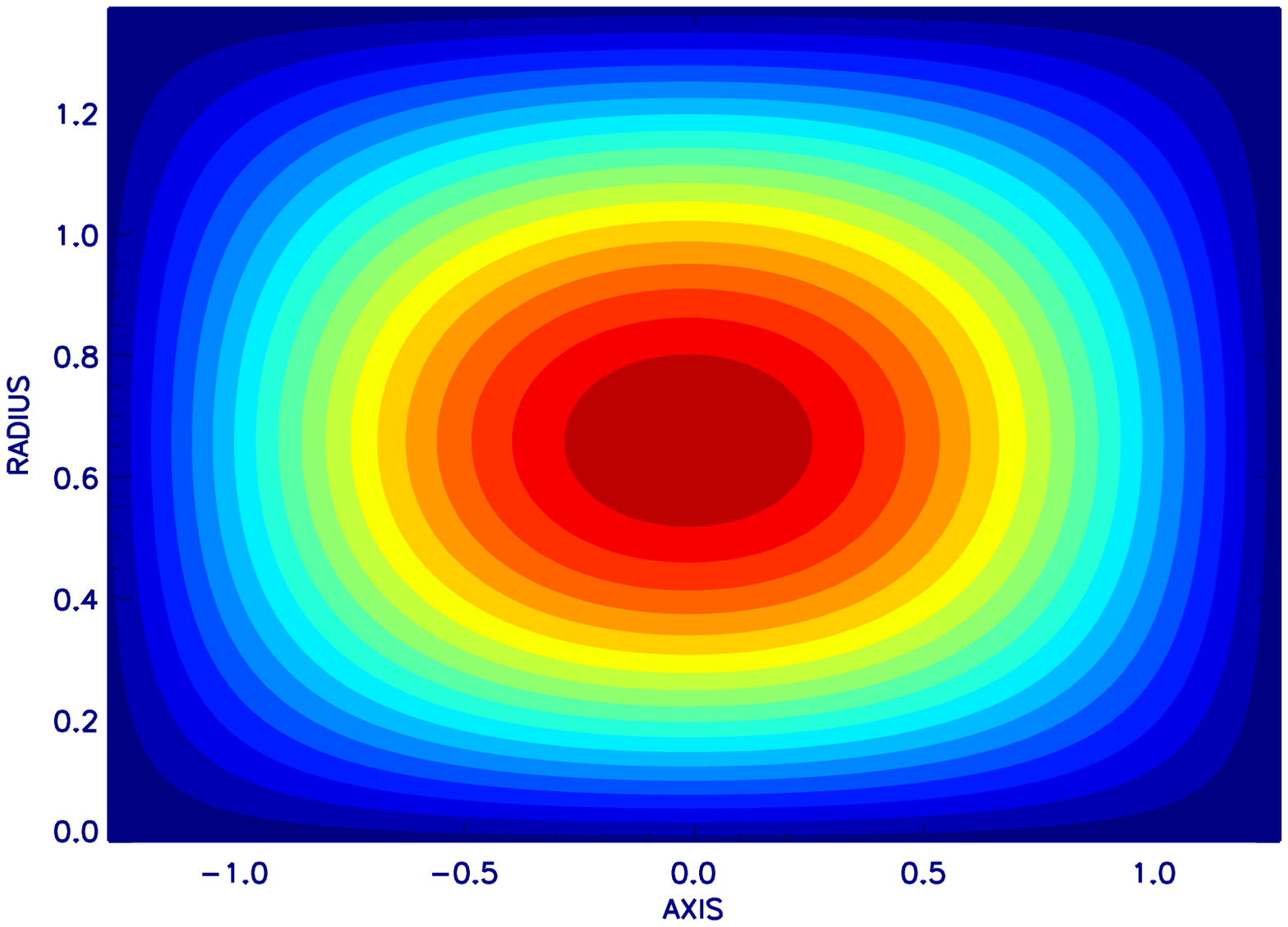}
\hspace*{0.1cm}
\includegraphics[height=3cm]{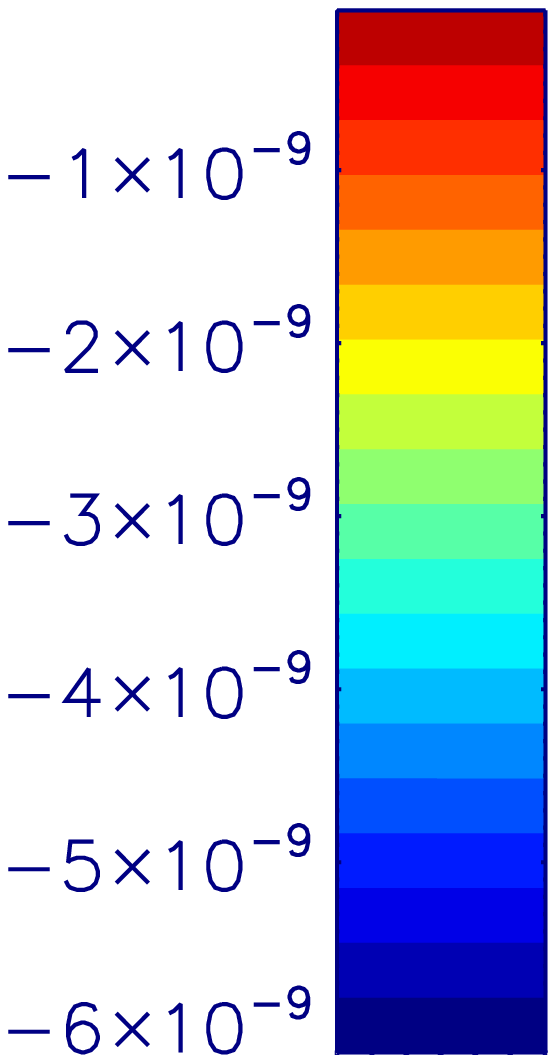}
\hspace*{-3.5cm}
\includegraphics[height=3cm]{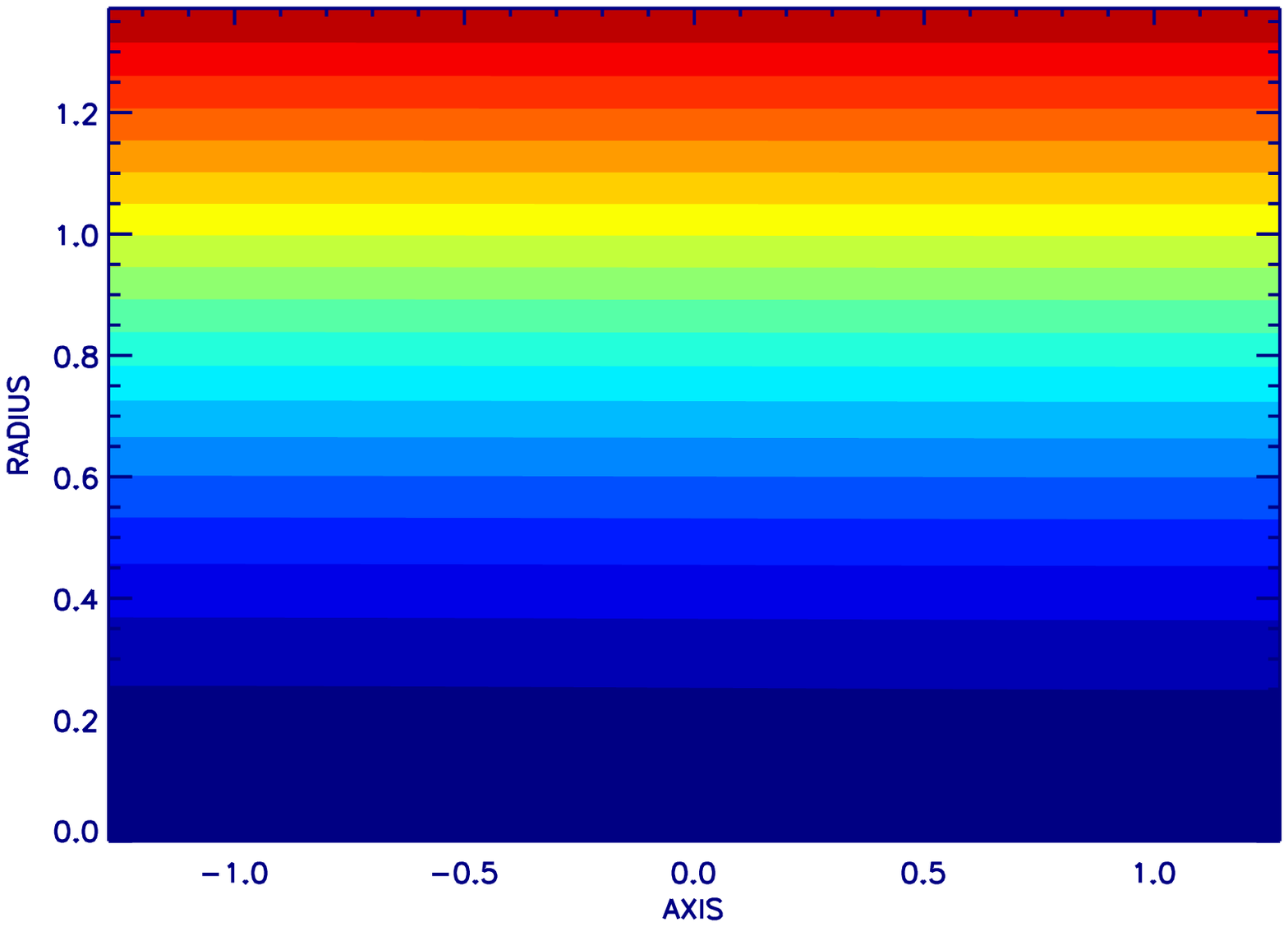}
\end{minipage}
\begin{minipage}{1.5cm}
\vspace*{-1cm}
\begin{eqnarray}
\mu_{\rm{r}}&=&1\nonumber\\
\mu_0\sigma&=&1\nonumber
\end{eqnarray}
\\[-1.4cm]
\begin{center}
VTF
\end{center}
\end{minipage}
\\
\begin{minipage}{16cm}
\includegraphics[height=3cm]{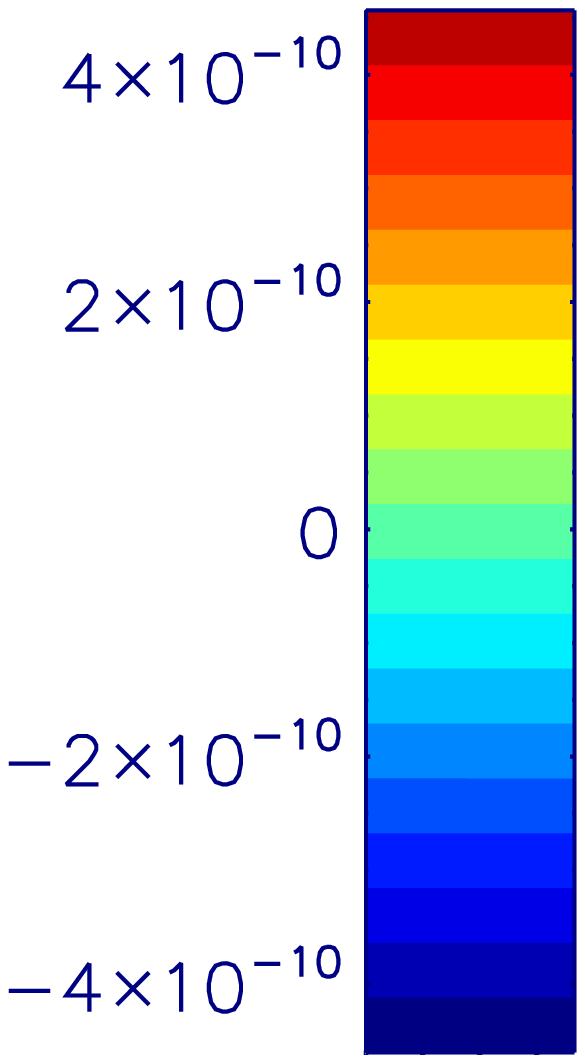}
\hspace*{-3.5cm}
\includegraphics[height=3cm]{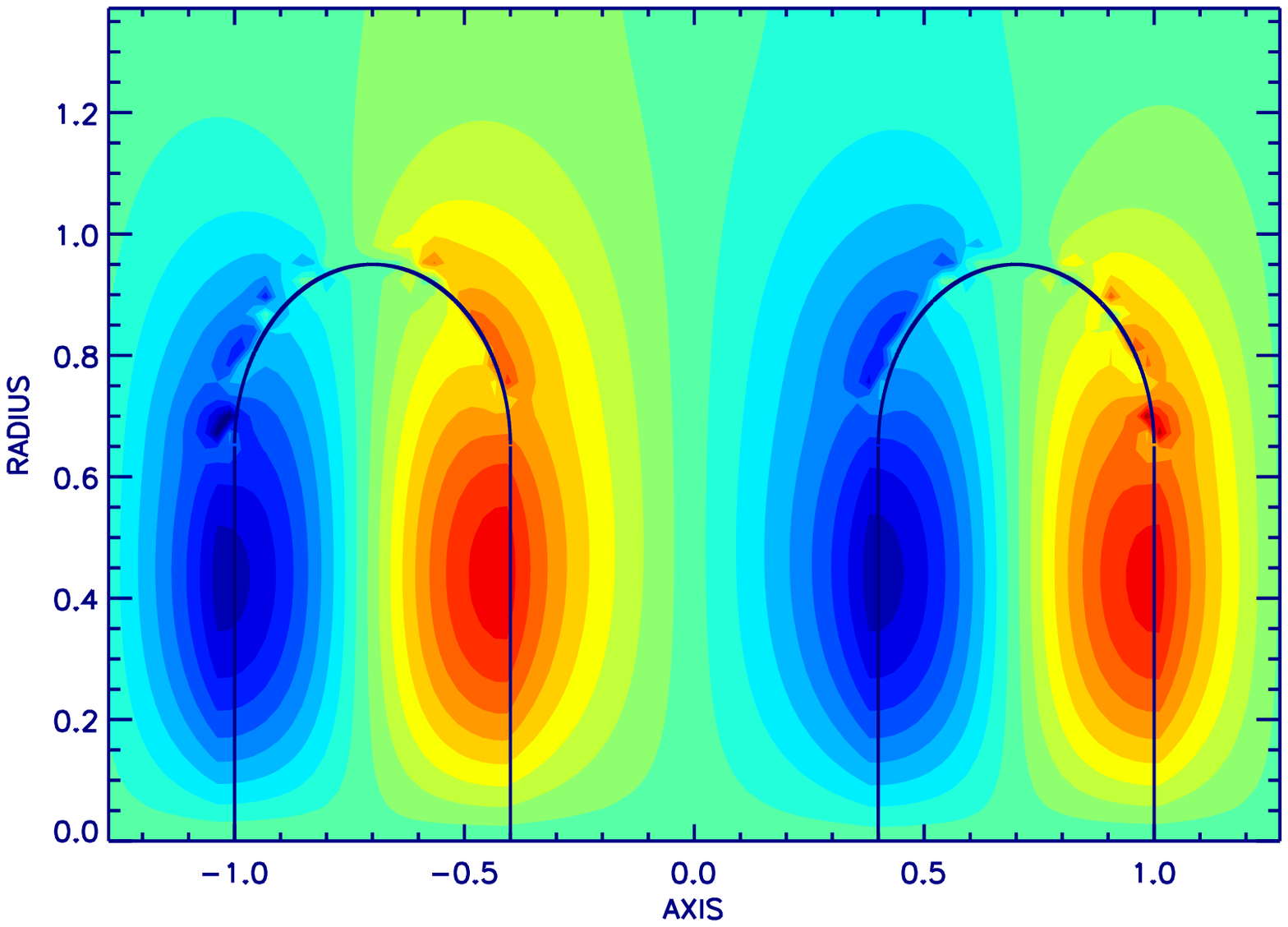}
\hspace*{0.1cm}
\includegraphics[height=3cm]{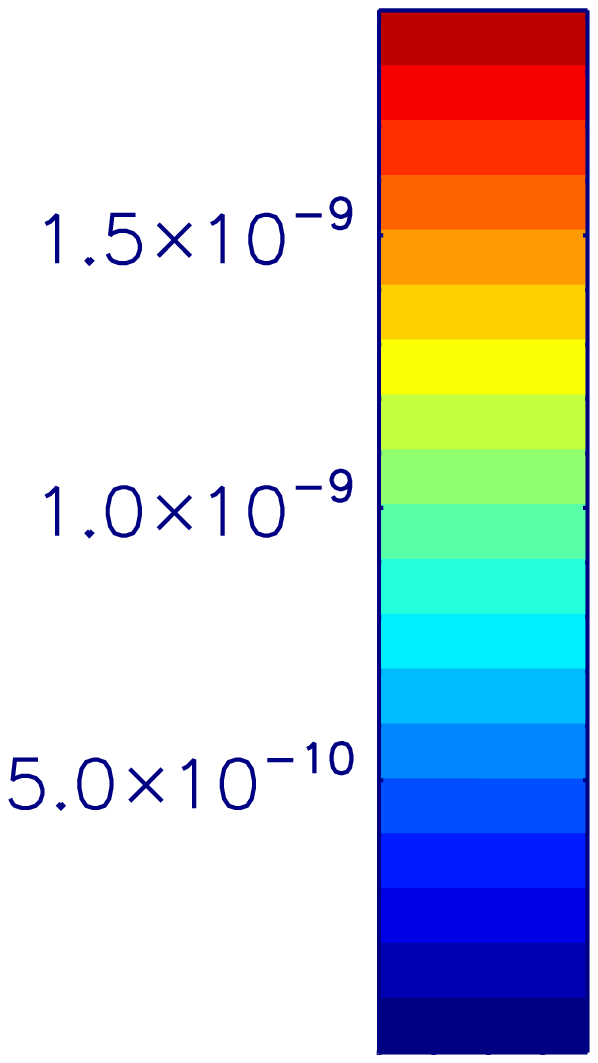}
\hspace*{-3.5cm}
\includegraphics[height=3cm]{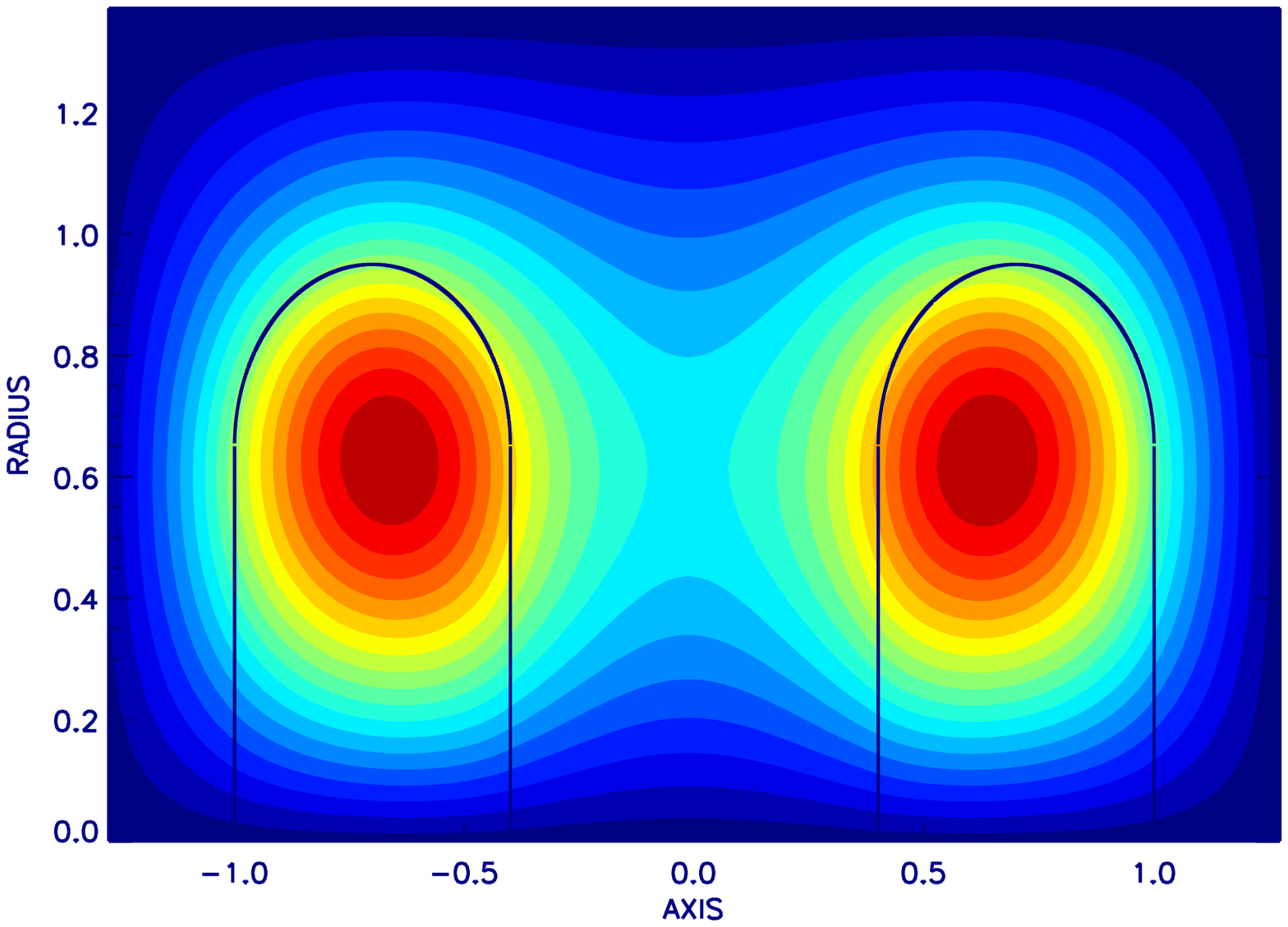}
\hspace*{0.1cm}
\includegraphics[height=3cm]{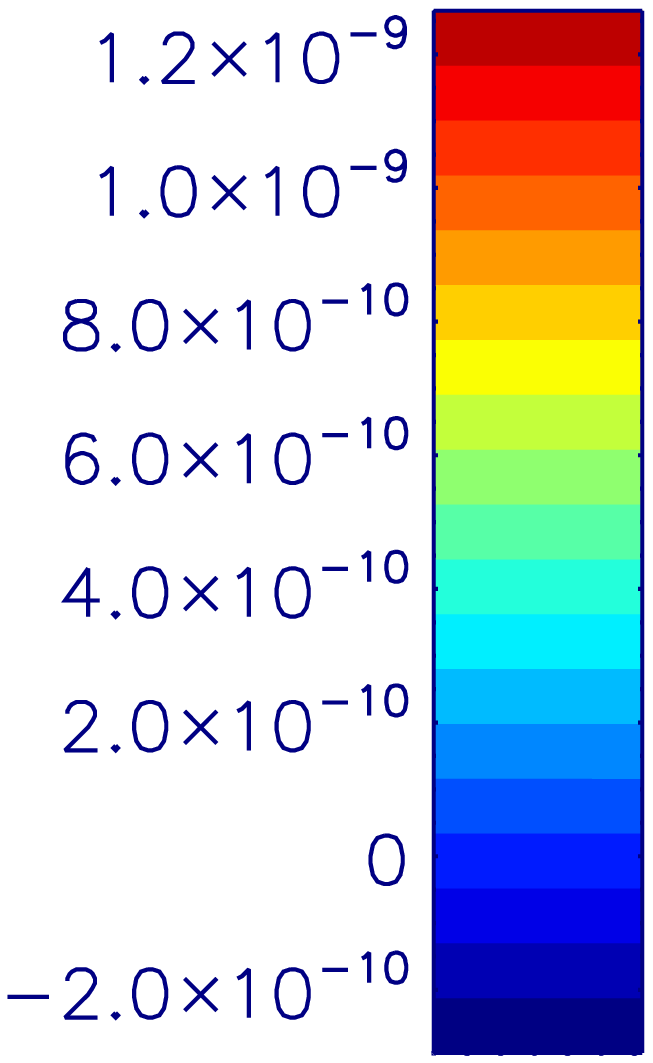}
\hspace*{-3.5cm}
\includegraphics[height=3cm]{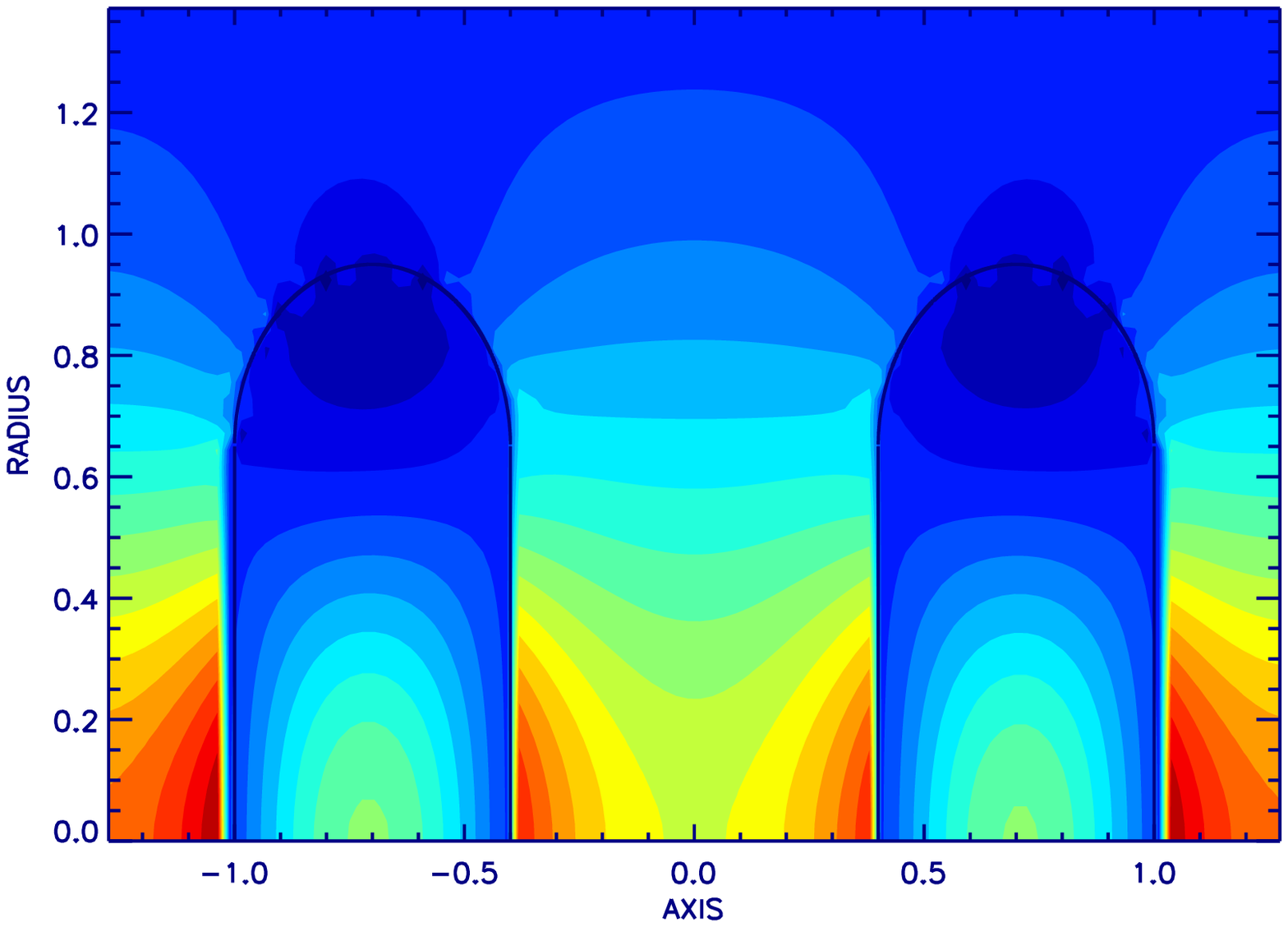}
\end{minipage}
\begin{minipage}{1.5cm}
\vspace*{-1cm}
\begin{eqnarray}
\mu_{\rm{r}}&=&100\nonumber\\
d&=&0.6\nonumber
\end{eqnarray}
\\[-1.4cm]
\begin{center}
VTF
\end{center}
\end{minipage}
\\
\begin{minipage}{16cm}
\includegraphics[height=3cm]{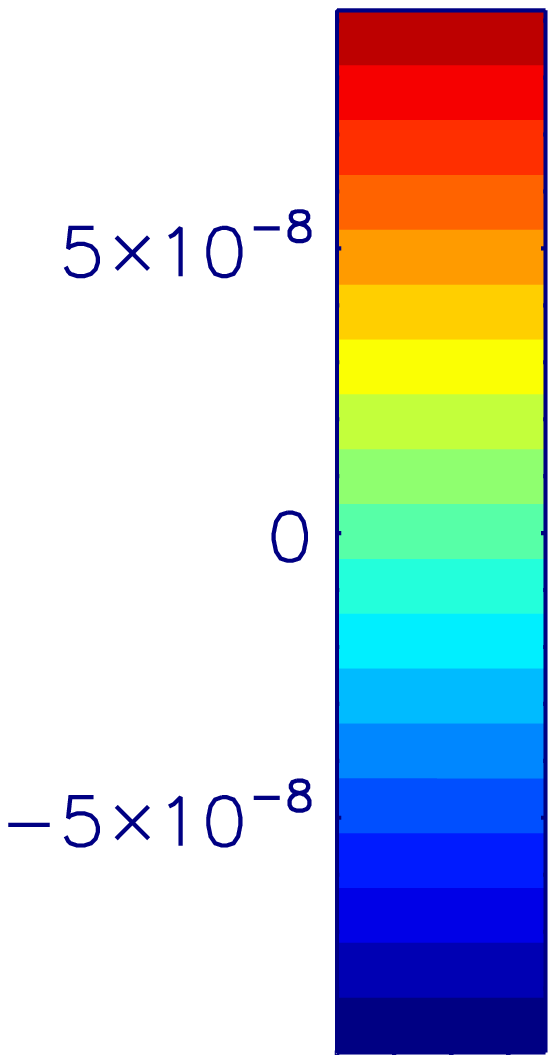}
\hspace*{-3.5cm}
\includegraphics[height=3cm]{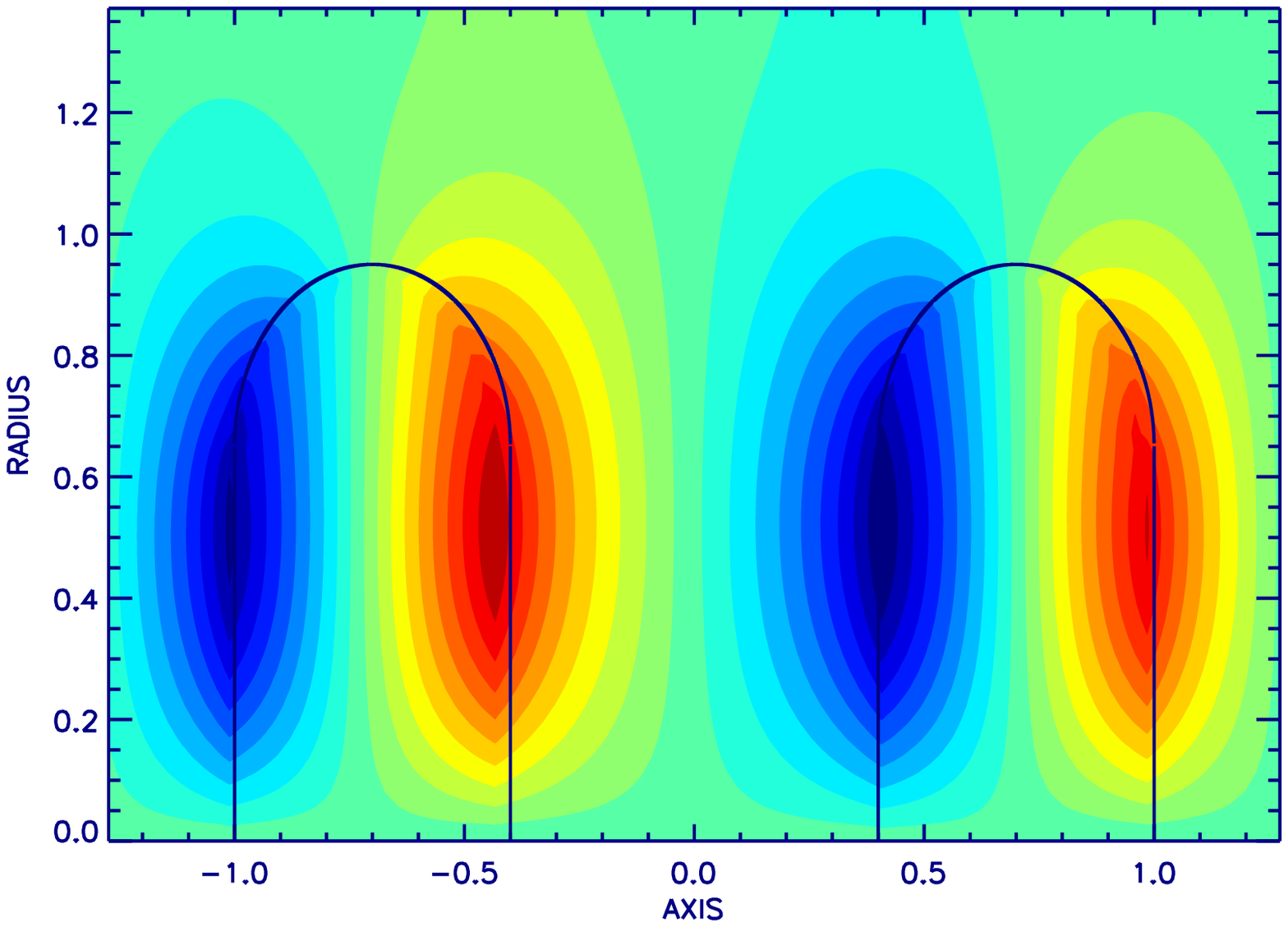}
\hspace*{0.1cm}
\includegraphics[height=3cm]{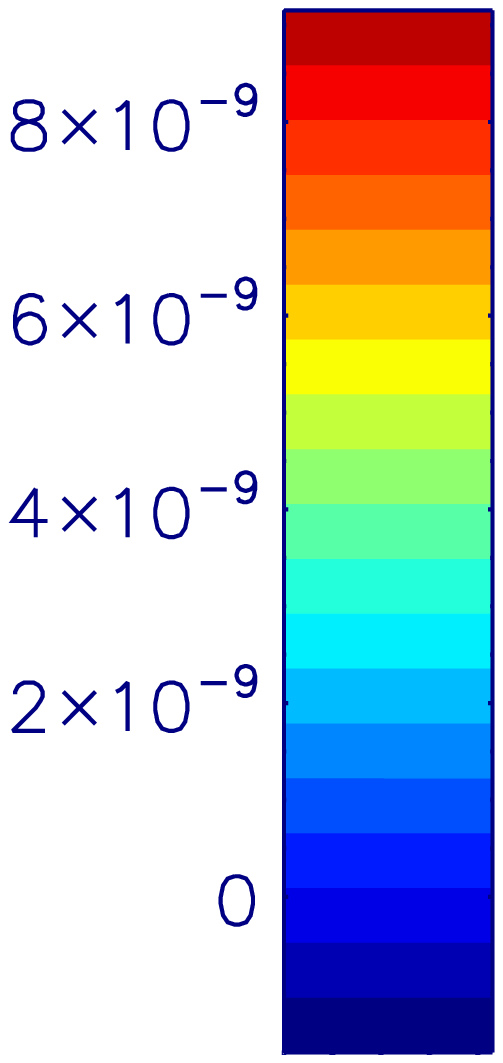}
\hspace*{-3.5cm}
\includegraphics[height=3cm]{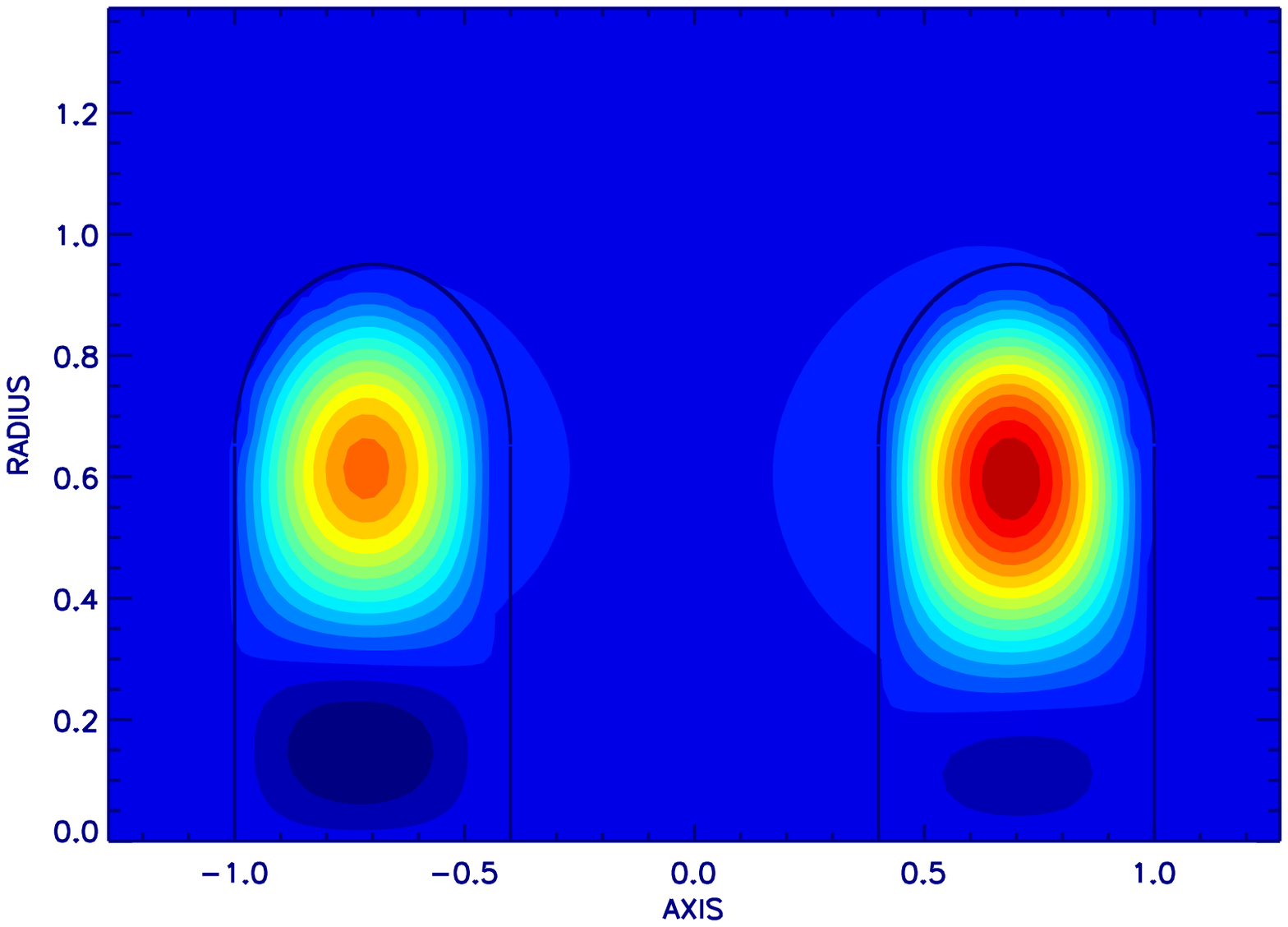}
\hspace*{0.1cm}
\includegraphics[height=3cm]{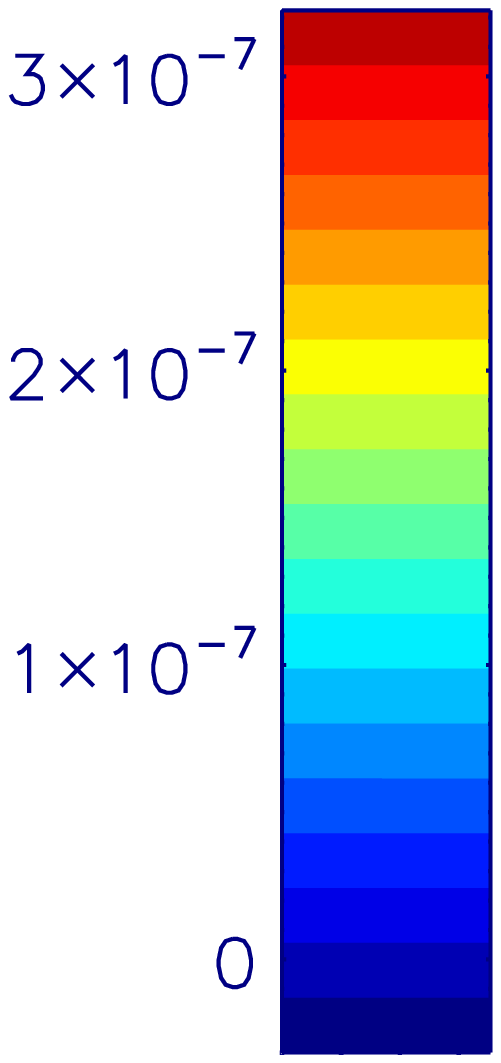}
\hspace*{-3.5cm}
\includegraphics[height=3cm]{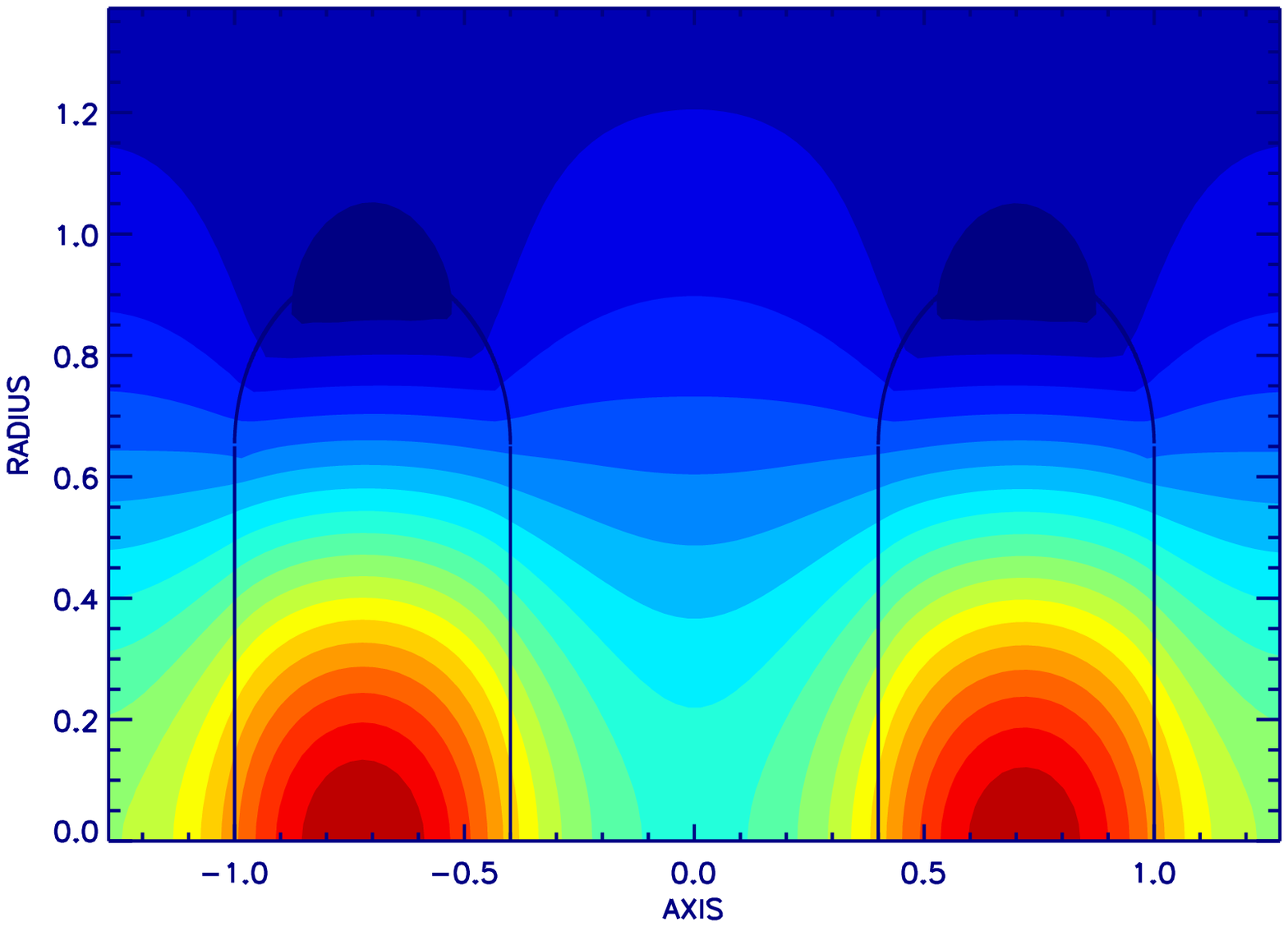}
\end{minipage}
\begin{minipage}{1.5cm}
\vspace*{-1cm}
\begin{eqnarray}
\mu_0\sigma&=&100\nonumber\\
d&=&0.6\nonumber
\end{eqnarray}
\\[-1.4cm]
\begin{center}
VTF
\end{center}
\end{minipage}
\caption{Ohmic decay. Axisymmetric eigenmodes of the magnetic field $\vec{H}=\mu_{\rm{r}}^{-1}\vec{B}$ (from left to
  right: $H_r, H_{\varphi}, H_z$); From top to bottom: $\mu_r=\mu_0\sigma=1$
  (no disks), $\mu_r=100$, $\mu_0\sigma=100$ (all with insulating boundary
  conditions and $d=0.6$), $\mu_r=\mu_0\sigma=1$
  (no disks), $\mu_r=100$, $\mu_0\sigma=100$ (all with vanishing tangential
  field  boundary conditions and $d=0.6$).}
\label{fig::pattern}
\end{figure}

\begin{figure}[h!]
\hspace*{3cm}$H_r$\hspace*{5cm}$H_{\varphi}$\hspace*{5cm}$H_z$
\\
\begin{minipage}{16cm}
\includegraphics[height=3cm]{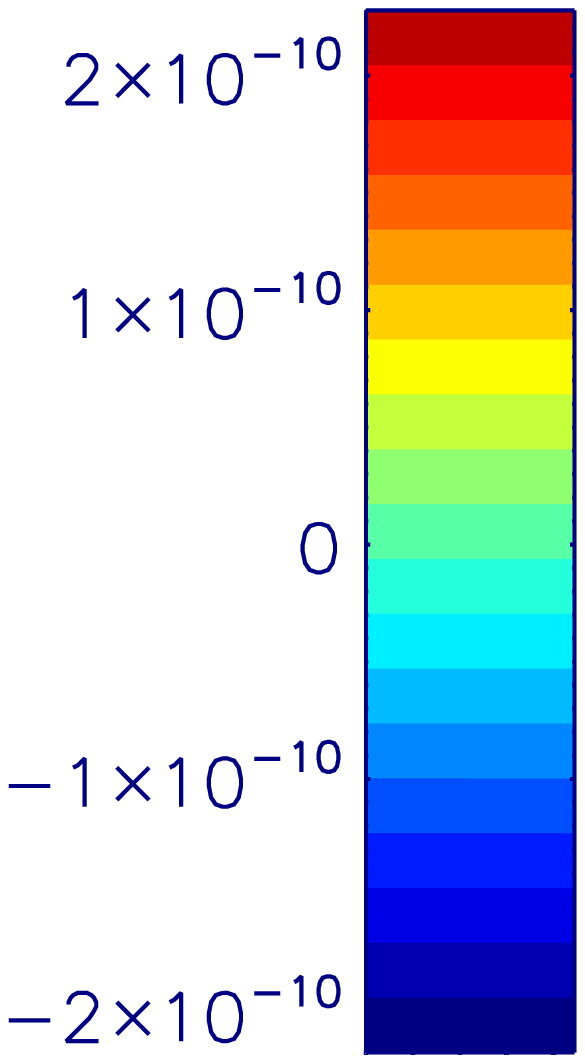}
\hspace*{-3.5cm}
\includegraphics[height=3cm]{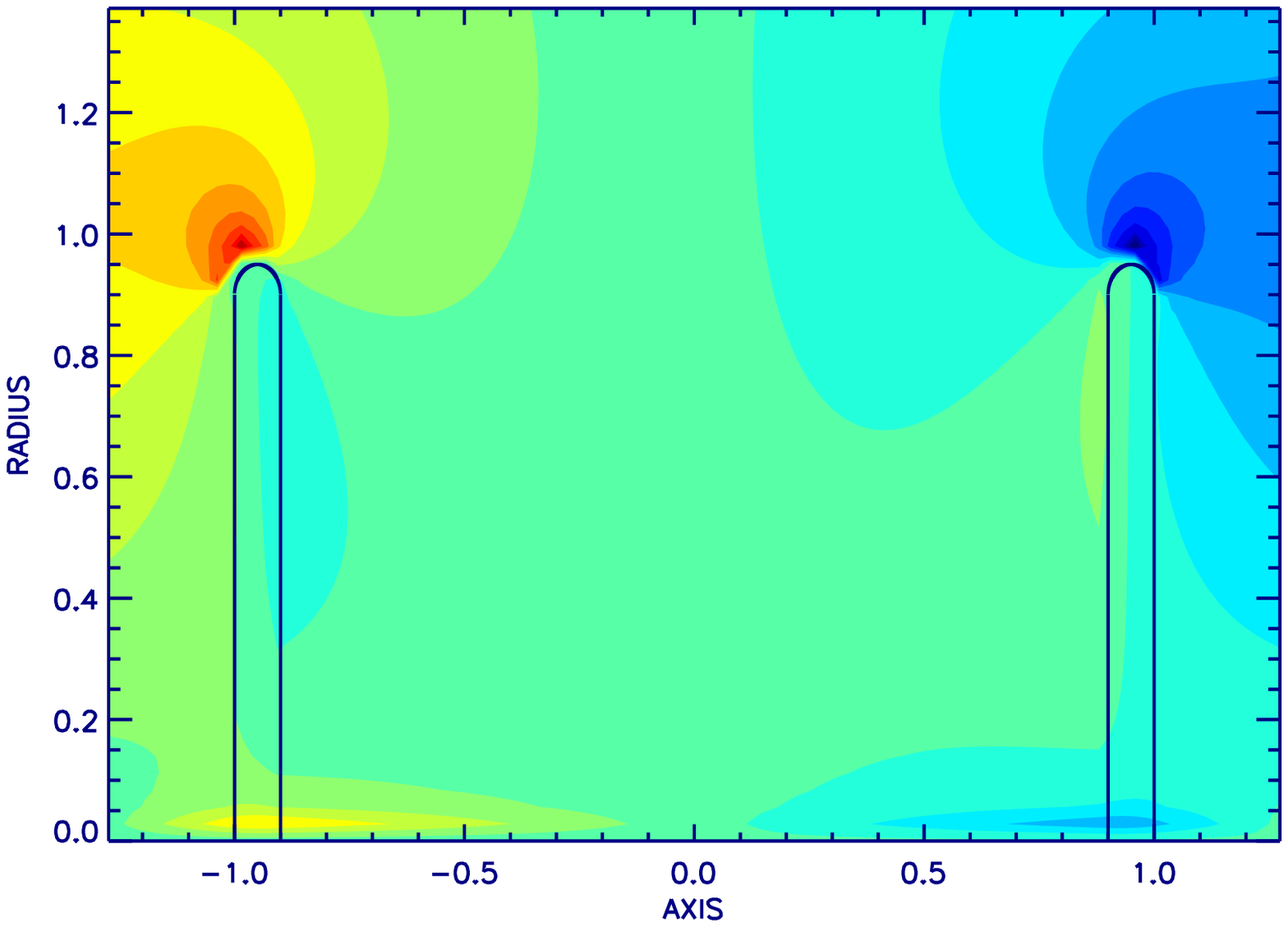}
\hspace*{0.1cm}
\includegraphics[height=3cm]{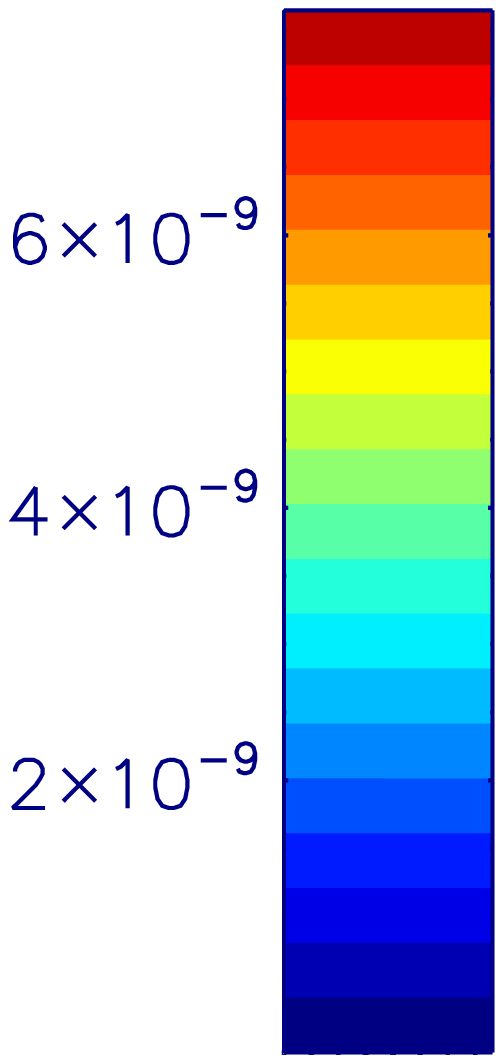}
\hspace*{-3.5cm}
\includegraphics[height=3cm]{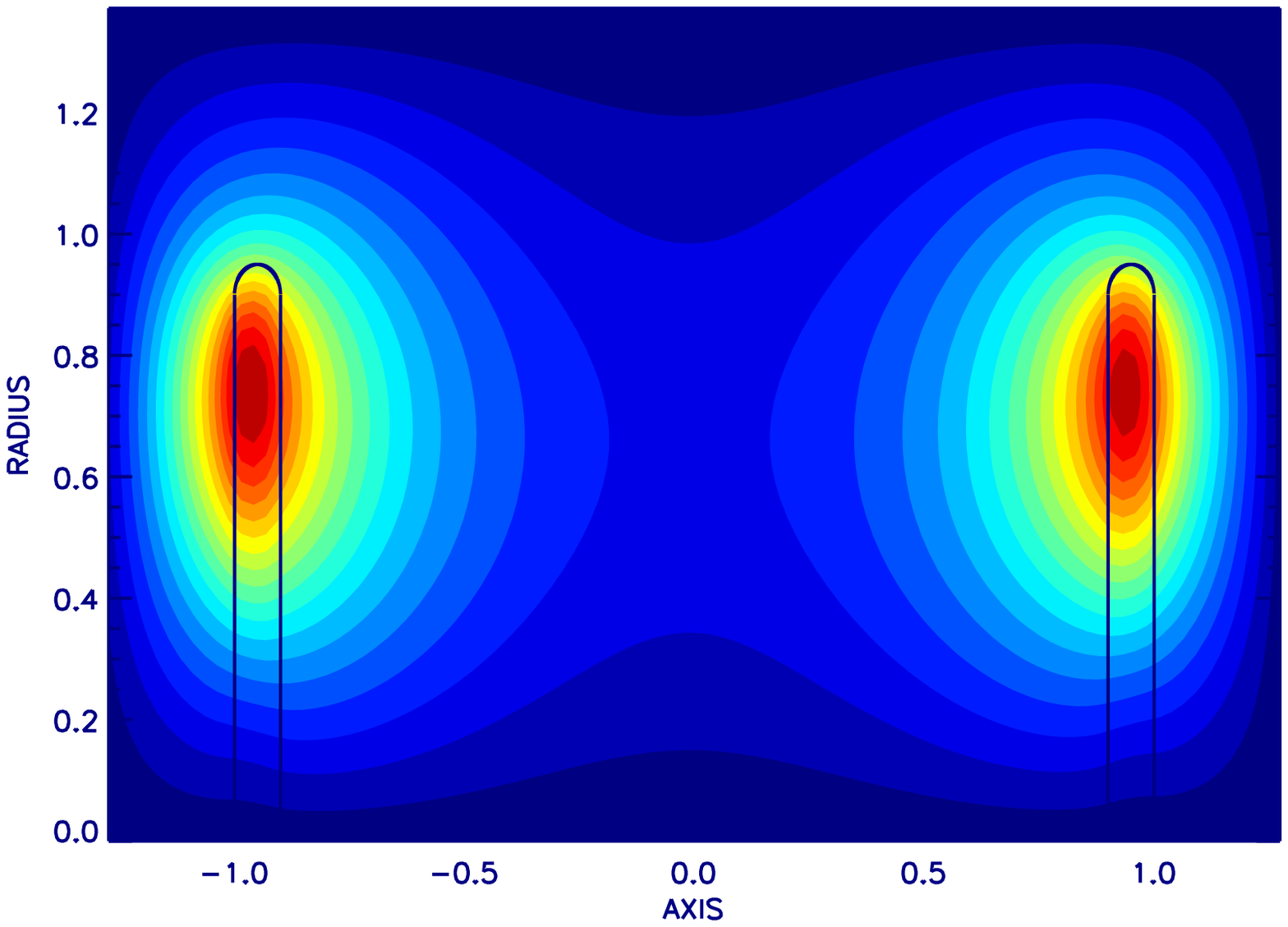}
\hspace*{0.1cm}
\includegraphics[height=3cm]{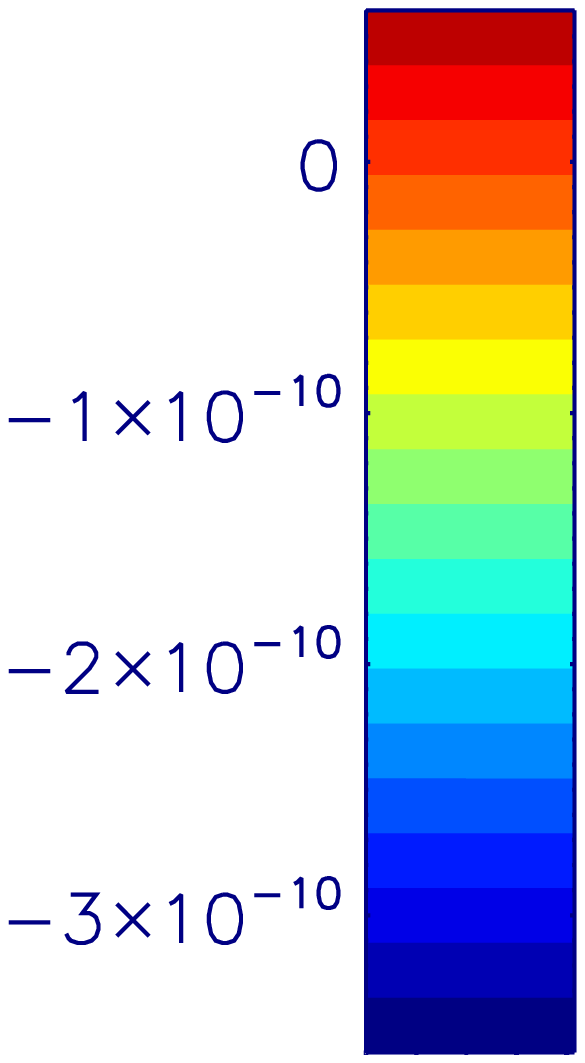}
\hspace*{-3.5cm}
\includegraphics[height=3cm]{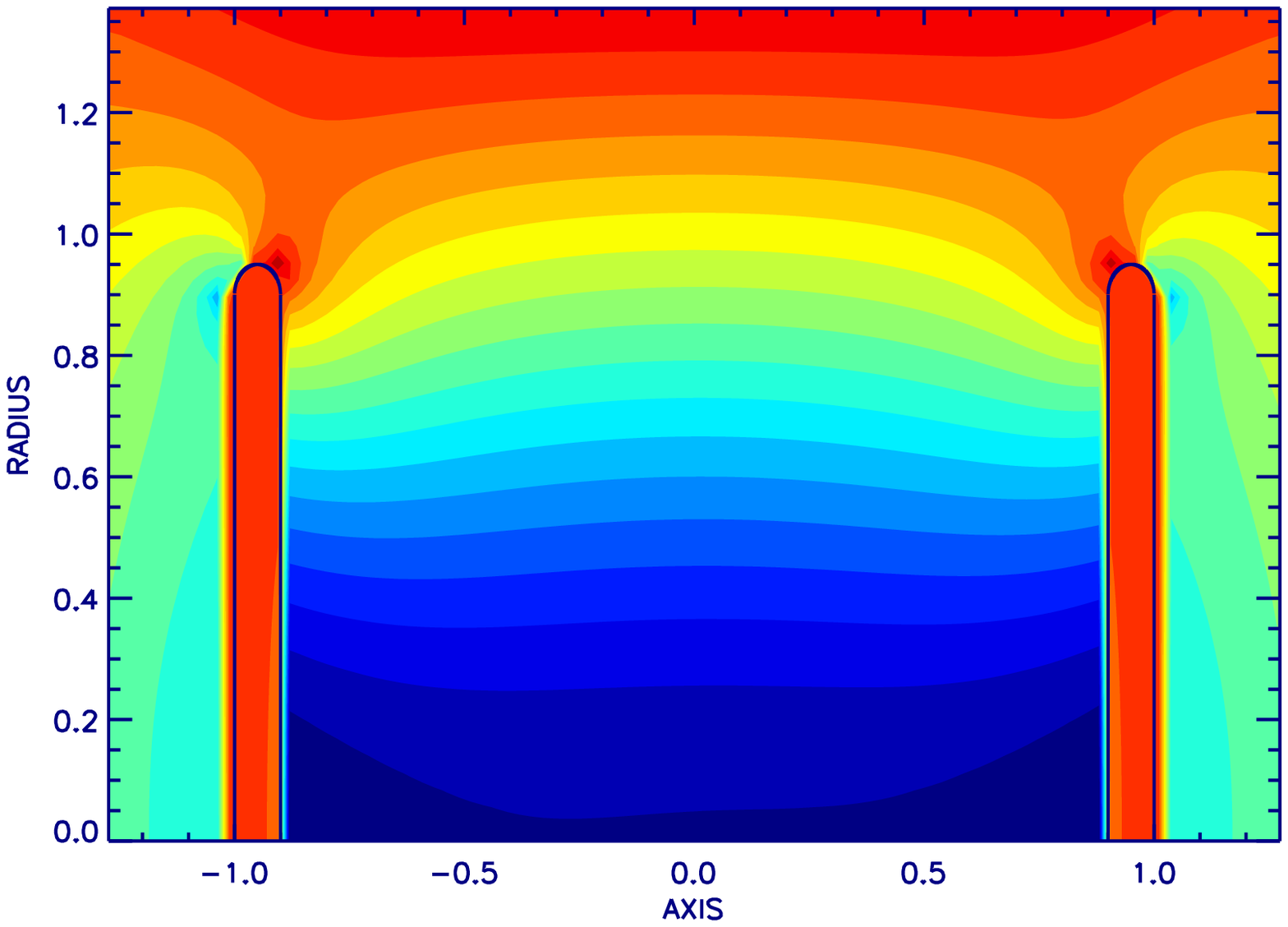}
\end{minipage}
\begin{minipage}{1.5cm}
\vspace*{-1cm}
\begin{eqnarray}
\mu_{\rm{r}}&=&100\nonumber\\
d&=&0.1\nonumber
\end{eqnarray}
\\[-1.4cm]
\begin{center}
Vacuum
\end{center}
\end{minipage}
\\
\begin{minipage}{16cm}
\includegraphics[height=3cm]{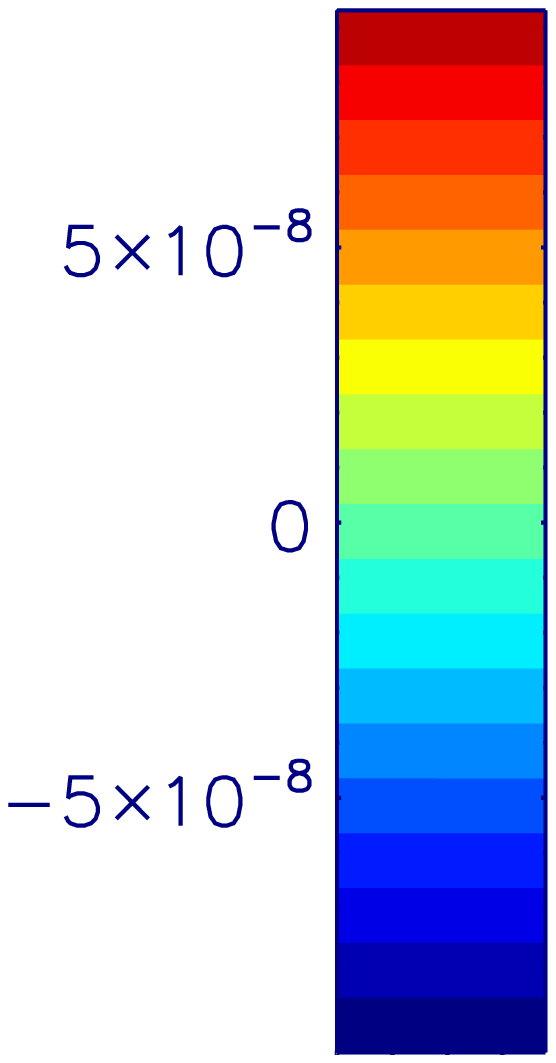}
\hspace*{-3.5cm}
\includegraphics[height=3cm]{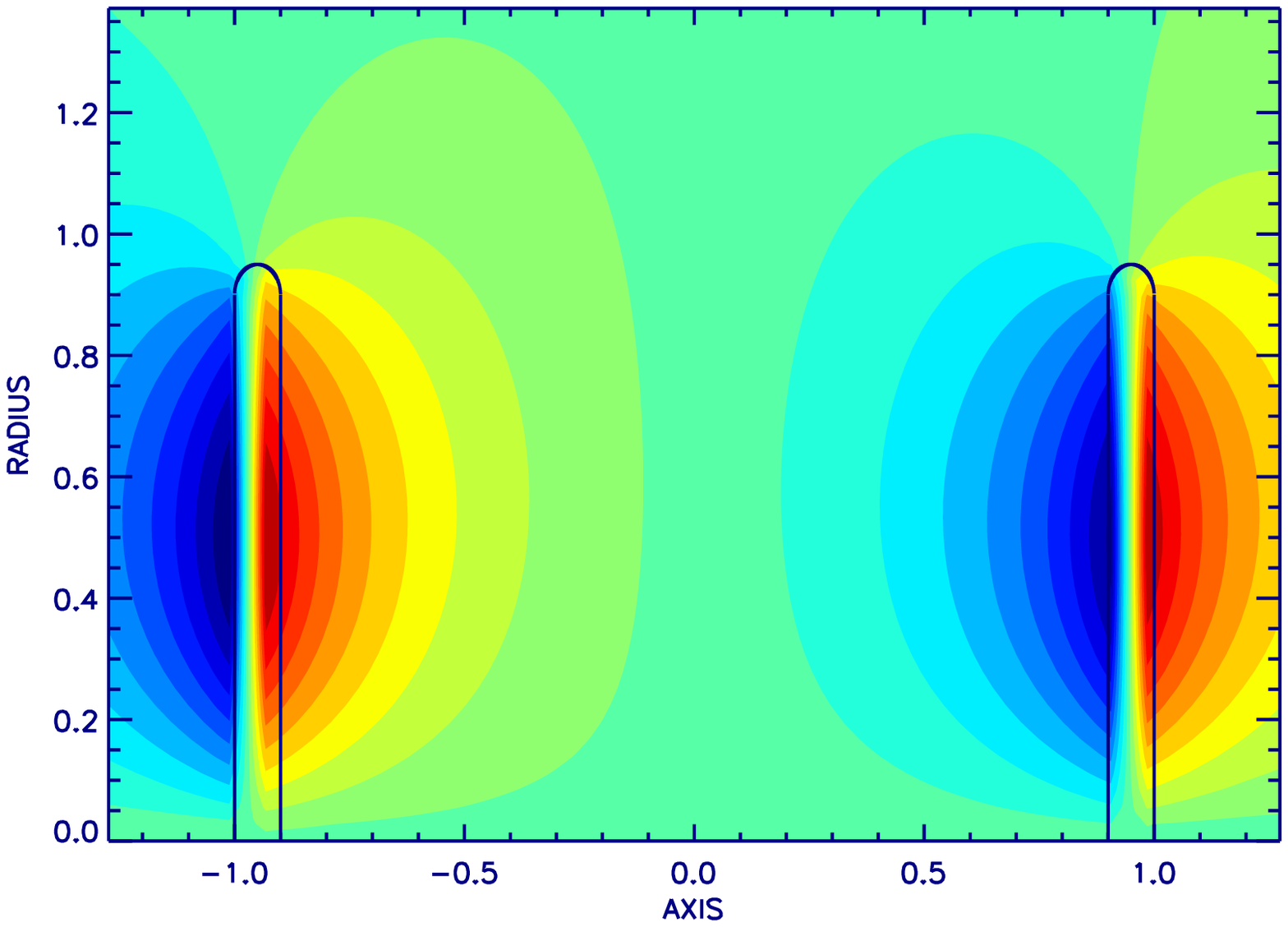}
\hspace*{0.1cm}
\includegraphics[height=3cm]{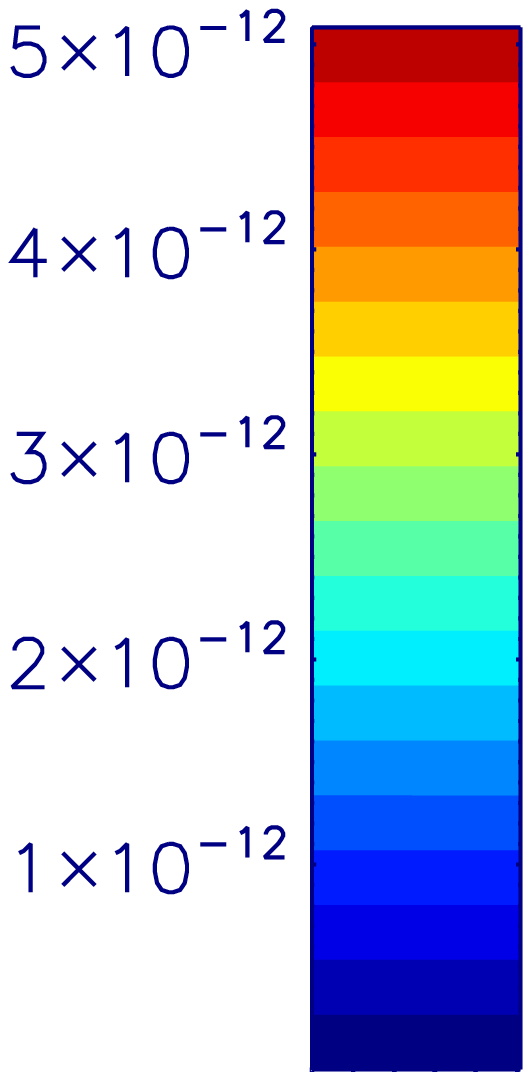}
\hspace*{-3.5cm}
\includegraphics[height=3cm]{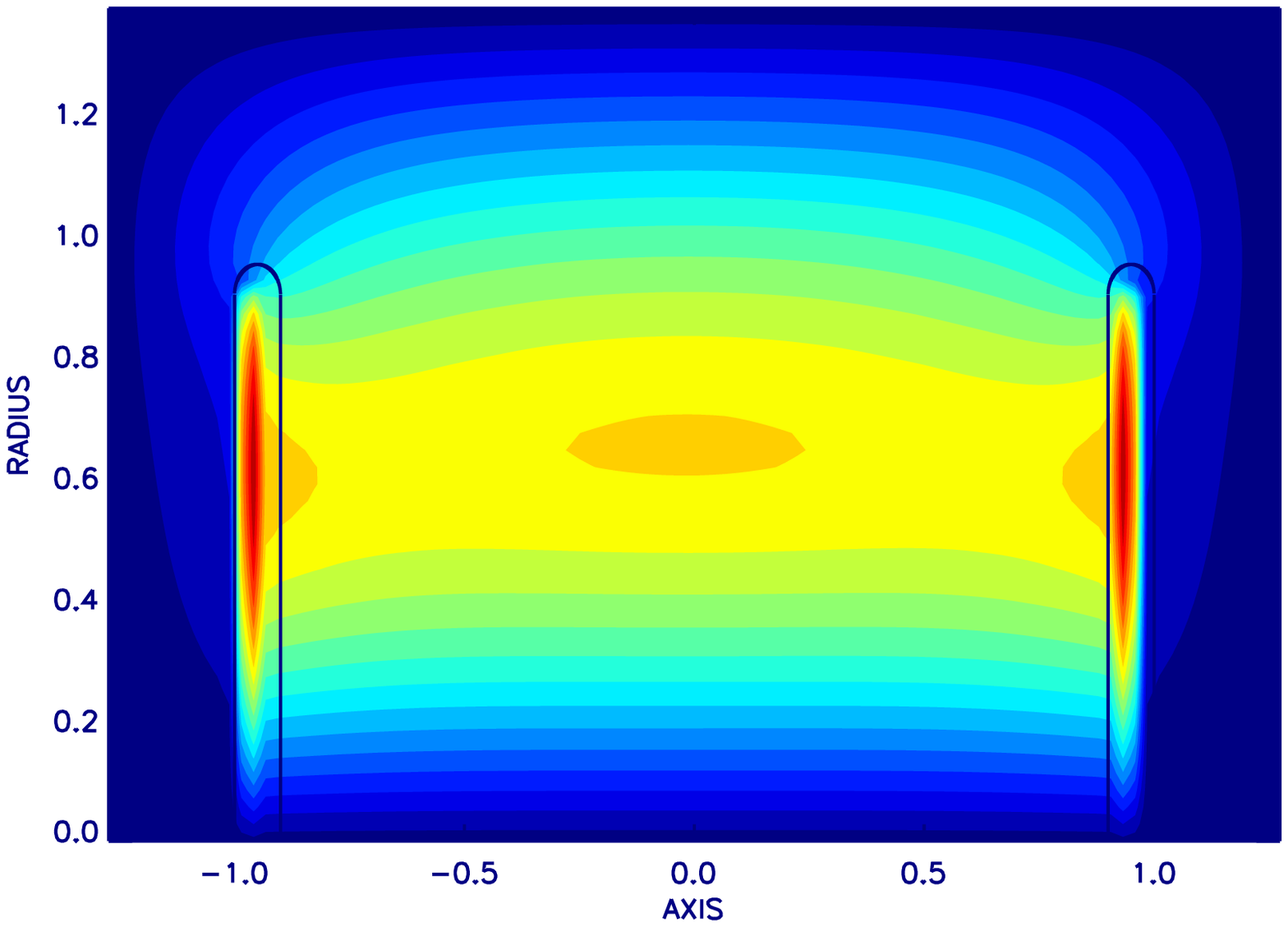}
\hspace*{0.1cm}
\includegraphics[height=3cm]{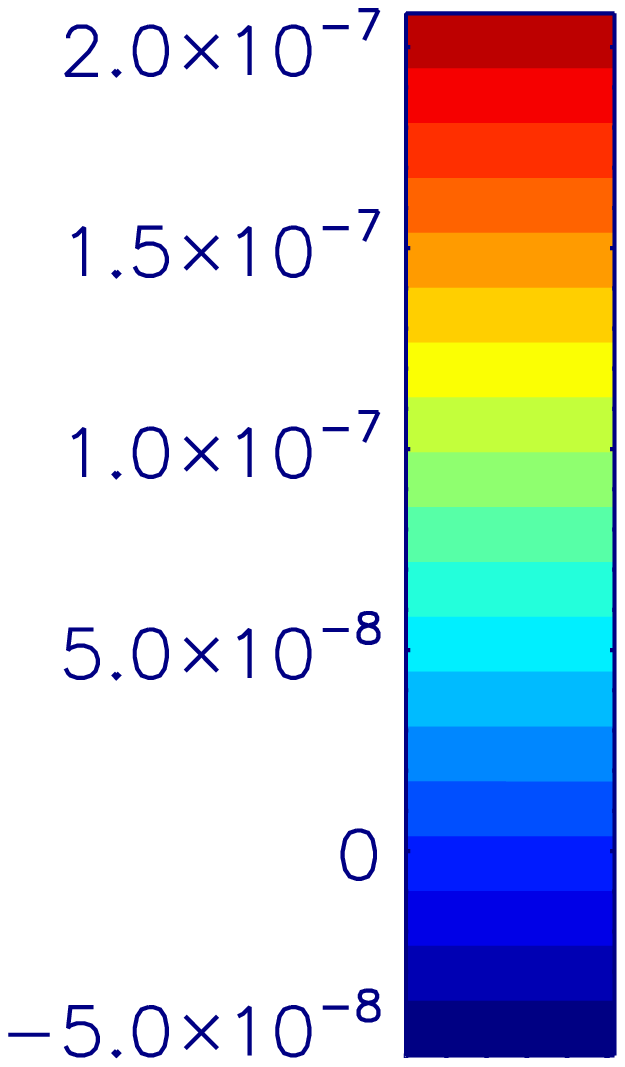}
\hspace*{-3.5cm}
\includegraphics[height=3cm]{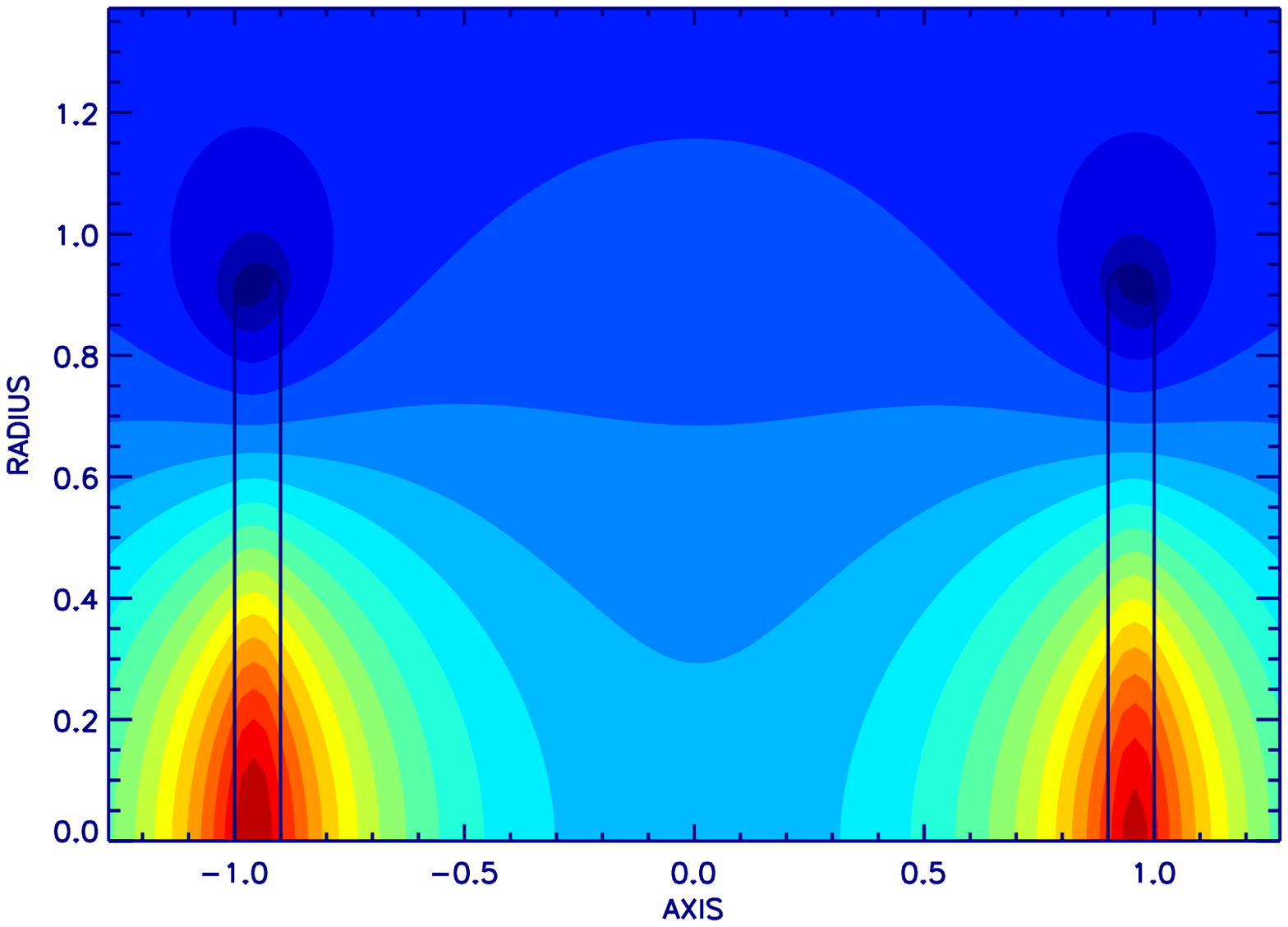}
\end{minipage}
\begin{minipage}{1.5cm}
\vspace*{-1cm}
\begin{eqnarray}
\mu_0\sigma&=&100\nonumber\\
d&=&0.1\nonumber
\end{eqnarray}
\\[-1.4cm]
\begin{center}
Vacuum
\end{center}
\end{minipage}
\caption{Ohmic decay. Axisymmetric field $\vec{H}=\mu_{\rm{r}}^{-1}\vec{B}$ for the thin
  disk case ($d=0.1$, from left to
  right: $H_r, H_{\varphi}, H_z$); Top row: $\mu_r=100$, bottom row:
  $\mu_0\sigma=100$. Insulating boundary conditions.}
\label{fig::pattern_d0p1}
\end{figure}

\begin{figure}[h!]
  \includegraphics[width=4cm]{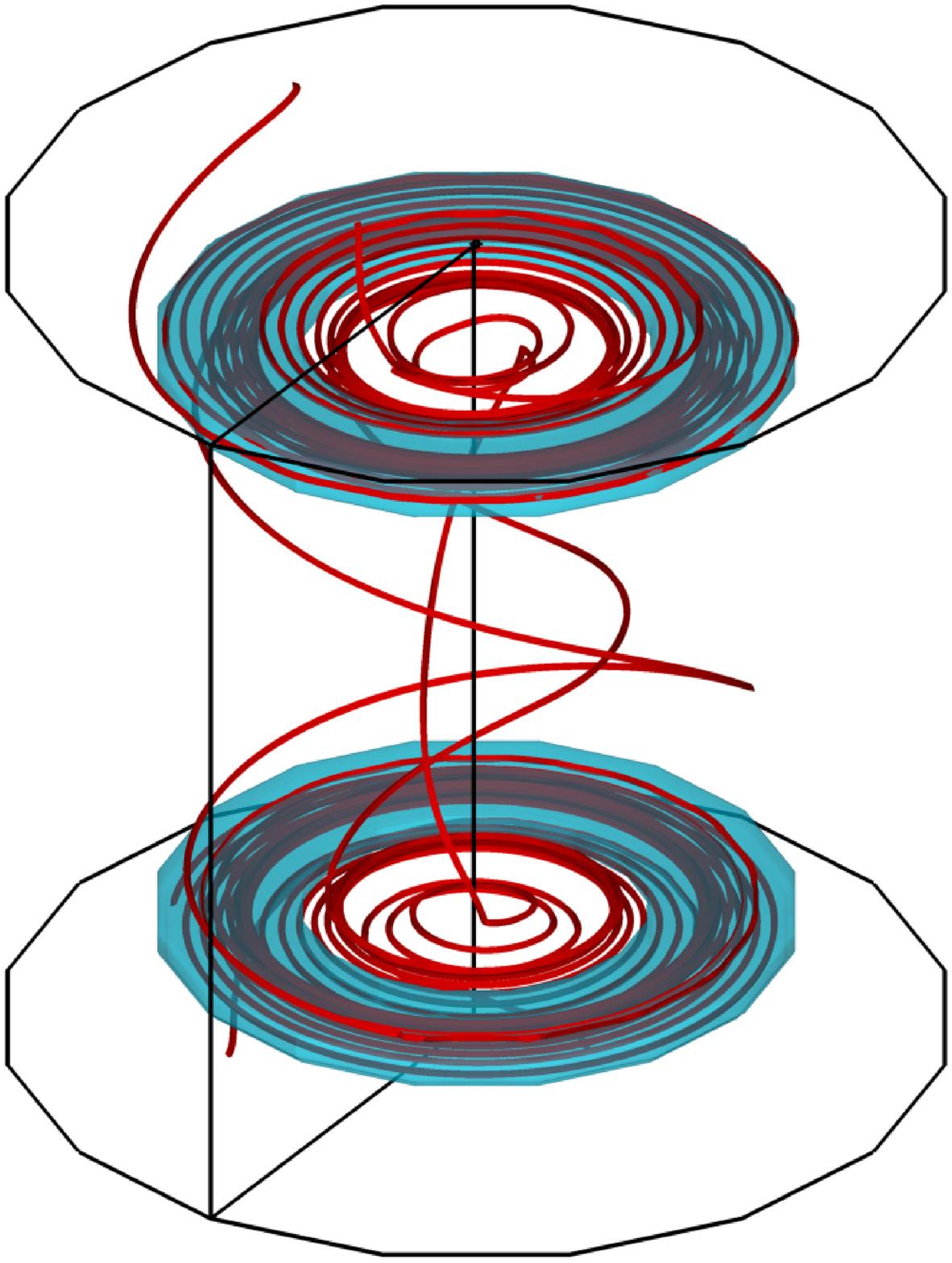}
  \includegraphics[width=4cm]{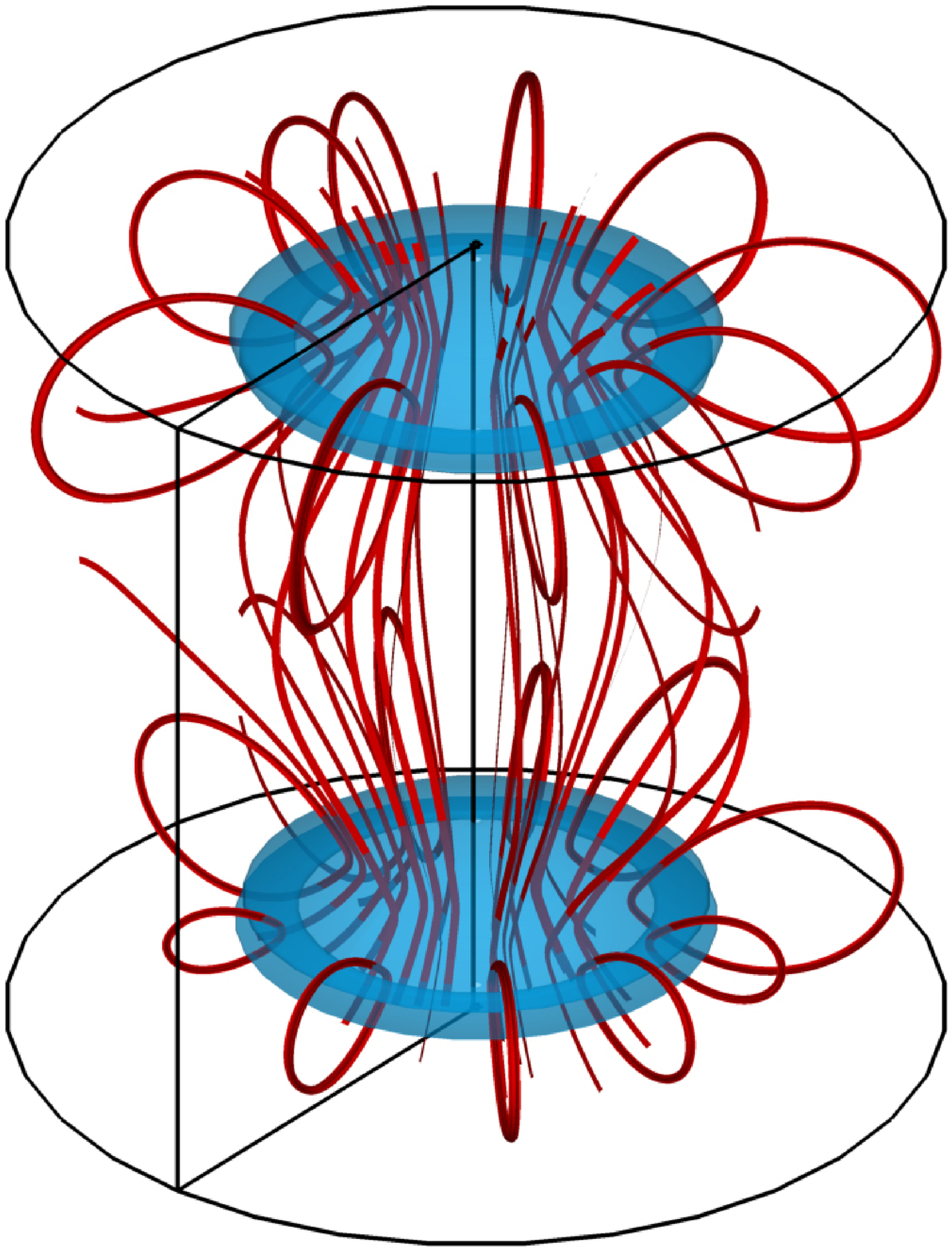}
\caption{Ohmic decay for thin disks ($d=0.1$). Left panel: $\mu_{\rm{r}}=100$,
  right panel: $\mu_0\sigma=100$). The isosurfaces present the magnetic energy
density at 25\% of its maximum value.}\label{fig::3d_thinn_disk}
\end{figure}

\begin{figure}[h!]
\includegraphics[width=4cm,angle=0]{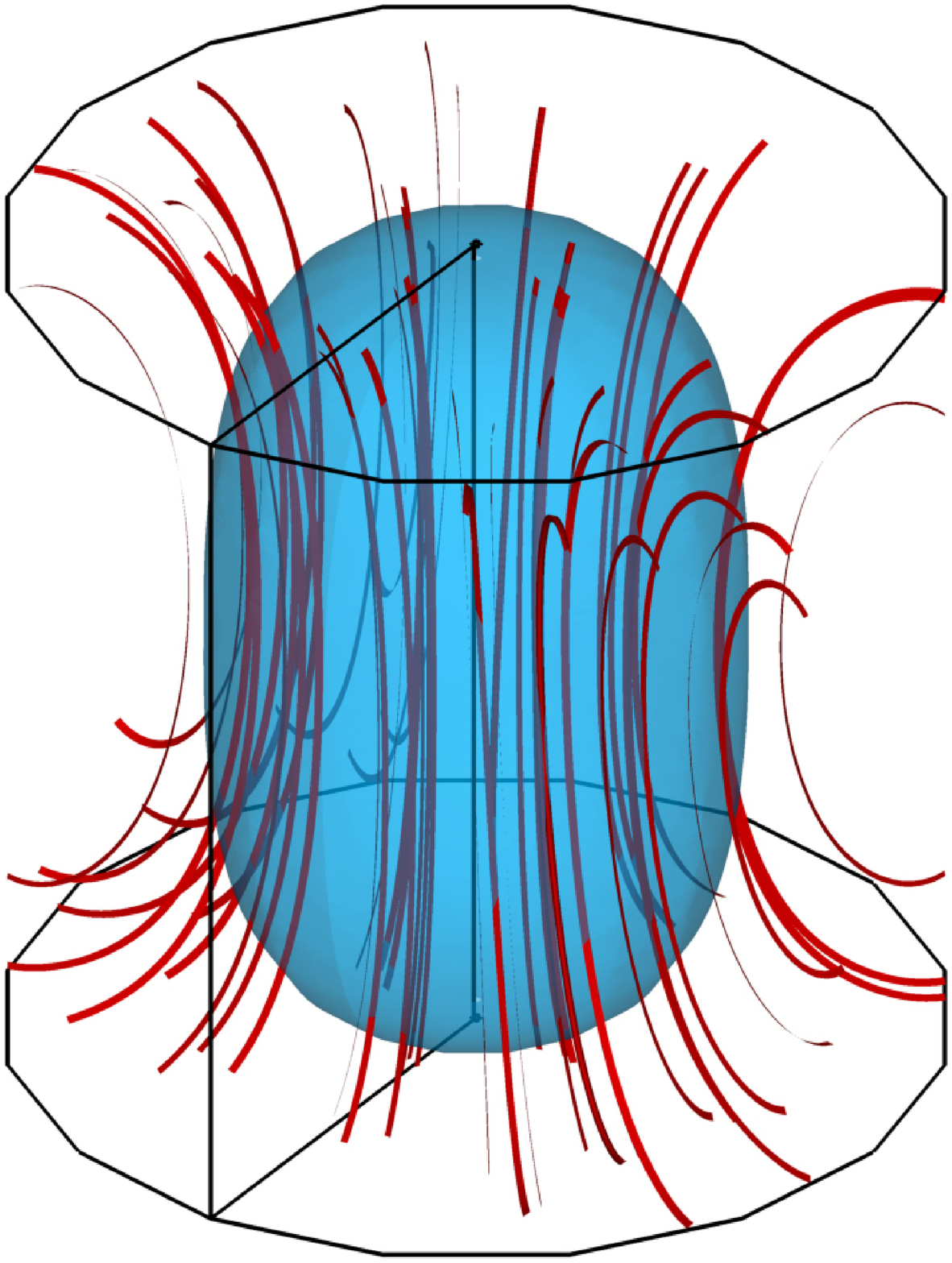}
\nolinebreak[4!]
\hspace*{0.3cm}
\includegraphics[width=4cm,angle=0]{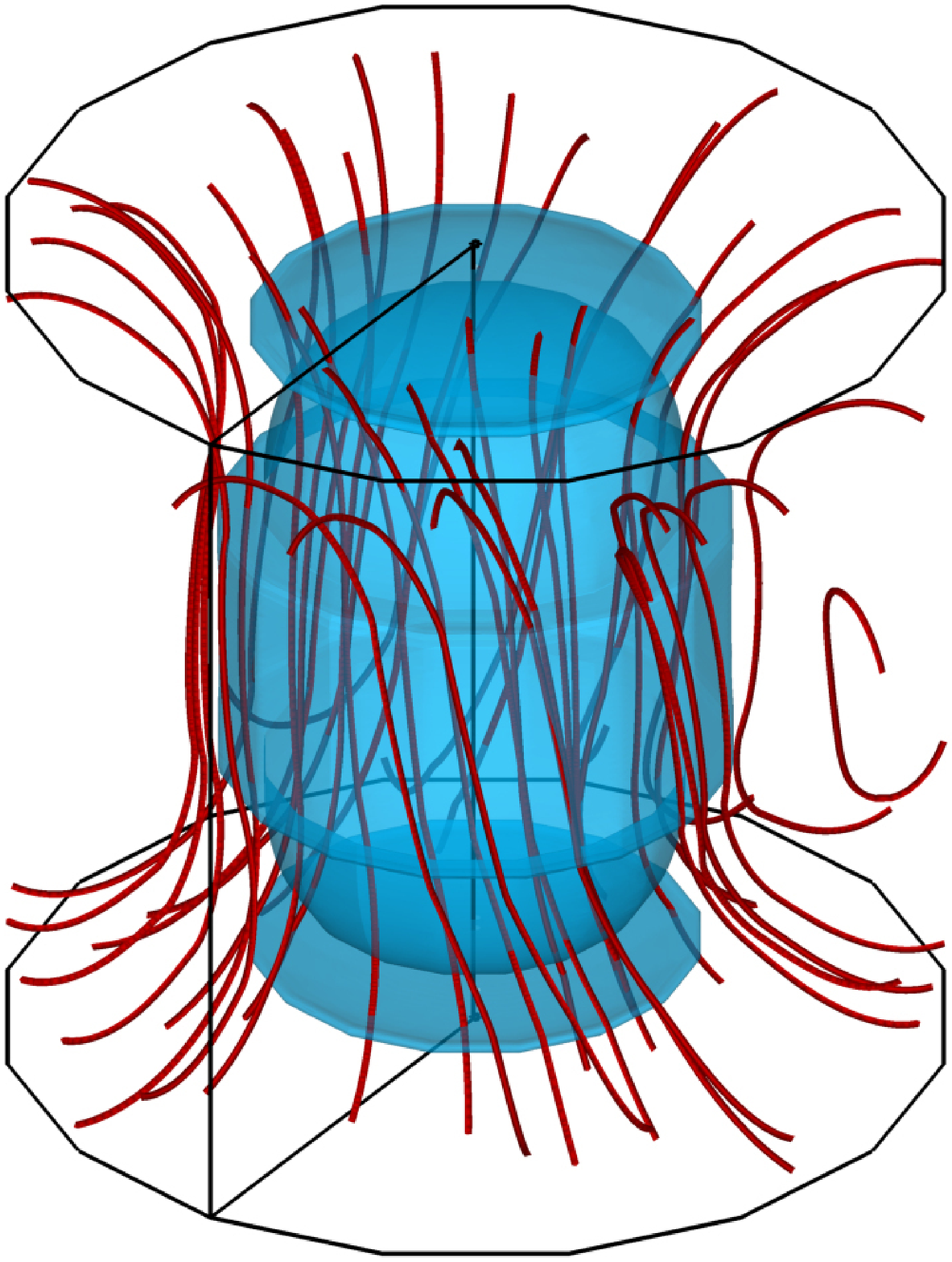}
\nolinebreak[4!]
\hspace*{0.3cm}
\includegraphics[width=4cm,angle=0]{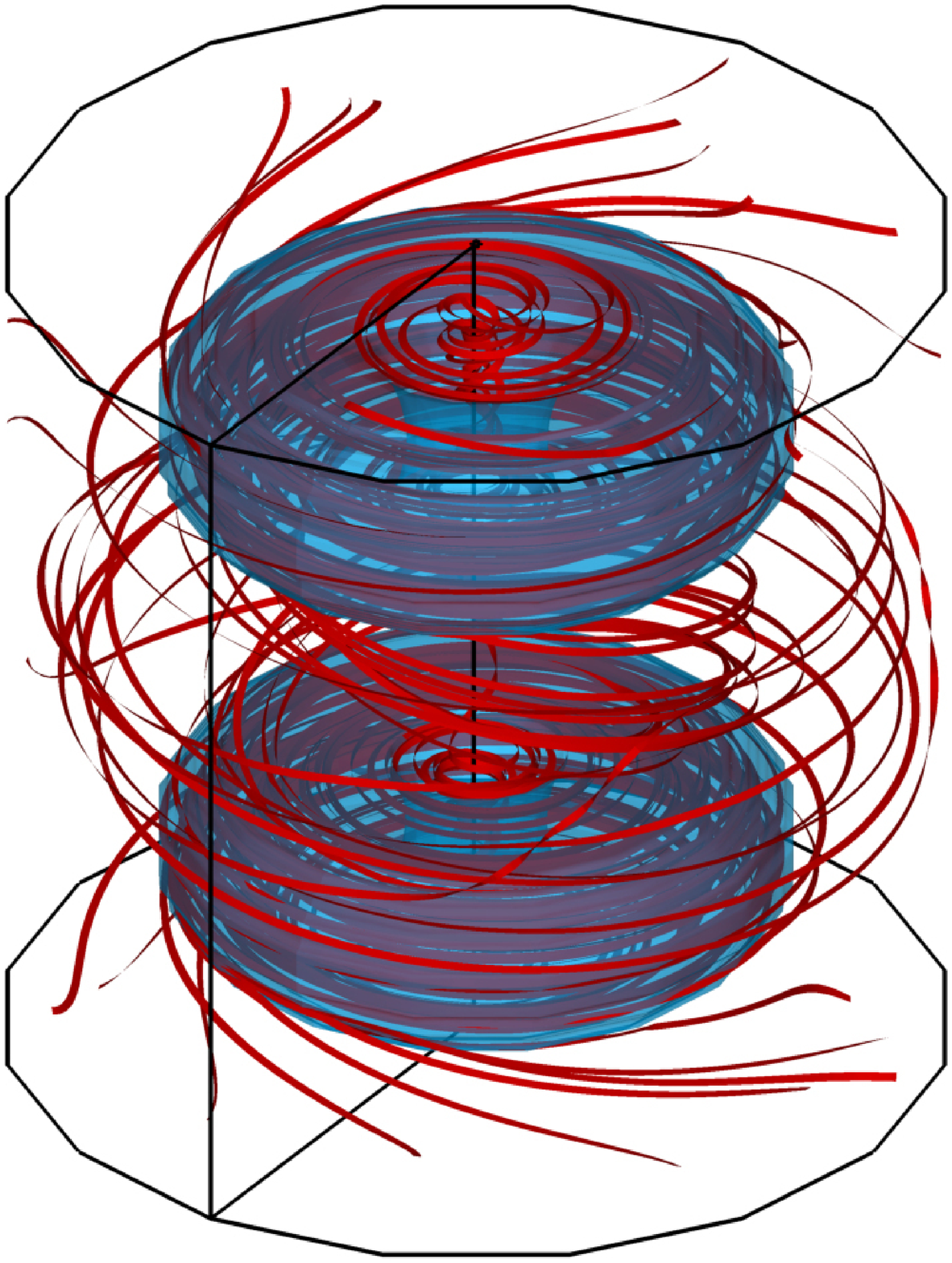}
\nolinebreak[4!]
\hspace*{0.3cm}
\includegraphics[width=4cm,angle=0]{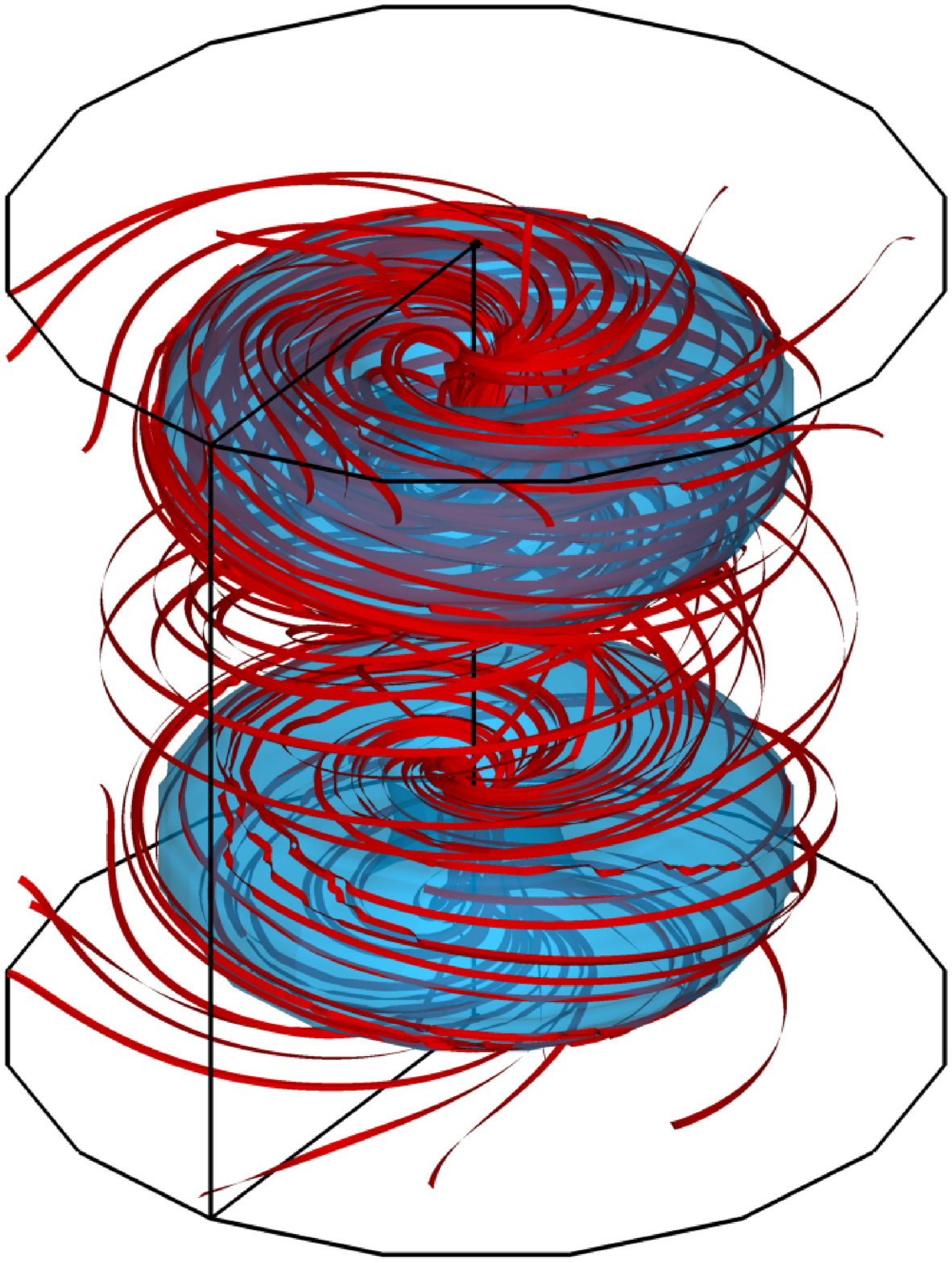}
\caption{Ohmic decay. The blue transparent isosurfaces present the magnetic
  energy density at 25\% of the maximum value and 
  the red streamlines show the field structure for $d=0.6$ and (from left to
  right): $\mu_{\rm{r}}=1, 2, 10, 100$ (corresponding to
  $\mu_{\rm{r}}^{\rm{eff}}=1, 1.2, 2.7, 19.5$).}\label{fig::mu_d} 
\end{figure}

%
%
%
A couple of simulations have been performed utilizing vanishing
tangential field (VTF) boundary conditions (sometimes also called Pseudo
Vacuum) which enforce a field geometry on the outer boundary that resembles
the behavior in case of external materials with infinite permeability. 
Figure~\ref{fig::pattern} shows the structure of the field geometry with the
container embedded in vacuum (upper part) and for VTF boundary conditions
(lower part).  
Whereas a significant impact occurs without disks the influence of the
external boundary conditions on the field geometry 
remains negligible if the disk permeability or conductivity is large enough. 
A more noticeable difference between the field distribution results from the
comparison of high permeability disks with high conductive disks. 
In the first case the field structure is dominated by two distinct annular
accumulations of azimuthal magnetic field -- essentially located within the
disks.
The high conducting disk case is characterized by the domination of
the axial field with a slab like structure
concentrated around the axis in which the high conducting medium is
embedded.   
A remarkable change in the field structure is obtained for the thin
disk case ($d=0.1$, see Fig.~\ref{fig::pattern_d0p1} \&
\ref{fig::3d_thinn_disk}). 
In case of high $\mu_{\rm{r}}$ the azimuthal field is dominated by two ring
like structures centered on the outer part of both disks.
The radial field is concentrated within two highly localized paths on
the outer edge of the disk whereas the axial field has become nearly
independent from $z$ except close to the disks because the jump
conditions inhibit the
constitution of $H_z$ within the disks.  
For the high conductive disks, the differences in the field pattern
between $ d=0.6$ and $d=0.1$ are less significant and
the torus-like structure of the poloidal field component is
always dominating (see right panel in Fig.~\ref{fig::3d_thinn_disk})'
For increasing permeability, the axisymmetric mode changes from
a poloidal dominant structure to a toroidal dominant structue
(see Fig.~\ref{fig::mu_d} for $d=0.6$). The mode crossing occurs for
$\mu_{\rm{r}}^{\rm{eff}} \approx 1.5$. 
\subsection{Decay rates and dominating mode}
\begin{figure}[h!]
\includegraphics[width=8.2cm]{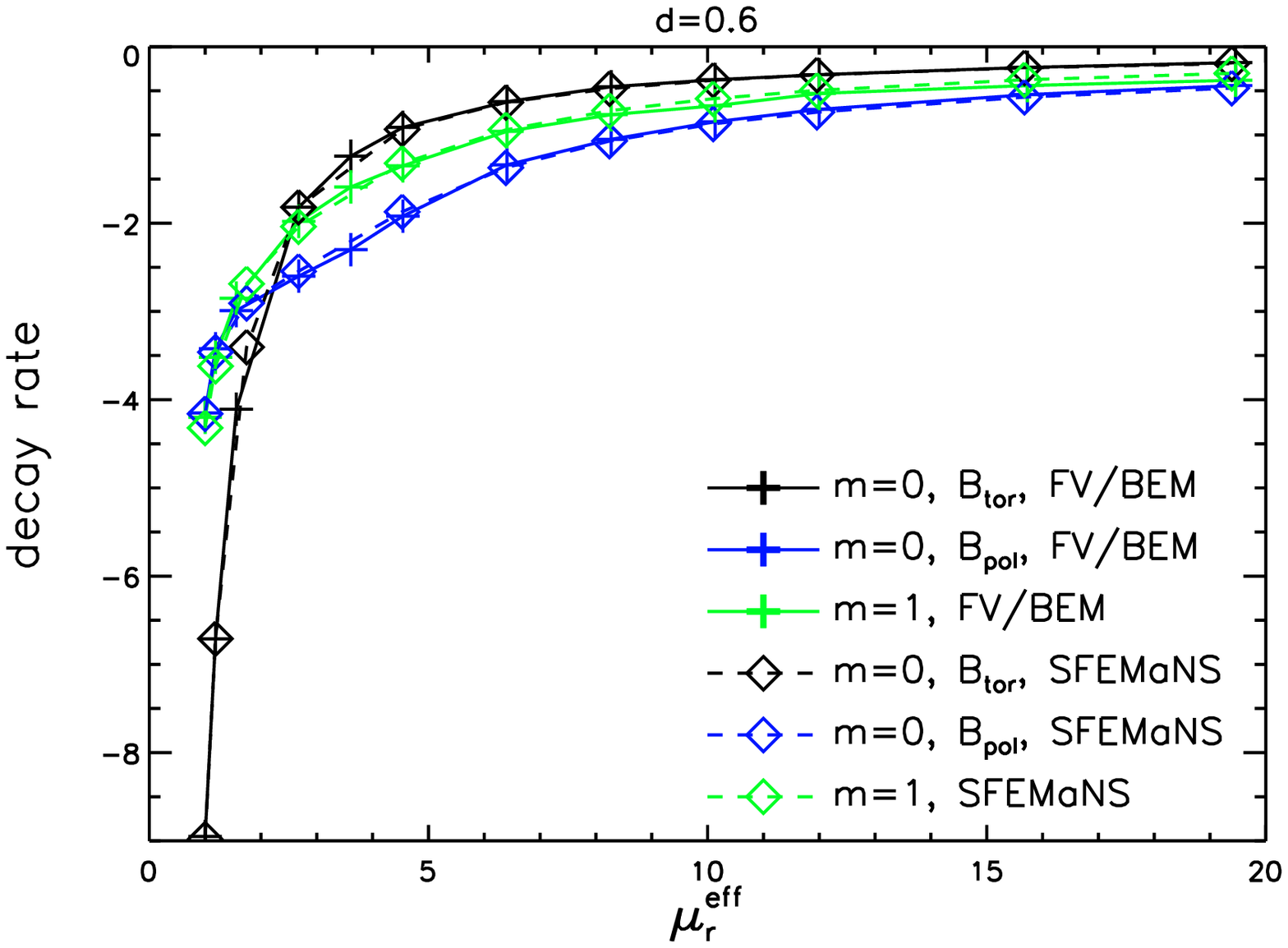}
\includegraphics[width=8.2cm]{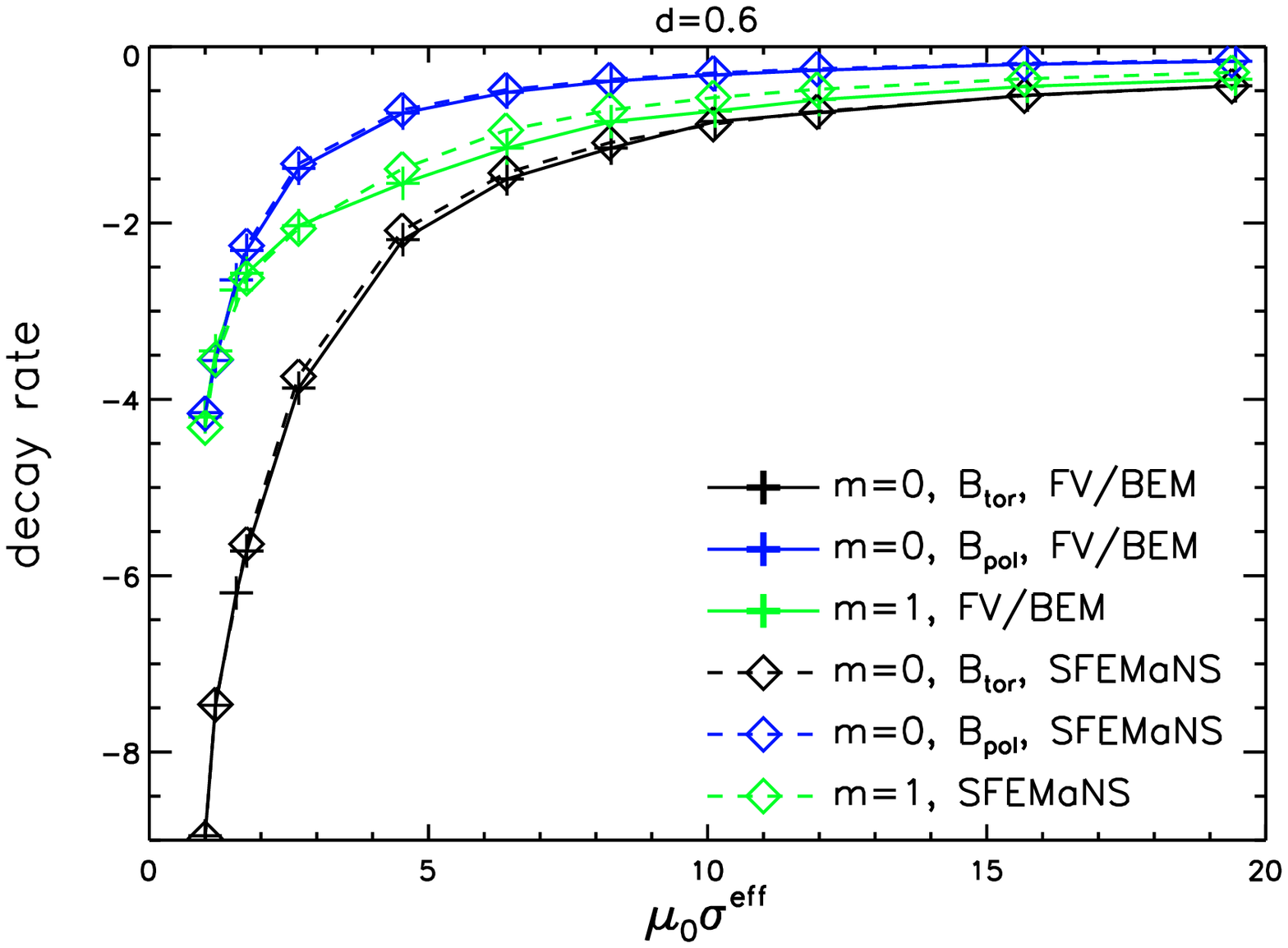}
\\
\includegraphics[width=8.2cm]{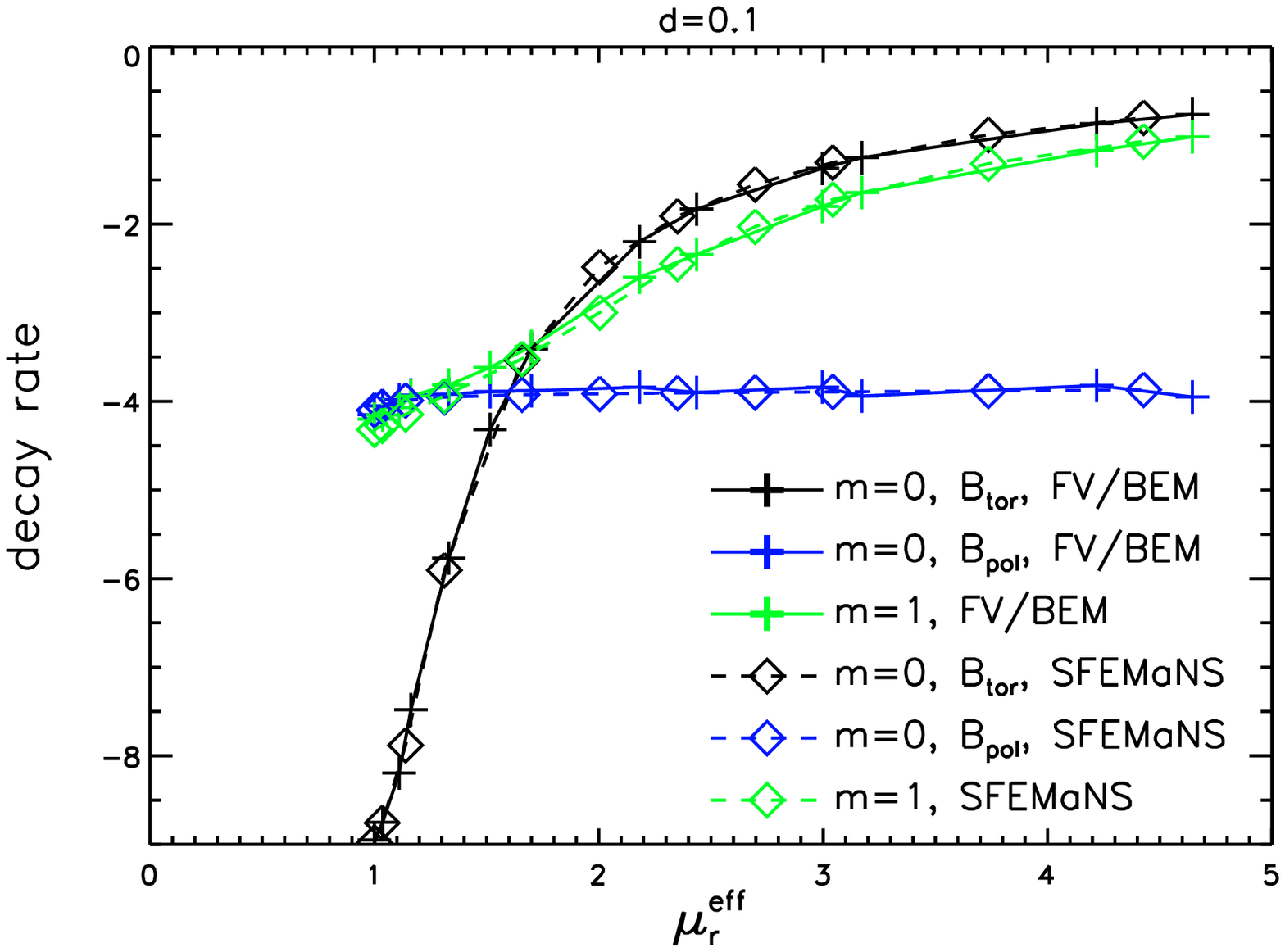}
\includegraphics[width=8.2cm]{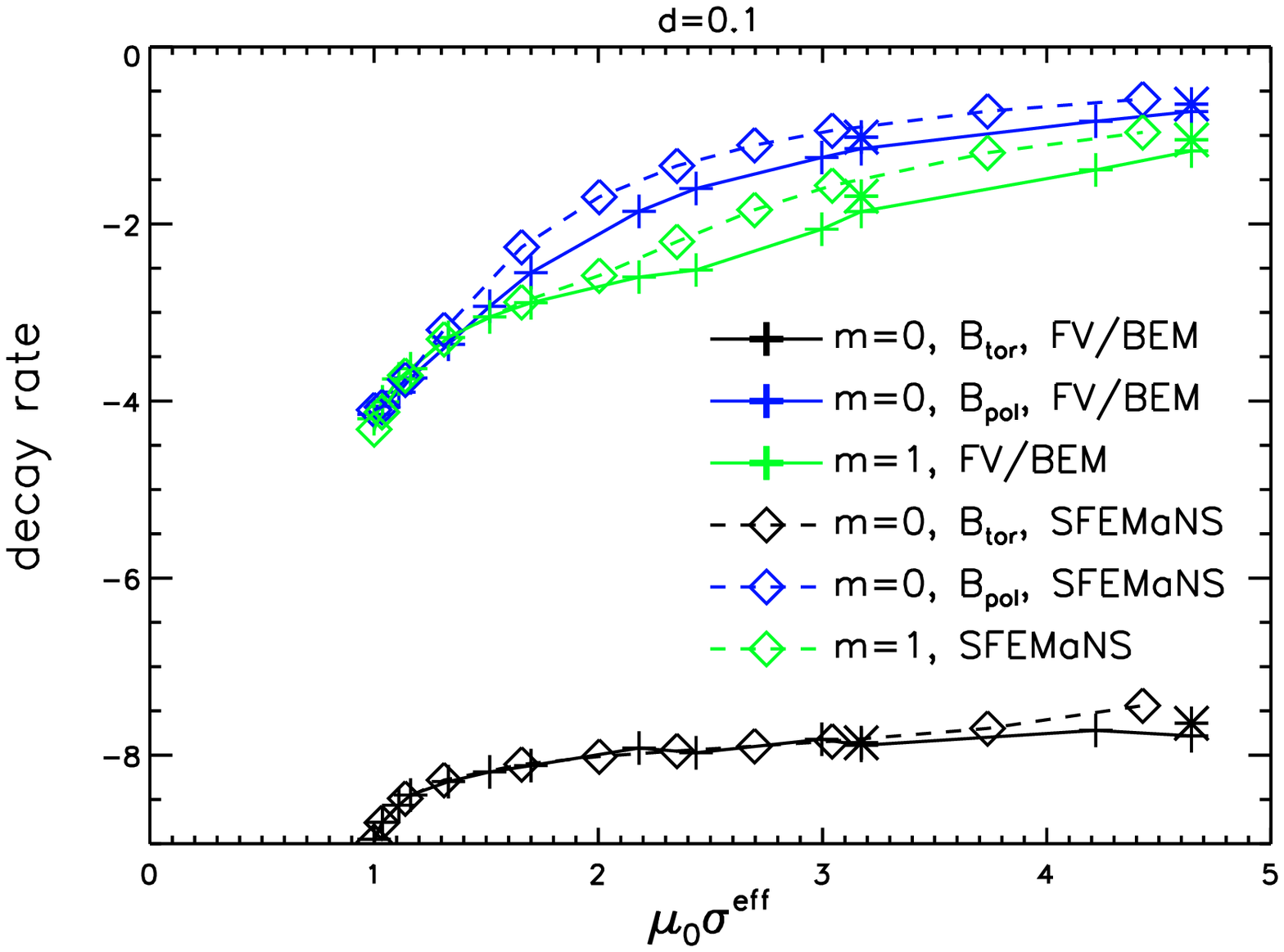}
\caption{Decay rates with vacuum BC against $\mu_{\rm{r}}^{\rm{eff}}$ (left column) and against $\mu_0\sigma^{\rm{eff}}$
(right column) for $d=0.6$ (top row) and $d=0.1$ (bottom row).
The solid curves show the results obtained from the hybrid FV/BEM scheme and
the dashed curves denote the results from the SFEMaNS scheme. The stars in the
lower right panel present the results of a FV/BEM run with higher resolution demonstrating
that the FV/BEM algorithm might approach the SFEMaNS data.
}
\label{fig::decay_rates_vac}
\end{figure}

The temporal behavior of the magnetic eigenmodes follows an exponential law
$B\propto e^{\gamma t}$ where $\gamma$ denotes the growth or decay rate.
Figure~\ref{fig::decay_rates_vac} shows the magnetic field decay rates for a thick disk
($d=0.6$) and a thin disk ($d=0.1$) against 
$\mu_{\rm{r}}^{\rm{eff}}$ (left column) and against $\sigma^{\rm{eff}}$ (right
column) where $\mu_{\rm{r}}^{\rm{eff}}$ and $\sigma^{\rm{eff}}$ denote effective values for
permeability and conductivity that are defined as
$\mu_{\rm{r}}^{\rm{eff}}={V}^{-1}\int\mu_{\rm{r}}(\vec{r})dV$
and $\sigma^{\rm{eff}}={V}^{-1}\int\sigma(\vec{r})dV$ 
with $V$ the volume of the cylindrical domain.
The essential properties of the field behavior can be summarized as follows:
The presence of high permeability/conductivity material enhances axisymmetric
and $(m=1)$ field modes. However, for decreasing thickness the enhancement
works selectively for the axisymmetric toroidal field (in case of high $\mu_{\rm{r}}$),
respectively for the poloidal axisymmetric mode (in case of high $\sigma$). 
For the thin disk the decay rate of the poloidal (respectively toroidal)
field component remains nearly independent of the permeability
(respectively conductivity). 
Note the changeover of a dominating axisymmetric poloidal mode to the
dominating axisymmetric toroidal mode for the high permeability disks which
occurs irrespective of the disk thickness around
${\mu_{\rm{r}}}\approx 1.5$ (see also Fig.~\ref{fig::mu_d}).
%
%
%
%
%
%
%
The mode crossing does not occur for a high conducting disk, where the
$(m=1)$ mode and the poloidal axisymmetric mode have nearly the same decay rate for $\mu_0\sigma^{\rm{eff}}
\la 1.5$ and the axisymmetric poloidal mode dominates for large $\sigma$. 

Small deviations between both algorithms occur in case of thin disks
($d=0.1$) for the axisymmetric poloidal mode and for the $(m=1)$ mode.
A couple of simulations with higher resolution in axial direction (marked by
the blue and the green stars in the lower right panel of Fig.~\ref{fig::decay_rates_vac}) show that
these deviations
are most probably the result of poor resolution in case of the
FV/BEM scheme because only few grid cells are available to resolve the
vertical structure of the disk 
(namely 40 mesh points for SFEMaNS against 6 mesh points for FV/BEM). 
\begin{figure}[h!]
%
%
\includegraphics[width=6cm]{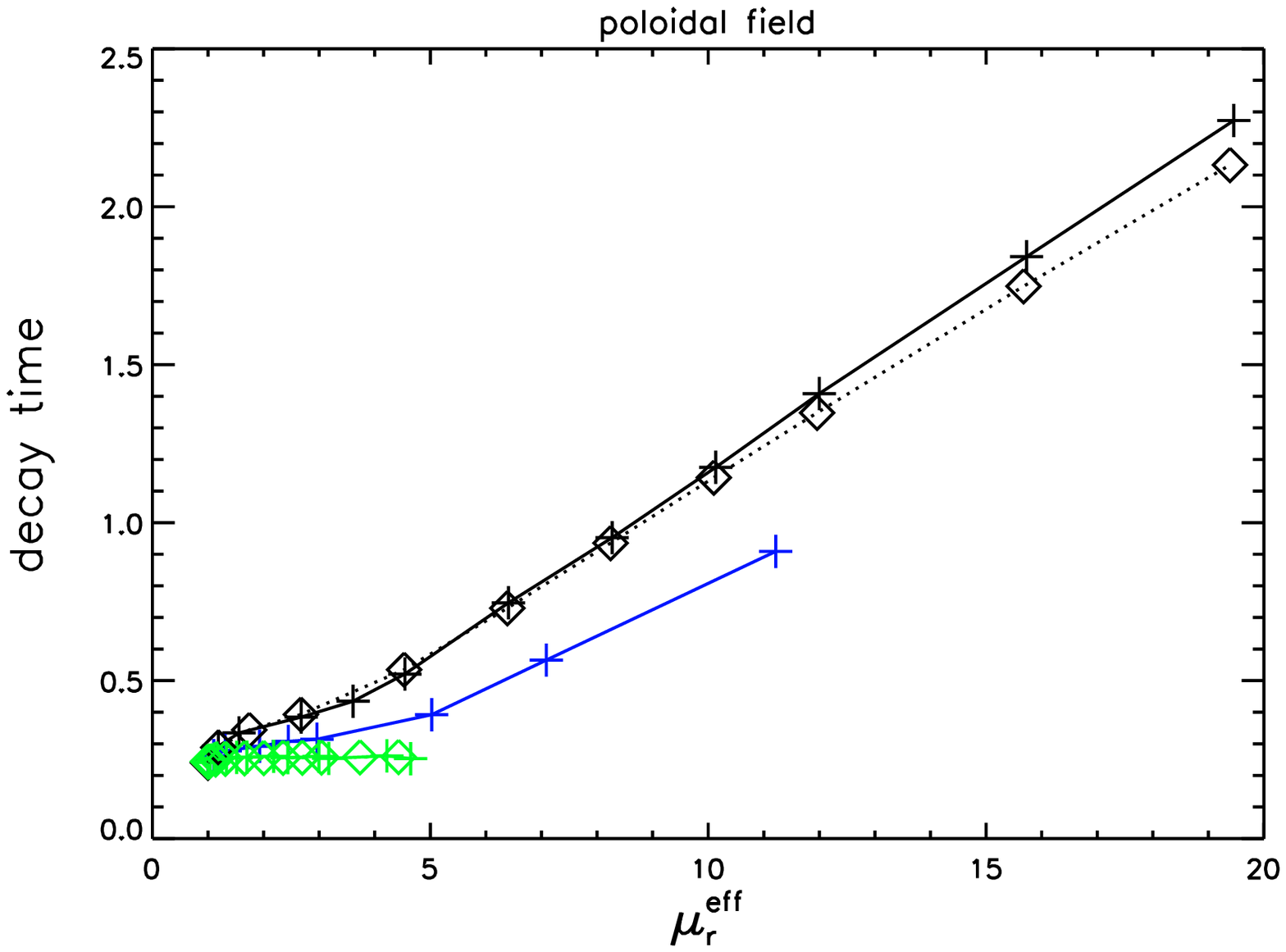}
\includegraphics[width=6cm]{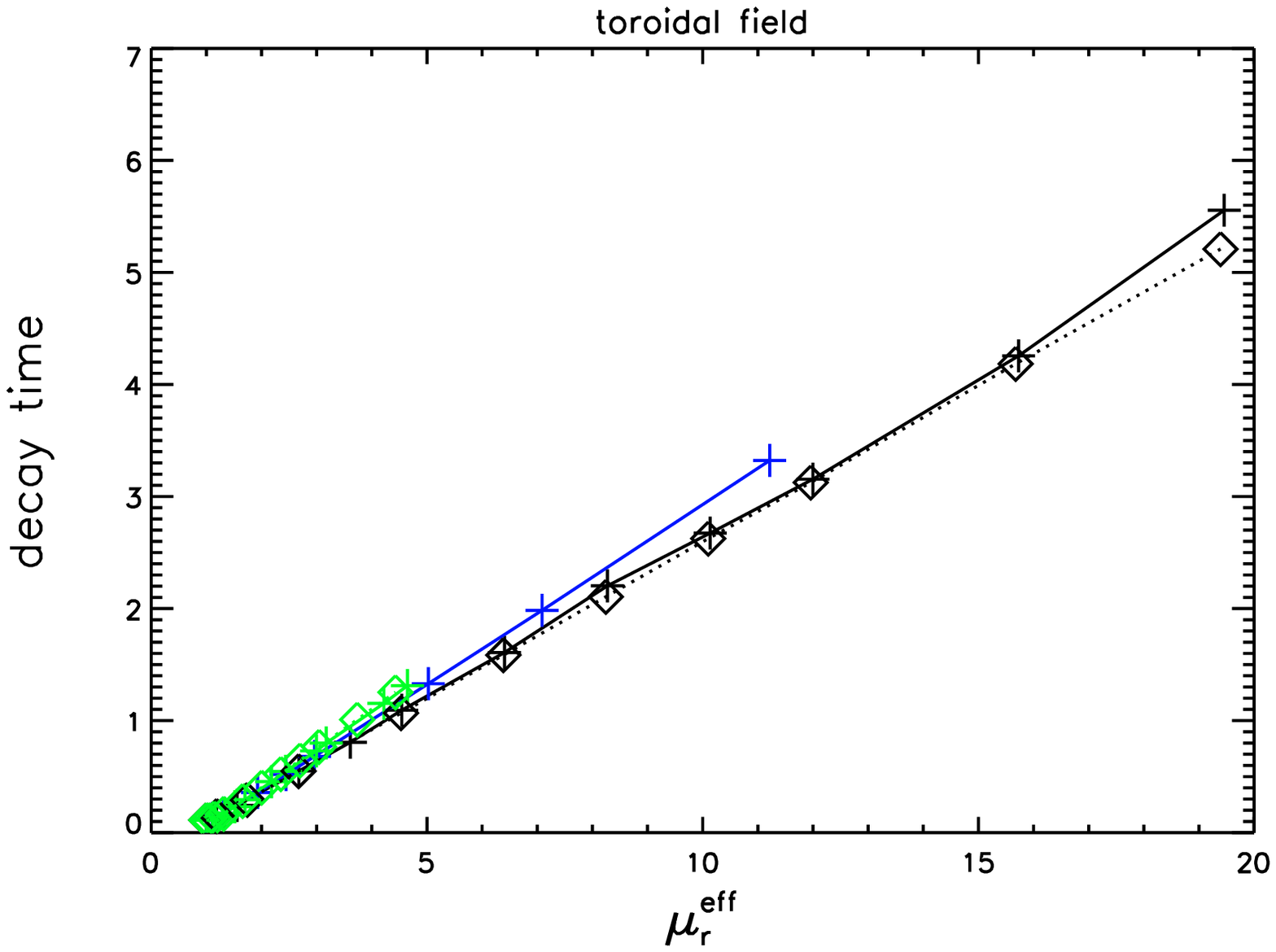}
\includegraphics[width=6cm]{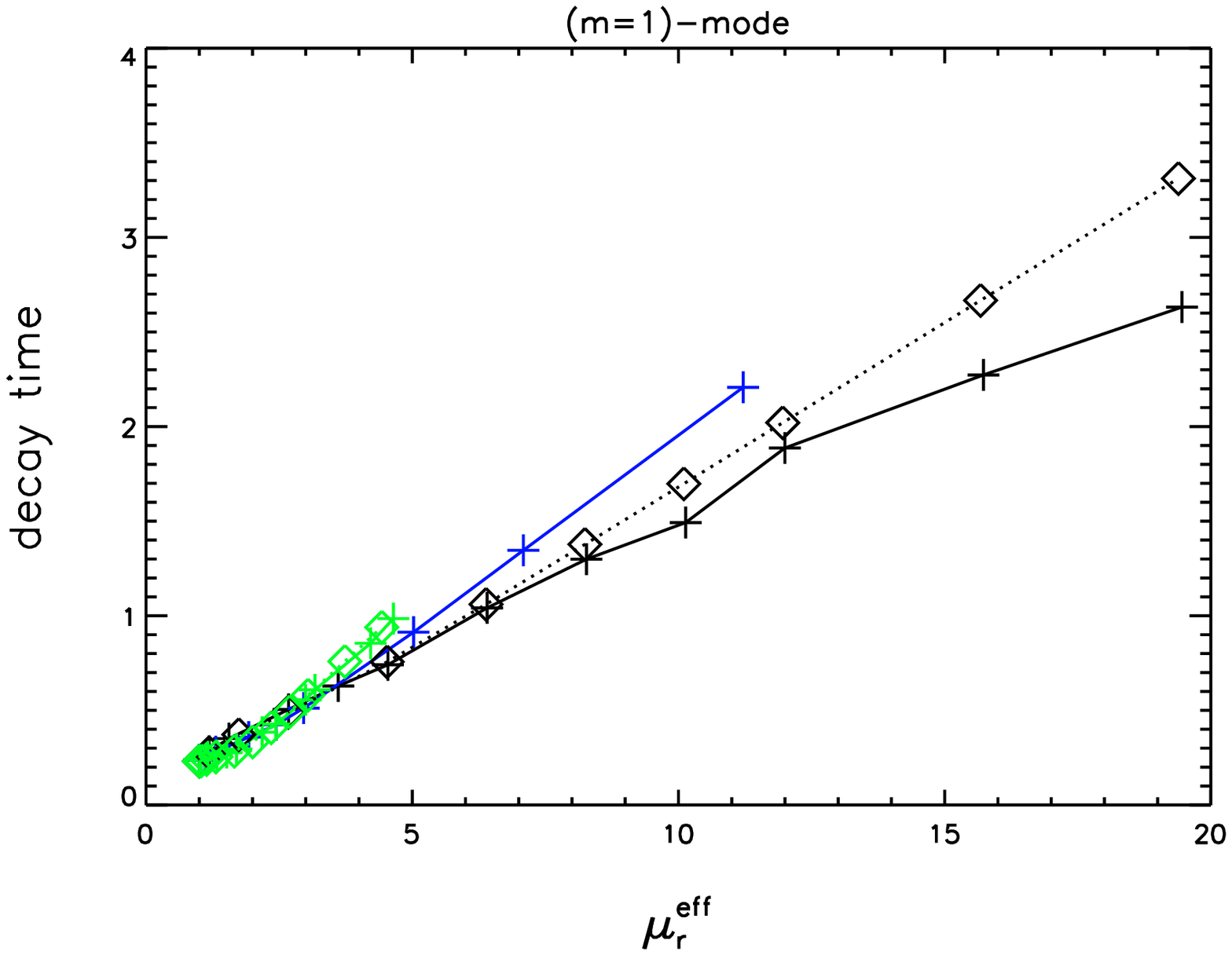}
\\
\includegraphics[width=6cm]{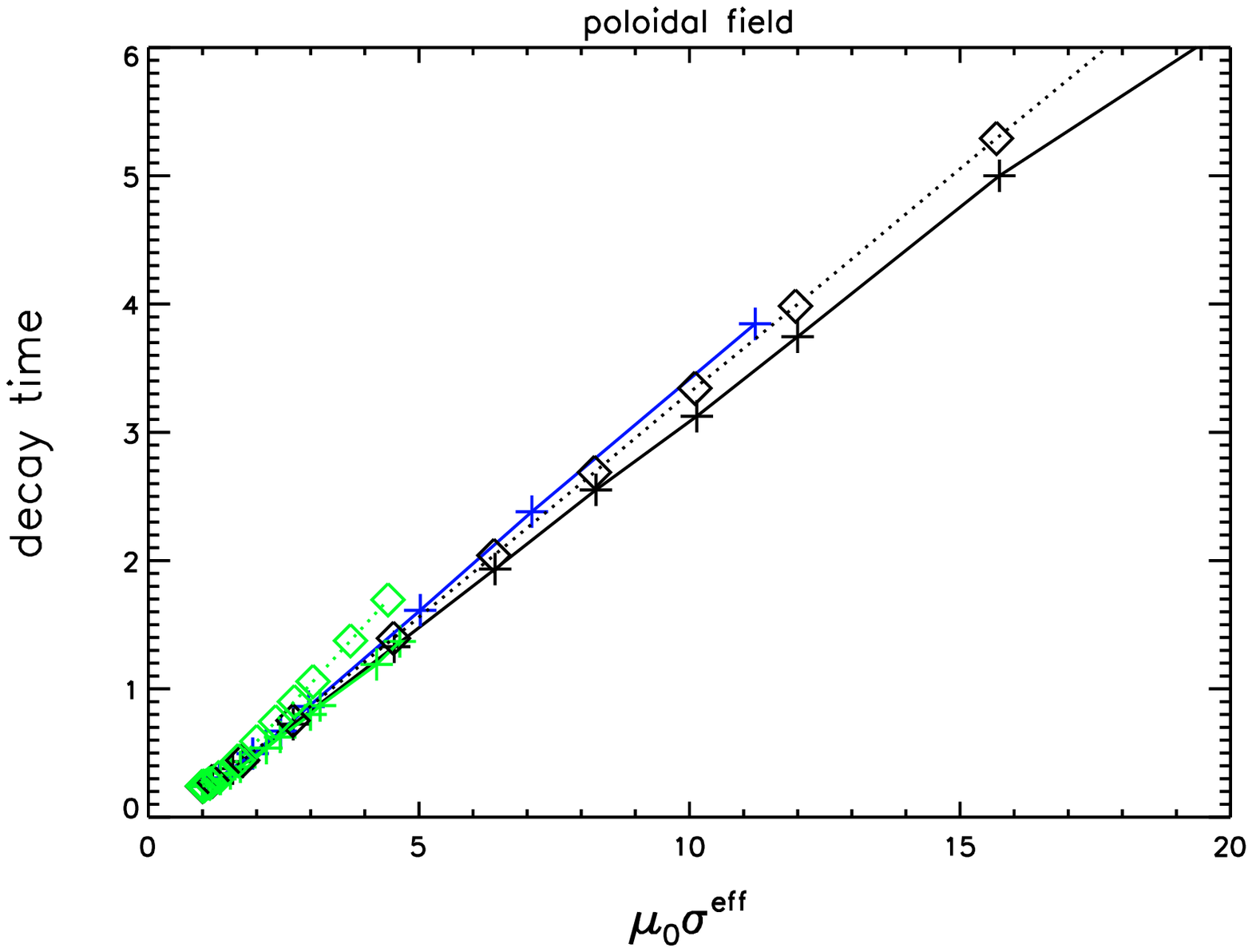}
\includegraphics[width=6cm]{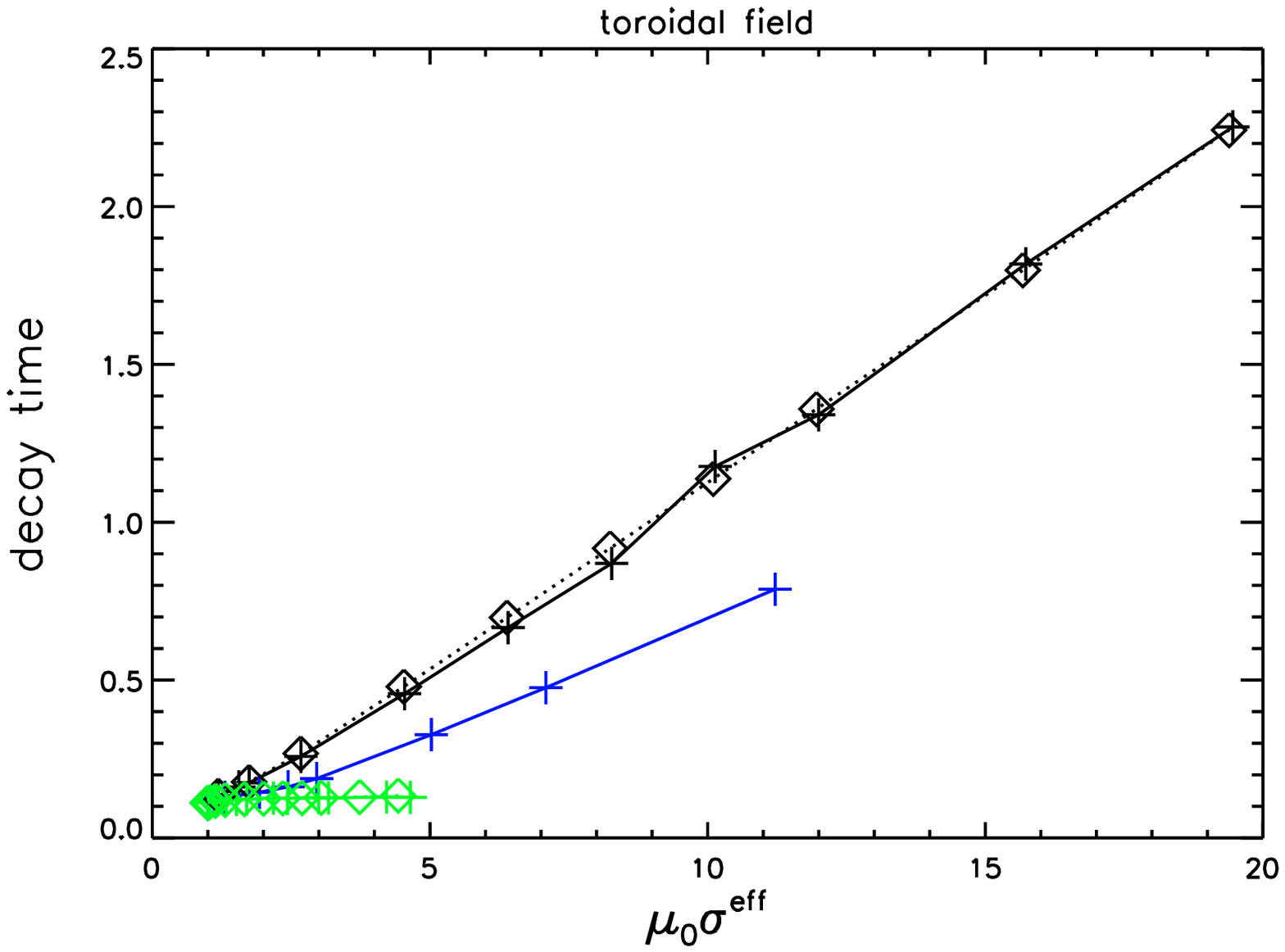}
\includegraphics[width=6cm]{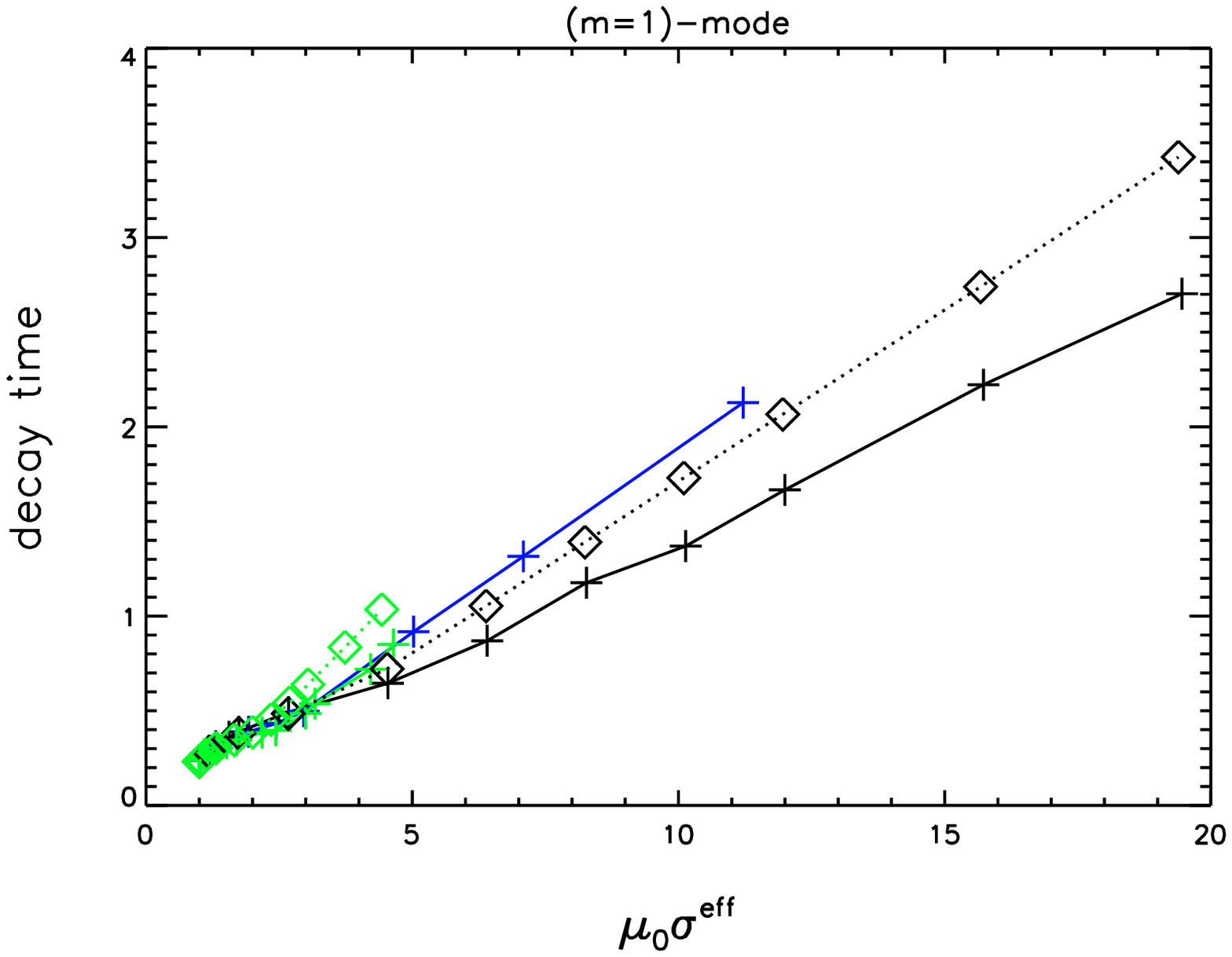}
\caption{Ohmic decay. Decay times against $\mu_{\rm{r}}^{\rm{eff}}$ (top row) and against $\mu_0\sigma^{\rm{eff}}$
(bottom row) for three disk thicknesses $d=0.6, 0.3, 0.1$ (black, blue, green).
The solid curves show the results obtained from the hybrid FV/BEM scheme and
the dotted curves denote the results from the SFEMaNS scheme.
}
\label{fig::decay_times_vac}
\end{figure}
A more systematic discrepancy between both algorithms becomes obvious by means
of the behavior of the decay time $\tau$ defined by the reciprocal value of the decay
rate (see Fig.~\ref{fig::decay_times_vac}).
For sufficient large $\mu_{\rm{r}}^{\rm{eff}}$ (respectively $\sigma^{\rm{eff}}$), 
$\tau$ varies following a scaling law $\tau \propto c
\mu^{\rm{eff}}_{\rm{r}}$ (respectively $\propto c \sigma^{\rm{eff}}$)
as reported in table~\ref{tab:vary_mu_sigma}.
For increasing $\mu_{\rm{r}}^{\rm{eff}}$ the decay time of the $(m=0)$
toroidal mode slightly increases with decreasing $d$ whereas
the axisymmetric poloidal mode exhibits an opposite behavior.
The variation of the decay time with $\sigma^{\rm{eff}}$ for the $(m=0)$
components (toroidal and poloidal) is the opposite to the ones with varying $\mu_{\rm{r}}^{\rm{eff}}$.
These variations suggest that the decay time
scaling law is not only due to the ferromagnetic volume
of the impellers but also to the geometric constraints associated
with the jump conditions~(\ref{eq::jumpconditions}).

\begin{figure}[h!]
\includegraphics[width=8.2cm]{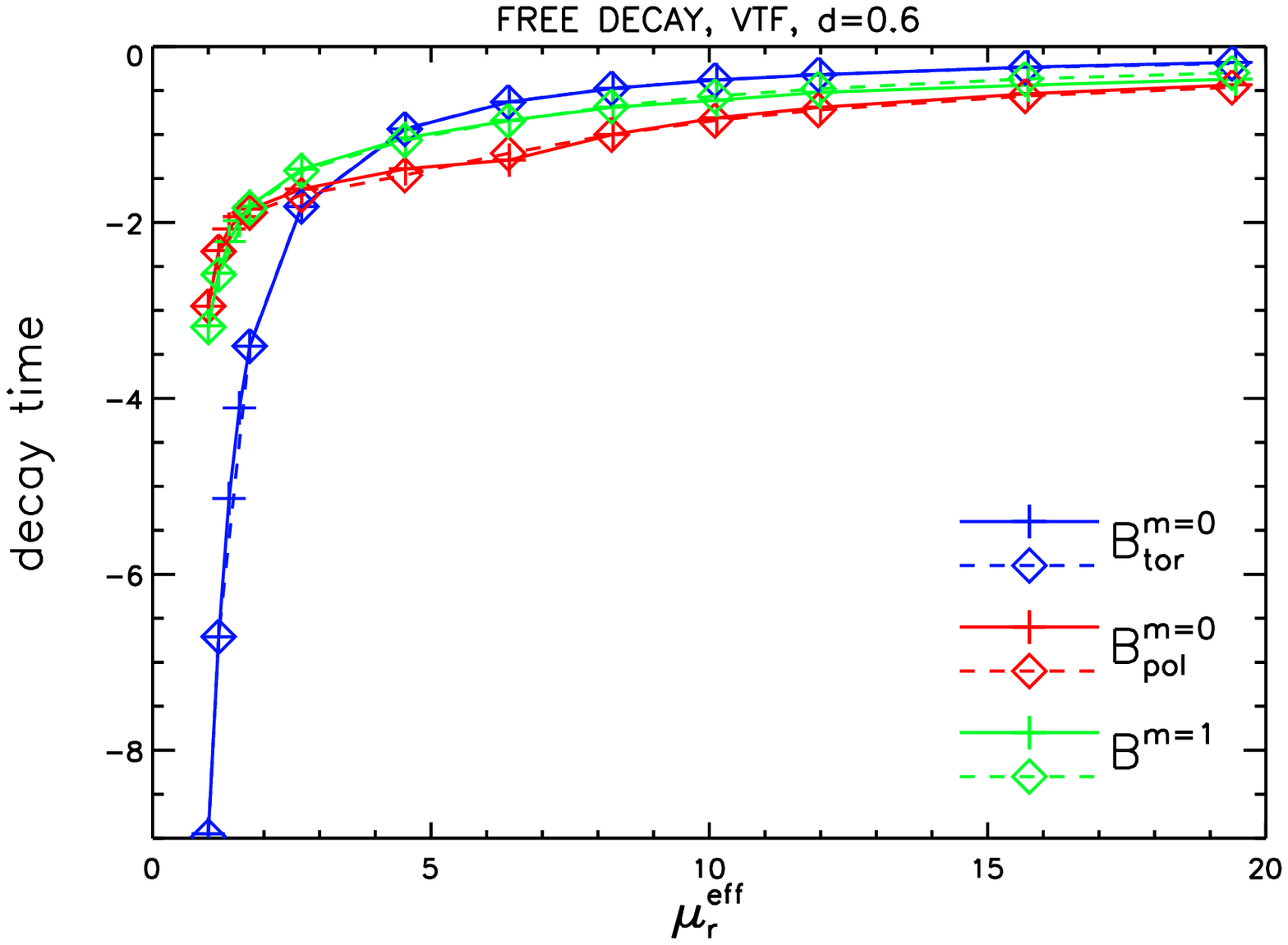}
\includegraphics[width=8.2cm]{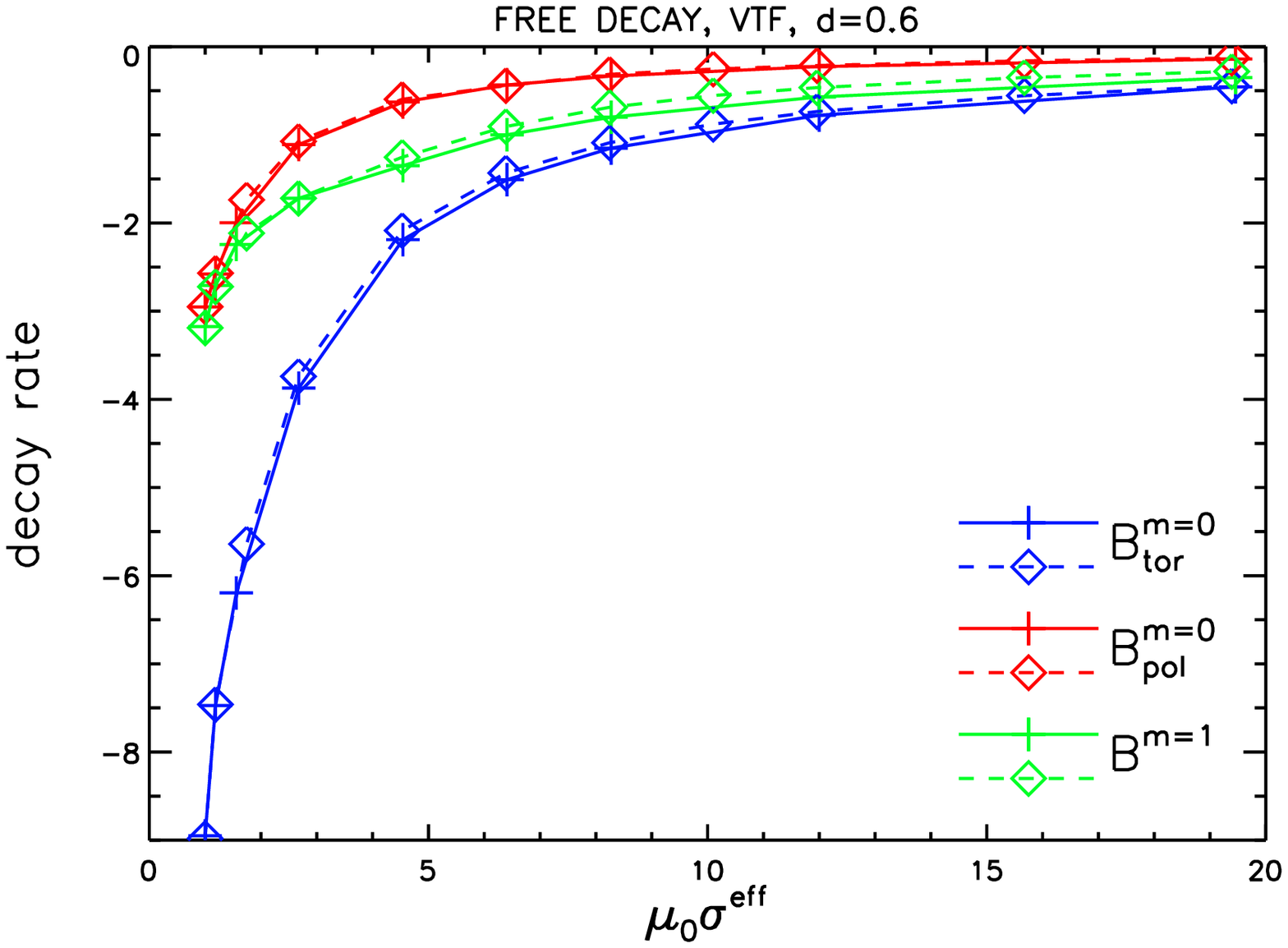}
\\
\includegraphics[width=8.2cm]{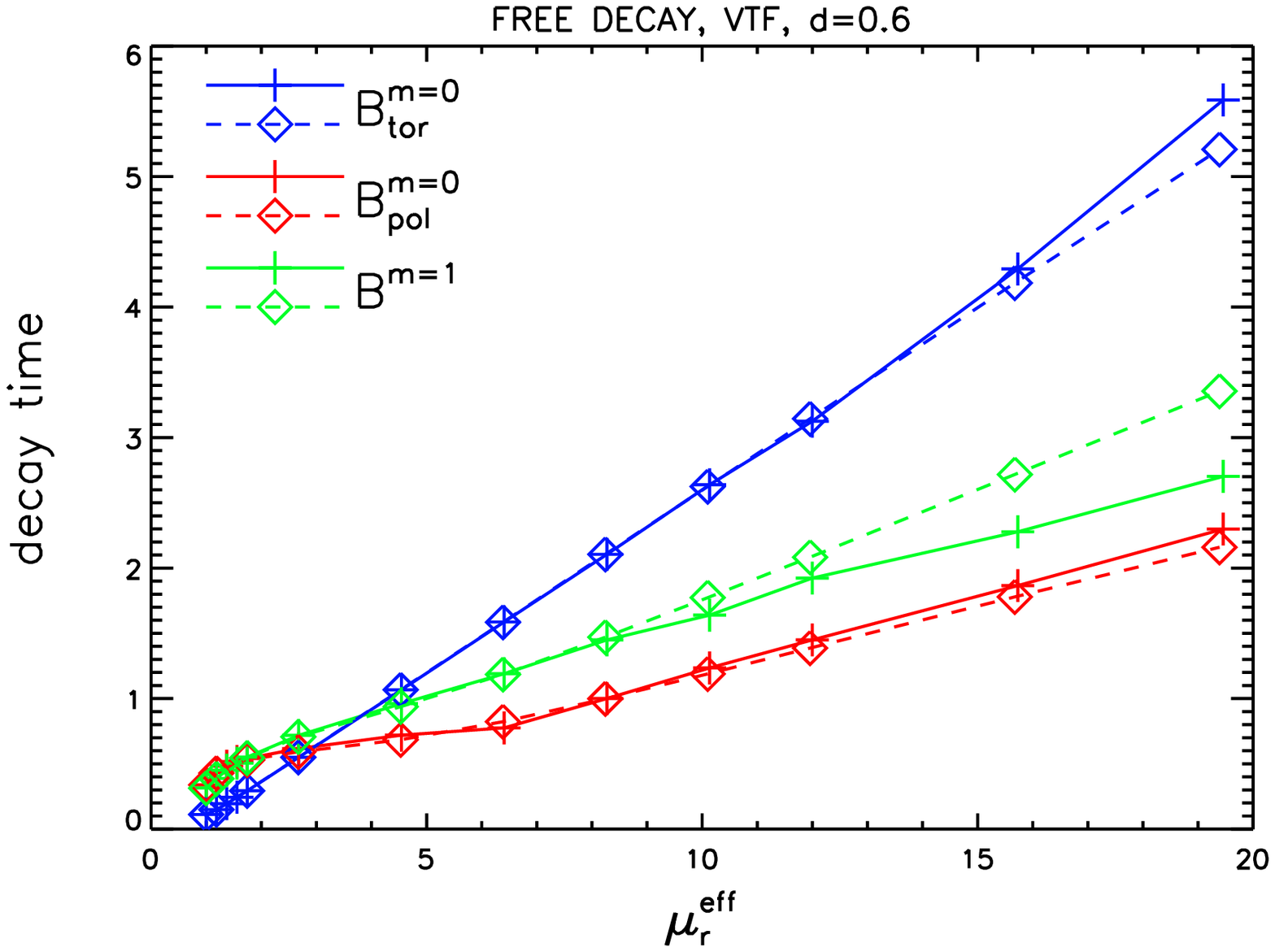}
\includegraphics[width=8.2cm]{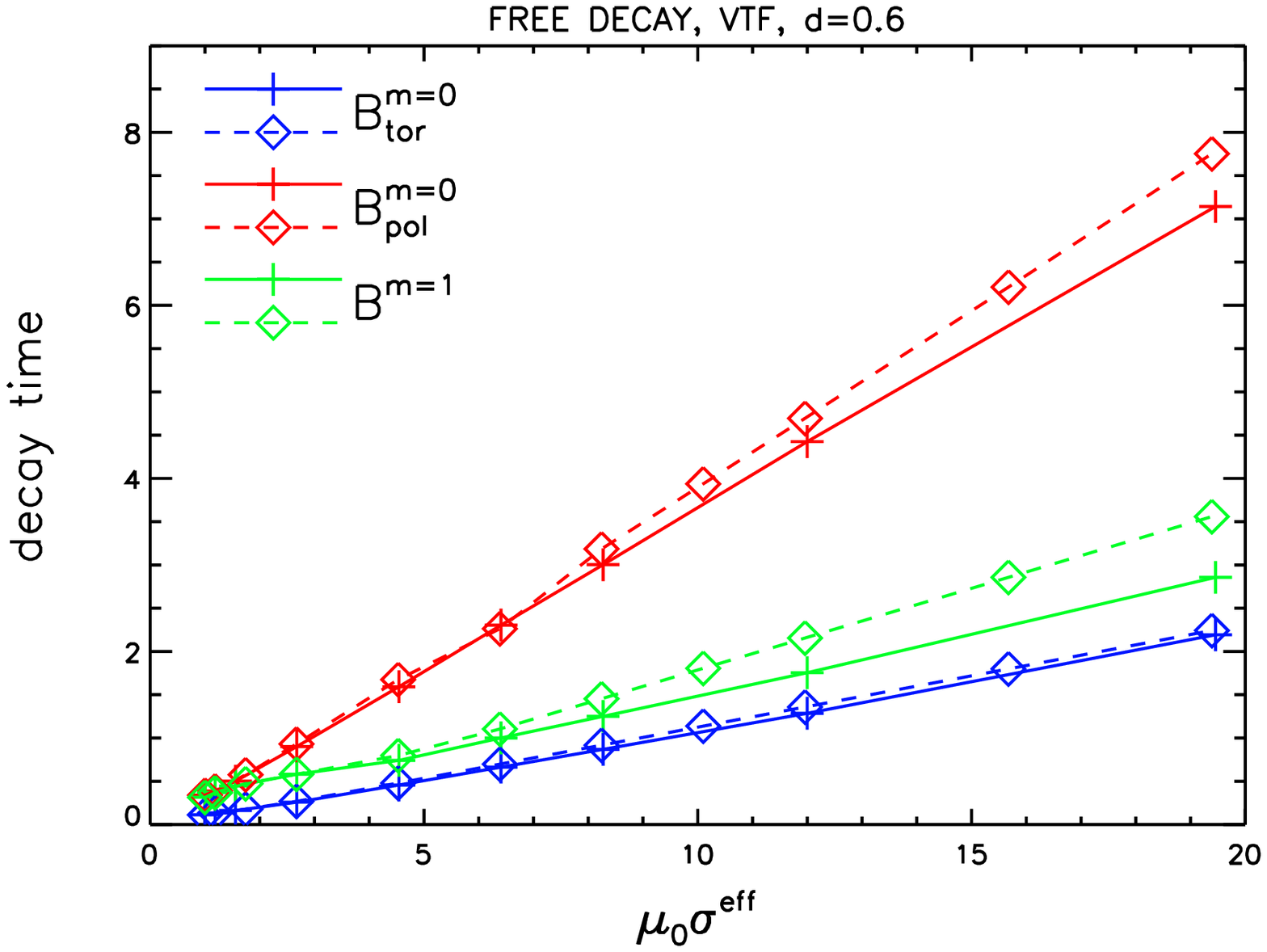}
\caption{Decay rates and decay times against ${\mu_{\rm{r}}^{\rm{eff}}}$ (left
  column) and against 
  $\mu_0\sigma^{\rm{eff}}$ (right column) for vanishing tangential fields
  boundary conditions. $d=0.6$ The solid (dashed) curves denote the results
  from the FV/BEM (SFEMaNS) scheme.}\label{fig::decay_rates_vtf}
 \end{figure}

\begin{figure}[h!]
\includegraphics[width=8.2cm]{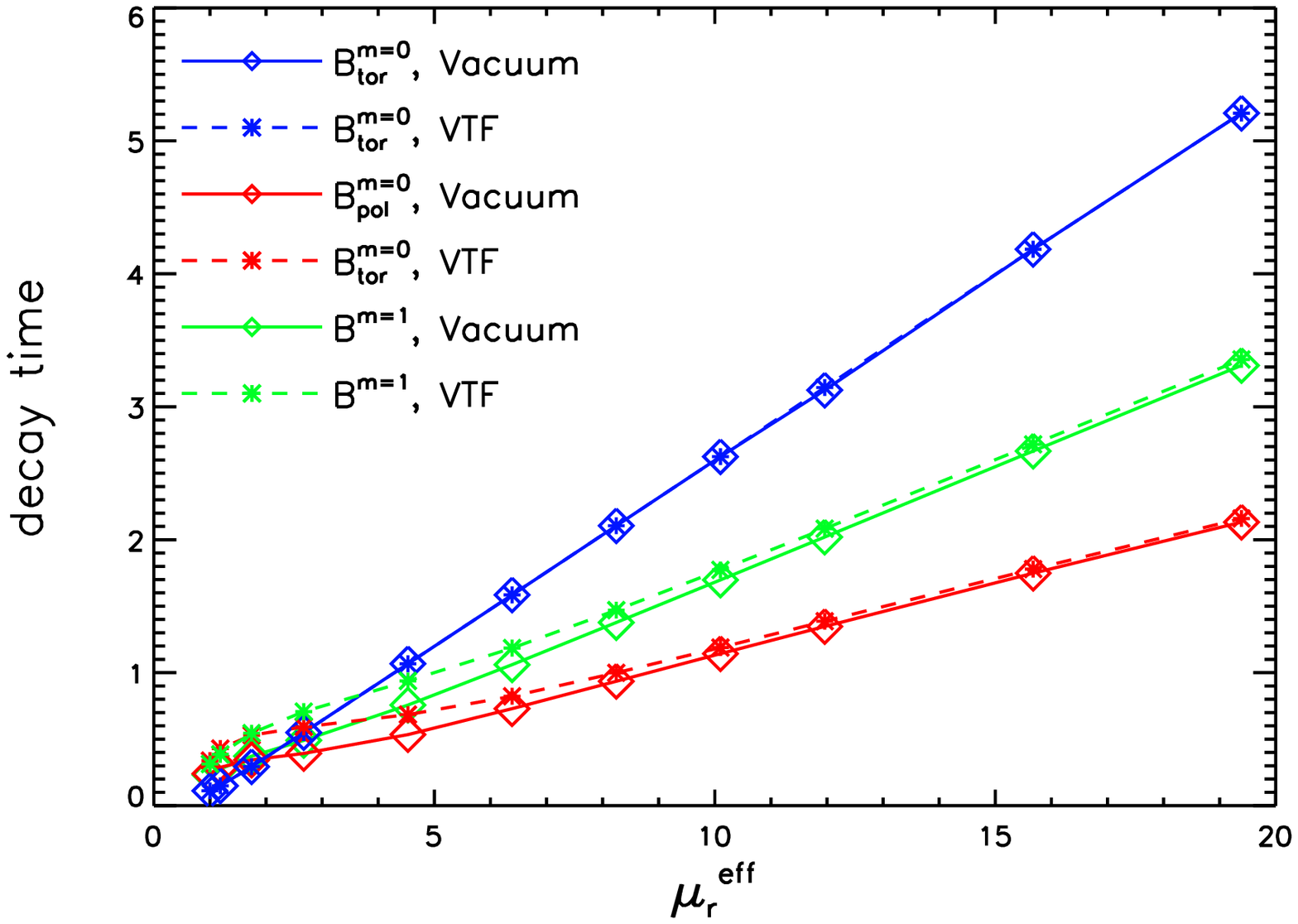}
\includegraphics[width=8.2cm]{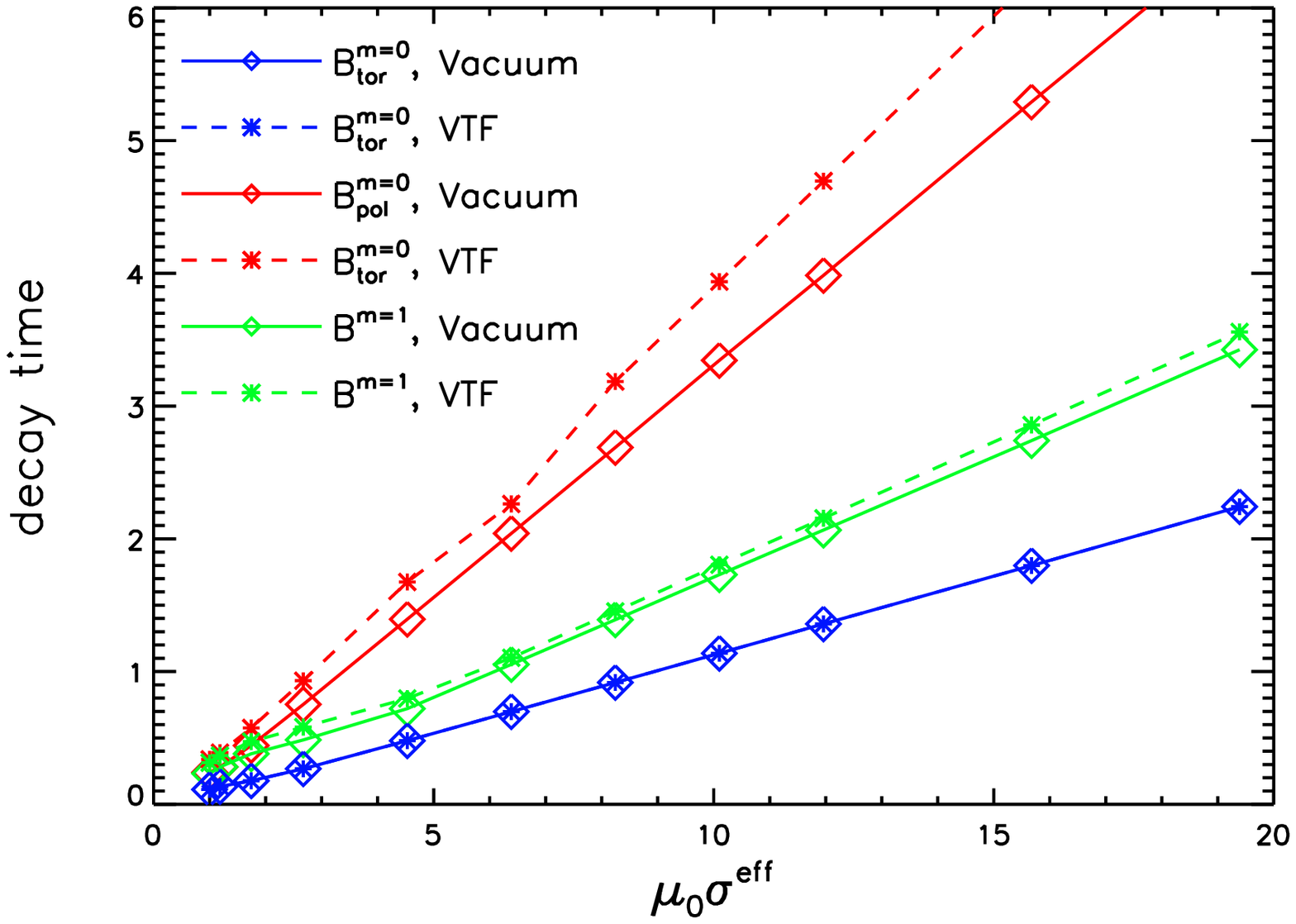}
\caption{Comparison of boundary conditions. Decay times against ${\mu_{\rm{r}}^{\rm{eff}}}$ (left
  panel) and against 
  $\mu_0\sigma^{\rm{eff}}$ (right panel) for vacuum BC (solid
  curves) and VTF boundary conditions (dashed curves). $d=0.6$. All data results
  from the SFEMaNS scheme.}\label{fig::comp_decay_times}
 \end{figure}

The evaluation of the discrepancies in the scaling behavior obtained by both
numerical schemes remains difficult
because this would require larger values for $\mu_{\rm{r}}^{\rm{eff}}$ and/or
$\sigma^{\rm{eff}}$ which is not
possible without significantly improving the numerical schemes. 
In particular for the thin disk case ($d=0.1$) the achievable values for
$\mu_{\rm{r}}$ and/or $\sigma$ are 
restricted to $\mu_{\rm{r}}^{\rm{eff}}$ (respectively $\mu_0\sigma^{\rm{eff}}$)
$\la 5$ and, it is not obvious if the available data already belongs to the
region that follows a linear scaling. 
In any case the absolute
values for the decay rates obtained by both algorithms are close, giving
confidence that the results imply a sufficient accurate description of the
magnetic field behavior in the presence of non-heterogenous materials.
\begin{center}
\begin{table}[!h]
\begin{tabular}{r|r|r|r||r|r|r|l}
& \multicolumn{3}{|c||}{$\mu_{\rm{r}}^{\rm{eff}}$} & \multicolumn{3}{|c|}{$\sigma^{\rm{eff}}$}&\\
\hline
$d$ & 0.6 & 0.3 & 0.1 & 0.6 & 0.3 & 0.1 & {\bf{Algorithm}}\\
\hline
$\tau(B^{\rm{tor}}_{m=0})$ & 0.29 & 0.32 & 0.33  & 0.12 & 0.07 & 0.00 & FV/BEM\\
& 0.28 &   -- & 0.34  & 0.12 & --   & 0.00 & SFEMaNS\\
\hline
$\tau(B^{\rm{pol}}_{m=0})$ & 0.12 & 0.08 & 0.00  & 0.32 & 0.36 & 0.33 & FV/BEM\\ 
& 0.11 &   -- & 0.00  & 0.35 & --   & 0.45 & SFEMaNS\\
\hline
$\tau(B_{m=1})$            & 0.12 & 0.21 & 0.25  & 0.14 & 0.20 & 0.20 & FV/BEM\\
& 0.17 & --   & 0.25  & 0.18 & --   & 0.28 & SFEMaNS\\ 
\end{tabular}
\caption{Scaling coefficient $c$ for the decay time as $\tau \propto c
  \mu^{\rm{eff}}_{\rm{r}}$ (respectively  $c\mu_0\sigma^{\rm{eff}}$)
for different $m=0$ and $m=1$ modes as indicated (vacuum BC).}
\label{tab:vary_mu_sigma}
\end{table}
\end{center}
As already indicated by the marginal differences in the field pattern for both
examined boundary conditions, we find no qualitative change in the behavior of
the decay rates or decay times with vacuum boundary conditions or 
VTF boundary conditions (see Fig.~\ref{fig::decay_rates_vtf}).  
Although for small values of $\mu_{\rm{r}}^{\rm{eff}}$ and $\sigma^{\rm{eff}}$
the absolute values of the decay rates differ by 30\% the scaling behavior of the decay time is
nearly independent of the external boundary conditions (see Tab.~\ref{tab:vary_mu_vtf}). 
For increasing $\mu_{\rm{r}}$ the influence of these boundary conditions is
further reduced. Whereas the decay rates (for the thick disks) differ by
approximately 30\% for $\mu_{\rm{r}}\la 5$ there are nearly no differences in
$\gamma$ for higher values  of the permeability. This behavior is less obvious
in case of a high conductivity disk where the
poloidal axisymmetric field exhibits differences in the decay rates of 15\%
even at the highest available conductivity (see Fig.~\ref{fig::comp_decay_times}). 
Note that the axisymmetric toroidal field behaves exactly in the same way for
both kinds of boundary conditions because insulating boundary conditions and
vanishing tangential field conditions are identical for the axisymmetric part
of $B_{\varphi}$. 
\begin{center}
\begin{table}[!h]
\begin{tabular}{r|r||r|l}
& \multicolumn{1}{|c||}{$\mu_{\rm{r}}^{\rm{eff}}$} & \multicolumn{1}{|c|}{$\sigma^{\rm{eff}}$}&\\
\hline
$\tau(B^{\rm{tor}}_{m=0})$ & 0.29  & 0.12 & FV/BEM VTF\\
& 0.28  & 0.12  & SFEMaNS VTF\\
\hline
$\tau(B^{\rm{pol}}_{m=0})$ & 0.12 & 0.37 & FV/BEM VTF\\
& 0.10 & 0.42 & SFEMaNS VTF\\
\hline
$\tau(B_{m=1})$ & 0.11 & 0.14 & FV/BEM VTF\\
& 0.17 & 0.19 & SFEMaNS VTF\\
\end{tabular}
\caption{Scaling coefficient $c$ for the decay time as $\tau \propto c
  \mu^{\rm{eff}}_{\rm{r}}$ 
for thick disks ($d=0.6$) and VTF boundary conditions.}
\label{tab:vary_mu_vtf}
\end{table}
\end{center}
\section{Kinematic Dynamo}\label{sec::mnd}
\begin{figure}[h!]
\includegraphics[height=8cm]{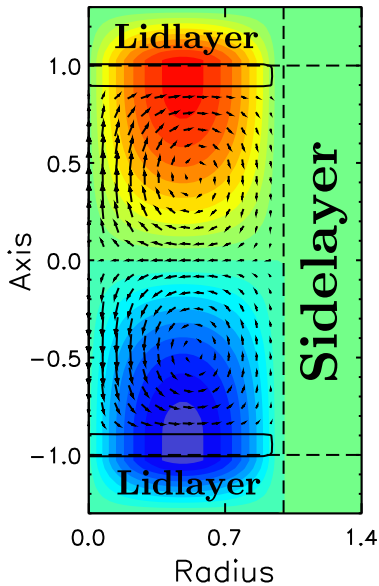}
\caption{Structure of the prescribed axisymmetric velocity field. The color
  coded pattern represents the azimuthal velocity and the arrows show the
  poloidal velocity field. The black solid lines represent the shape of the
  impeller disk.}\label{fig::velfield}
\end{figure}

In the following, the kinematic induction equation is solved numerically with
${\rm{Rm}}> 0$ in order to examine if the behavior of the magnetic field
obtained in the free decay is maintained when interaction with a mean flow is allowed. 
With reference to the VKS experiment we apply the so called MND-flow
\citep{2004phfl} given by
\begin{eqnarray}
u_r(r,z)&=&-(\pi/H) \cos\!\left({{2\pi z/H}}\right)r(1-r)^2(1+2r),\nonumber\\
u_{\varphi}(r,z)&=&4\epsilon r(1-r)\sin\left({{\pi z/H}}\right),\label{eq::s2t2}\\
u_z(r,z)&=&(1-r)(1+r-5r^2)\sin\left({{2\pi z/H}}\right)\nonumber,
\end{eqnarray}
where $H=1.8$ denotes the distance between both impeller disks and
$\epsilon$ describes the relation between toroidal and poloidal component of
the velocity (here, $\epsilon=0.7259$ is chosen following previous work,
e.g. \citeauthor{2006EJMF...25..894S} \citeyear{2006EJMF...25..894S}). 
The flow magnitude is characterized by the magnetic Reynolds number which is
defined as
\begin{equation}
{\rm{Rm}}=\displaystyle\mu_0\sigma{U_{\rm{max}}R},
\end{equation}
where $U_{\rm{max}}$ is the maximum of the flow velocity and $\sigma$ denotes
the fluid conductivity.
Figure~\ref{fig::velfield} shows the structure of the velocity field
where Eqs.~(\ref{eq::s2t2}) are only applied in the region between the two
impellers.
The flow active region with radius $R=1$ (corresponding to
$20.5 \mbox{ cm}$ in the experiment) is surrounded by a layer of stagnant fluid with
a thickness of $0.4R$ (the side layer) which significantly reduces
${\rm{Rm}}^{\rm{c}}$ (\citeauthor{2006EJMF...25..894S} \citeyear{2006EJMF...25..894S}).
In the domain of the impellers a purely azimuthal velocity is assumed given by
the azimuthal velocity of the MND flow (Eq.~\ref{eq::s2t2}) at $z=\pm H/2$.
Behind each impeller disk a so called lid layer is added.  
Within these lid layers a purely rotating flow is assumed, modeled by a linear
interpolation of the azimuthal velocity at the outer 
side of the impeller disk towards to zero at the end cap of the cylindrical domain.
Similar to the simulations of free decay two disks are inserted into the
computational domain (see solid black lines in Fig.~\ref{fig::velfield}).
Here we limit our examinations to disks with a height $d=0.1$.
Note that the impellers are modeled only by the permeability and/or
conductivity distribution and no particular flow boundary conditions are
enforced on the (assumed) interface between impeller and fluid.
This setup is comparable to the configuration in \cite{PhysRevLett.104.044503}
except that the non axisymmetric permeability contribution representing the blade
structure has been dropped in the present study.

\begin{figure}[h!]
\includegraphics[height=6cm]{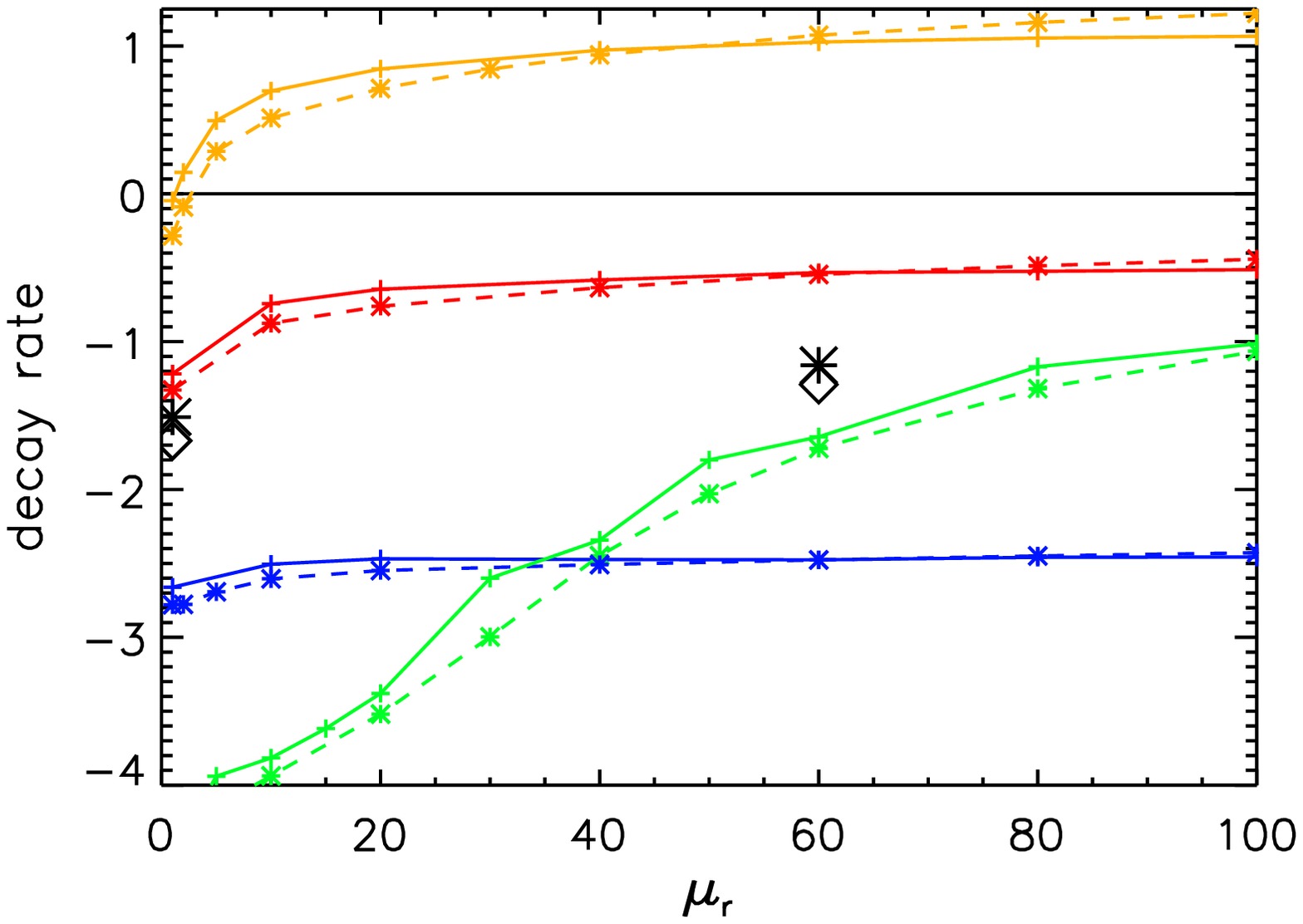}
\includegraphics[height=6cm]{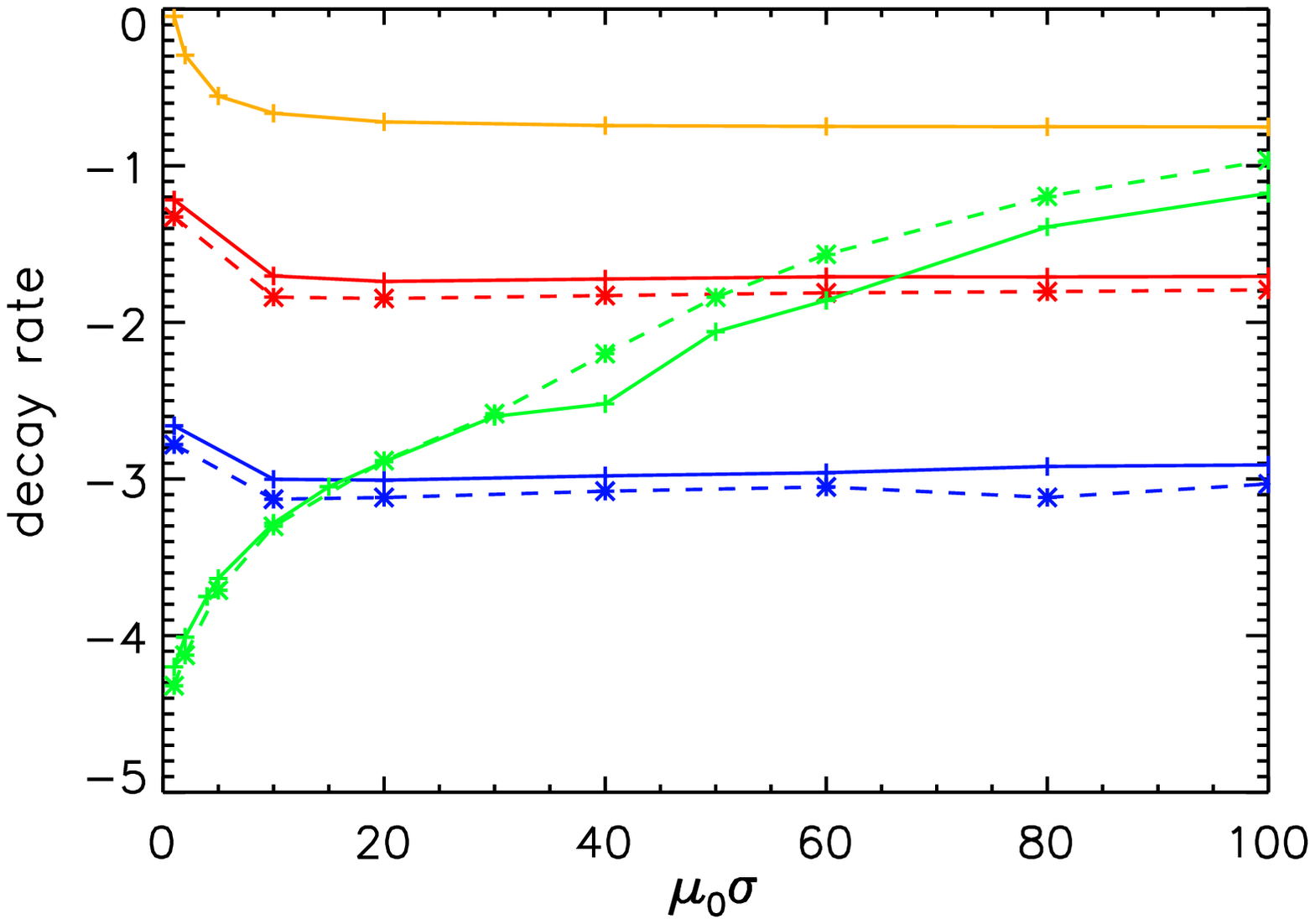}
\caption{Growth rates for the MND flow driven dynamo against $\mu_{\rm{r}}$
  (left panel) and against $\mu_0\sigma$ (right panel). Solid curves denote
  data obtained from the FV/BEM scheme, dashed curves denote the results from
  the SFEMaNS scheme. The green, blue, red, yellow colors denote the cases
  ${\rm{Rm}}=0, 30, 50, 70$. The black stars in the left panel show the results for the SMND
  flow at ${\rm{Rm}}=50$ (see
  text) as reported in Tab.~\ref{tab::m1_MND}.}\label{fig::mnd_gr}
\end{figure}

Figure~\ref{fig::mnd_gr} shows the growth rates for the
$(m=1)$ mode for different magnetic Reynolds numbers.
Compared to the free decay, we obtain a remarkable distinct behavior of the
growth rate if induction from a mean flow is added. 
A high permeability disk with ${\rm{Rm}}>0$ causes an enhancement of the
$(m=1)$ mode compared to the case $\mu_{\rm{r}}=1$.
The shift of the growth rate increases with increasing
${\rm{Rm}}$ resulting in a non-negligible impact on the critical magnetic
Reynolds number for the onset of dynamo action.
In order to get insight in the experimental values necessary to get dynamo
action, we have computed different thresholds for the $(m=1)$ mode:
${\rm{Rm}}^{\rm{c}}$ is reduced from around 76 at $\mu_{\rm{r}}=1$ to
${\rm{Rm}}^{\rm{c}}$ around $55$ at $\mu_{\rm{r}}=100$. 
%
%
%
%
The behavior of ${\rm{Rm}}^{\rm{c}}$ indicates a saturation around
${\rm{Rm}}^{\rm{c}}\approx 55$ for $\mu_{\rm{r}}\gg 1$ which still lies above
the experimental achievable value of approx 50. 
With respect to the Ohmic decay (${\rm{Rm}}=0$) the $(m=1)$-mode is clearly suppressed
with increasing $\mu_{\rm{r}}$ (see green curve in Fig.~\ref{fig::mnd_gr}).

For a conducting disk we obtain a reduction of the ($m=1$) growth rate.
In both cases the $(m=1)$ decay rate remains independent of $\mu_{\rm{r}}$
(respectively $\sigma$) for values exceeding approximately
$\mu_{\rm{r}}\approx 20$ (or $\mu_0\sigma \approx 20$).
%
%
%
The critical magnetic Reynolds number has also been computed for a different
set-up with the fluid restricted to the bulk region : $0 \le r \le 1.4, -0.9 \le z
\le 0.9$ with VTF conditions applied on the frontier of this region which
results in ${\rm{Rm}}^{\rm{c}} = 39$. Note
that this pseudo-vacuum set-up under-estimates the threshold by more than 30\%. 
%
%
This confirms that a
realistic description of the soft iron impellers is crucial to get correct
estimates.

The robustness of the solutions suffers from the rather delicate dependence of
the field behavior on the details of the flow distribution, in particular from
the flow in the lid layers.
Beside the well known dynamo killing influence of the lid flow
\citep{2006EJMF...25..894S} this is also true for the radial flow in the vicinity of
the inner side of the disks. 
A couple of simulations have been performed applying a slightly different
velocity field where the radial component is smoothed at the transition
between the bulk of the domain and the impeller disk (where $u_r=0$).
The resulting decay rates (black stars in the left panel of
Fig.~\ref{fig::mnd_gr} and Tab.~\ref{tab::m1_MND}) exhibit slight 
differences in case of $\mu_{\rm{r}}=1$ and a more moderate enhancement of the $(m=1)$
- mode for $\mu_{\rm{r}}=60$. 
\begin{center}
\begin{table}[!h]
\begin{math}
\begin{array}{|r|r|r|r|r|}
\hline
m=1, Rm=50 & \mu_{\rm{r}}=1 (\mbox{FV/BEM}) & \mu_{\rm{r}}=1 (\mbox{SFEMaNS}) & 
\mu_{\rm{r}}=60 (\mbox{FV/BEM}) & \mu_{\rm{r}}=60 (\mbox{SFEMaNS}) \\
\hline
MND & -1.218 &  -1.327 & -0.550 & -0.655 \\
SMND & -1.51 & -1.667 & -1.16 & -1.291 \\
\hline
\end{array}
\end{math}
\caption{Decay rate for $m=1$ mode for 2 flows MND and a similar flow with
  slightly modified (smoothed) radial velocity component (SMND).}
\label{tab::m1_MND}
\end{table}
\end{center}
\section{Conclusions}
Although soft iron is strongly connected to magnetostatics, experimental
dynamos have shown that this material may  also find important applications in
the field of  magnetohydrodynamics. 
For instance, at least one of the two
impellers of the Cadarache experiment must be made of soft iron in order to
achieve dynamo action. This is an  unexplained fact which suggests that one
may wonder if the role of this material is only to lower the critical magnetic
Reynolds number in the domain of experimental feasibility or if the dynamo
mechanism is fundamentally different  when the conducting medium is no longer
homogenous. 
This issue may be faced in principle numerically.
However, to face such problems with heterogenous domains, specific algorithms
must be implemented and validated and this is the aim of the present study
since analytical results are lacking.   
Our comparative runs of Ohmic decay problems proved in practice to be
extremely useful to optimize both codes and to select some numerical
coefficients occurring in the algorithms (such as in penalty terms). 

The problems which have been successively presented above are standard in MHD, 
but we were forced to reduce the dimension of the parameter space to
configurations more or less related to the Cadarache experiment, where the
impellers may be treated as disks in a conducting flow bounded by a cylinder
of a given aspect ratio. 
We have thus considered axisymmetric domains only 
(see \citet{PhysRevLett.104.044503} for non-axisymmetric cases), and
azimuthal modes of low order $(m=0$ and $1)$. 

We have first studied Ohmic decay  problems, with disk impellers of various
thicknesses to investigate scaling laws and the impact of the spatial
resolution.  
Internal assemblies of high permeability material within the fluid container
are different from the problem of an enhanced, but homogenous fluid
permeability because of inner boundary conditions for the magnetic field (in
case of high permeability material), respectively for the electric
field/current (in case of conductivity jumps).
In the free decay problem with thin high permeability disks a selective
enhancement of the axisymmetric toroidal field 
and the $(m=1)$ mode is observed whereas the axisymmetric poloidal field
component is preferred in case of high conductive disks. 

We have also shown that pseudo-vacuum boundary conditions, which are  easier
to implement on the cylinder walls than the jump conditions on the impellers,
have only a slight influence on the decay rates. 
The impact of the outer container boundaries on the field behavior
is limited to a shift of the decay/growth rates. This is surprising, insofar as
pseudo vacuum boundary conditions resemble the conditions that correspond to
an external material with infinite permeability.
Nevertheless, the presence of high permeability/conductivity disks within the liquid occlude
the influence of outer boundary conditions, and the simplifying approach
applying vanishing tangential field conditions at the end 
caps of the cylinder in order to mimic the effects of the high permeability
disks in the VKS experiment is not sufficient to describe the correct field
behavior. 
The consideration of lid layers and disks with (large but finite)
permeability remains indispensable in order to obtain the influence of the
material properties on growth rates as well as on the field geometry.

For completeness, we have also considered conductivity domains. 
From the experimental point of view the utilization of disks with a
conductivity that is 100 times larger than the conductivity of liquid
sodium remains purely academic.  
Nevertheless, the simulations show a crucial difference between heterogeneous permeabilities
and conductivities: even if these two quantities may appear in the definition
of an effective Reynolds number ${\rm{Rm}}^{\rm{eff}} = \mu_0 \mu^{\rm{eff}}_{\rm{r}}
\sigma^{\rm{eff}} U L$,  
they do not play the same role and they select different
geometries of the dominant decaying mode. It is not only a change of magnetic diffusivity
that matters.

We considered then kinematic dynamo action, using analytically defined
flows in accordance with the setting of the VKS mean flow.
Since these flows are axisymmetric, the azimuthal modes are decoupled. The most
important is the ($m=1$) mode which will be excited eventually through dynamo
action. We have shown that  our codes give comparable growth rates for this
mode.  
We have examined also the growth rate of the ($m=0$) magnetic field in presence of
soft iron impellers and the axisymmetric  MND flow. 
Since convergence of results is not achieved in all the cases considered, this
comparative study is 
still in progress and it has thus not been included in the present
paper. 
We recall that the main surprise of the Cadarache experiment was
perhaps the occurrence of the mode ($m=0$), which pointed out the possible role of
the non-axisymmetric flow fluctuations.  
%
%
Non-axisymmetric velocity contributions might be considered in terms of an $\alpha$-effect as
it has been proposed in \citet{2007GApFD.101..289P} and
\citet{2008PhRvL.101j4501L, 2008PhRvL.101u9902L}. Preliminary examinations 
applying simple $\alpha$-distributions are presented in \citet{andregafd} and
\citet{PhysRevLett.104.044503}. 
However, there is still a lack of knowledge on
the details and physical justification on a precise $\alpha$-distribution
which requires a non-linear hydrodynamic code.   
The questions related to this
empirical fact represent a main issue of the experimental  and numerical
approaches of the fluid dynamo problem and deserve a dedicated study. 
%
%
%
%
%
So far our model is
not capable to explain the main features of the VKS experiment, which are the
dominating axisymmetric field mode and the surprising low critical
magnetic Reynolds number of ${\rm{Rm}}\approx 32$. 
However, our results give a hint why the $(m=1)$ mode remains
absent in the experiment.
Dynamo action may occur when coupling terms between the magnetic field
components are present and antidynamo theorems are derived when such terms are
lacking: this is in particular the case for the ($m=0$) mode with an axisymmetric
flow (Cowling's theorem). 
Conversely, a source term on this mode appears when
the flow axisymmetry is broken. 
Although the relative amplitude of this source
cannot be discussed here, we note that the decay time of the $(m=0)$ toroidal mode
become the smallest when the effective permeability is high enough (see for
example Fig.~\ref{fig::decay_rates_vac}). 
It may thus appears as the dominant mode of the dynamo, as it
seems to be observed in the VKS experiment. 
Otherwise stated, the impact of
soft-iron impellers on the critical magnetic Reynolds number of the ($m=1$)-mode
could be rather low (decrease from $\sim 76$ to $\sim 55$ in the MND case) and could remain
unobservable, while it could be strong for the ($m=0$) mode (down to 32 in the VKS
geometry) when conjugated to a slight departure from flow
axisymmetry. 
Numerical evidences of this picture are growing.
\section*{Acknowledgments}
Financial support from Deutsche Forschungsgemeinschaft (DFG) in frame of the
Collaborative Research Center (SFB) 609 is gratefully acknowledged and from
European Commission under contract 028679.
The computations using SFEMaNS were carried
out on the IBM SP6 computer of Institut du D\'eveloppement et des
Ressources en Informatique Scientifique (IDRIS) (project \# 0254).
\bibliographystyle{gGAF} 

\begin{thebibliography}{26}
\providecommand{\natexlab}[1]{#1}

\bibitem[\protect\citeauthoryear{Bonito {\itshape{et~al.}}}{2010}]{AB_JLG_FL}
Bonito, A., Guermond, J.L. and Luddens, F., Approximation of the Eigenvalue
  Problem for Time Harmonic Maxwell System by Continuous Lagrange Finite
  Elements. {\itshape Math. Comp.} 2010 Under review.

\bibitem[\protect\citeauthoryear{{Busse} and
  {Wicht}}{1992}]{1992gafd...64..135B}
{Busse}, F.H. and {Wicht}, J., {A simple dynamo caused by conductivity
  variations}. {\itshape Geophys.\ Astrophys.\ Fluid\ Dyn.} 1992, \textbf{64},
  135--144.

\bibitem[\protect\citeauthoryear{Costabel}{1991}]{Costabel_1991}
Costabel, M., A coercive bilinear form for {M}axwell's equations. {\itshape J.
  Math. Anal. Appl.} 1991, \textbf{157}, 527--541.

\bibitem[\protect\citeauthoryear{{Dobler}
  {\itshape{et~al.}}}{2003}]{2003PhRvE..67e6309D}
{Dobler}, W., {Frick}, P. and {Stepanov}, R., {Screw dynamo in a time-dependent
  pipe flow}. {\itshape \pre} 2003, \textbf{67}, 056309--+.

\bibitem[\protect\citeauthoryear{{Frick}
  {\itshape{et~al.}}}{2002}]{2002EPJB...25..399F}
{Frick}, P., {Khripchenko}, S., {Denisov}, S., {Sokoloff}, D. and {Pinton},
  J.F., {Effective magnetic permeability of a turbulent fluid with
  macroferroparticles}. {\itshape Eur. Phys. J. B} 2002,
  \textbf{25}, 399--402.

\bibitem[\protect\citeauthoryear{{Giesecke}
  {\itshape{et~al.}}}{2008}]{2008giesecke_maghyd}
{Giesecke}, A., {Stefani}, F. and {Gerbeth}, G., {Kinematic simulations of
  dynamo action with a hybrid boundary-element/finite-volume method}. {\itshape
  Magnetohydrodynamics} 2008, \textbf{44}, 237--252.

\bibitem[\protect\citeauthoryear{{Giesecke}
  {\itshape{et~al.}}}{2010{\natexlab{a}}}]{andregafd}
{Giesecke}, A., {Nore}, C., {Plunian}, F., {Laguerre}, R., {Ribeiro}, A.,
  {Stefani}, F., {Gerbeth}, G., {Leorat}, J. and {Guermond}, J., {Generation of
  axisymmetric modes in cylindrical kinematic mean-field dynamos of VKS type}.
  {\itshape Geophys.\ Astrophys.\ Fluid\ Dyn.} 2010{\natexlab{a}}, \textbf{104}, 249--271.

\bibitem[\protect\citeauthoryear{Giesecke
  {\itshape{et~al.}}}{2010{\natexlab{b}}}]{PhysRevLett.104.044503}
Giesecke, A., Stefani, F. and Gerbeth, G., Role of Soft-Iron Impellers on the
  Mode Selection in the von K\'arm\'an--Sodium Dynamo Experiment. {\itshape
  Phys. Rev. Lett.} 2010{\natexlab{b}}, \textbf{104}, 044503.

\bibitem[\protect\citeauthoryear{Guermond {\itshape{et~al.}}}{2009}]{GLLN09}
Guermond, J.L., Laguerre, R., L{\'e}orat, J. and Nore, Nonlinear
  magnetohydrodynamics in axisymmetric heterogeneous domains using a
  Fourier/Finite Element technique and an Interior Penalty Method. {\itshape J.
  Comput. Phys.} 2009, \textbf{228}, 2739--2757.

\bibitem[\protect\citeauthoryear{Guermond {\itshape{et~al.}}}{2007}]{MR2290574}
Guermond, J.L., Laguerre, R., L{\'e}orat, J. and Nore, C., An interior penalty
  {G}alerkin method for the {MHD} equations in heterogeneous domains. {\itshape
  J. Comput. Phys.} 2007, \textbf{221}, 349--369.

\bibitem[\protect\citeauthoryear{{Haber} and
  {Ascher}}{2001}]{2001siam...22..1943H}
{Haber}, E. and {Ascher}, U.M., {Fast Finite Volume Simulation of 3d
  electromagnetic problems with highly discontinuous coefficients}. {\itshape
  SIAM J. Sci. Comput.} 2001, \textbf{22}, 1943--1961.


\bibitem[\protect\citeauthoryear{{Iskakov} and
  {Dormy}}{2005}]{2005GApFD..99..481I}
{Iskakov}, A.B. and {Dormy}, E., {On magnetic boundary conditions for
  non-spectral dynamo simulations}. {\itshape Geophys.\ Astrophys.\ Fluid\
  Dyn.} 2005, \textbf{99}, 481--492.

\bibitem[\protect\citeauthoryear{{Iskakov}
  {\itshape{et~al.}}}{2004}]{2004JCoPh.197..540I}
{Iskakov}, A.B., {Descombes}, S. and {Dormy}, E., {An integro-differential
  formulation for magnetic induction in bounded domains: boundary
  element-finite volume method}. {\itshape \jcp} 2004, \textbf{197}, 540--554.

\bibitem[\protect\citeauthoryear{{Jackson}}{1975}]{1975clel.book.....J}
{Jackson}, J.D., {\itshape {Classical electrodynamics}},  1975 (New
  York: Wiley, 1975, 2nd ed.).

\bibitem[\protect\citeauthoryear{{Laguerre}
  {\itshape{et~al.}}}{2008{\natexlab{a}}}]{2008PhRvL.101j4501L}
{Laguerre}, R., {Nore}, C., {Ribeiro}, A., {L{\'e}orat}, J., {Guermond}, J. and
  {Plunian}, F., {Impact of Impellers on the Axisymmetric Magnetic Mode in the
  VKS2 Dynamo Experiment}. {\itshape Physical Review Letters}
  2008{\natexlab{a}}, \textbf{101}, 104501--+.

\bibitem[\protect\citeauthoryear{{Laguerre}
  {\itshape{et~al.}}}{2008{\natexlab{b}}}]{2008PhRvL.101u9902L}
{Laguerre}, R., {Nore}, C., {Ribeiro}, A., {L{\'e}orat}, J., {Guermond}, J. and
  {Plunian}, F., {Erratum: Impact of Impellers on the Axisymmetric Magnetic
  Mode in the VKS2 Dynamo Experiment [Phys. Rev. Lett. 101, 104501 (2008)]}.
  {\itshape Phys. Rev. Lett.} 2008{\natexlab{b}}, \textbf{101},
  219902--+.

\bibitem[\protect\citeauthoryear{{Lowes} and
  {Wilkinson}}{1963}]{1963Natur.198.1158L}
{Lowes}, F.J. and {Wilkinson}, I., {Geomagnetic Dynamo: A Laboratory Model}.
  {\itshape \nat} 1963, \textbf{198}, 1158--1160.

\bibitem[\protect\citeauthoryear{{Lowes} and
  {Wilkinson}}{1968}]{1968Natur.219..717L}
{Lowes}, F.J. and {Wilkinson}, I., {Geomagnetic Dynamo: An Improved Laboratory
  Model}. {\itshape \nat} 1968, \textbf{219}, 717--718.

\bibitem[\protect\citeauthoryear{{Mari{\'e}}
  {\itshape{et~al.}}}{2006}]{2004phfl}
{Mari{\'e}}, L., {Normand}, C. and {Daviaud}, F., {Galerkin analysis of
  kinematic dynamos in the von K{\'a}rm{\'a}n geometry}. {\itshape Phys.
  Fluids} 2006, \textbf{18}, 017102--+.

\bibitem[\protect\citeauthoryear{{Monchaux {\it{et
  al.}}}}{2007}]{2007PhRvL..98d4502M}
{Monchaux {\it{et al.}}}, R., {Generation of a Magnetic Field by Dynamo Action
  in a Turbulent Flow of Liquid Sodium}. {\itshape Phys. Rev. Lett.} 2007,
  \textbf{98}, 044502.

\bibitem[\protect\citeauthoryear{{P{\'e}tr{\'e}lis}
  {\itshape{et~al.}}}{2007}]{2007GApFD.101..289P}
{P{\'e}tr{\'e}lis}, F., {Mordant}, N. and {Fauve}, S., {On the magnetic fields
  generated by experimental dynamos}. {\itshape Geophys.\ Astrophys.\ Fluid\
  Dyn.} 2007, \textbf{101}, 289--323.

\bibitem[\protect\citeauthoryear{{Stefani}
  {\itshape{et~al.}}}{2006}]{2006EJMF...25..894S}
{Stefani}, F., {Xu}, M., {Gerbeth}, G., {Ravelet}, F., {Chiffaudel}, A.,
  {Daviaud}, F. and {Leorat}, J., {Ambivalent effects of added layers on steady
  kinematic dynamos in cylindrical geometry: application to the VKS
  experiment}. {\itshape Eur. J. Mech. B} 2006, \textbf{25}, 894--908.

\bibitem[\protect\citeauthoryear{{Stone} and
  {Norman}}{1992{\natexlab{a}}}]{1992ApJS...80..753S}
{Stone}, J.M. and {Norman}, M.L., {ZEUS-2D: A radiation magnetohydrodynamics
  code for astrophysical flows in two space dimensions. I - The hydrodynamic
  algorithms and tests.}. {\itshape \apjs} 1992{\natexlab{a}}, \textbf{80},
  753--790.

\bibitem[\protect\citeauthoryear{{Stone} and
  {Norman}}{1992{\natexlab{b}}}]{1992ApJS...80..791S}
{Stone}, J.M. and {Norman}, M.L., {ZEUS-2D: A Radiation Magnetohydrodynamics
  Code for Astrophysical Flows in Two Space Dimensions. II. The
  Magnetohydrodynamic Algorithms and Tests}. {\itshape \apjs}
  1992{\natexlab{b}}, \textbf{80}, 791--+.

\bibitem[\protect\citeauthoryear{{Verhille}
  {\itshape{et~al.}}}{2010}]{2010NJPh...12c3006V}
{Verhille}, G., {Plihon}, N., {Bourgoin}, M., {Odier}, P. and {Pinton}, J.,
  {Induction in a von K{\'a}rm{\'a}n flow driven by ferromagnetic impellers}.
  {\itshape New J. Phys.} 2010, \textbf{12}, 033006--+.

\bibitem[\protect\citeauthoryear{{Ziegler}}{1999}]{1999CoPhC.116...65Z}
{Ziegler}, U., {A three-dimensional Cartesian adaptive mesh code for
  compressible magnetohydrodynamics}. {\itshape Comp. Phys.
  Comm.} 1999, \textbf{116}, 65--77.

\end{thebibliography}

\end{document}